\documentclass[]{spie}  
\usepackage[]{graphicx}
\usepackage{color}
\usepackage{cancel}
\usepackage{amsmath}
\usepackage{mathtools}
\usepackage{verbatim}
\definecolor{linkcolor}{cmyk}{1, 0.65, 0, 0.3}			
\definecolor{citecolor}{cmyk}{0.79, 0, 0.87, 0.56}		

\usepackage[
	colorlinks,
	linkcolor=linkcolor,
	urlcolor=linkcolor,
	citecolor=linkcolor,
	pdftex,
	pdfstartview=FitH,
]{hyperref}

\usepackage[all]{hypcap}

\usepackage{footnote}

\usepackage[font={small,it}]{caption}

\renewcommand{\figurename}{Figure}

\usepackage{listings}

\usepackage{helvet}

\newcommand\smallurl[1]{{\small{\url{#1}}}}
\newcommand\tablespace{\vspace{2.5mm}}
\newcommand\fig{supplementary figure }
\newcommand\Fig{Supplementary figure }
\newcommand\F{\textbf{Fig. }}
\newcommand\SF{\textbf{Suppl. Fig. }}
\newcommand\ST{\textbf{Suppl. Table }}
\newcommand\SN{\textbf{Suppl. Note Chapter }}
\newcommand\B{\textbf}
\title{Efficient Bayesian-based Multi-View Deconvolution} 

\author{Stephan Preibisch\supit{1,2,3,}*, Fernando Amat\supit{2}, Evangelia Stamataki\supit{1},\\ Mihail Sarov\supit{1}, Robert H. Singer\supit{2,3}, Gene Myers\supit{1,2} and Pavel Tomancak\supit{1,}*
\skiplinehalf
\small{
\supit{1}Max Planck Institute of Molecular Cell Biology and Genetics, Pfotenhauerstrasse 108, Dresden, Germany \\
\supit{2}HHMI Janelia Farm Research Campus, 19700 Helix Drive, Ashburn, VA 20147, USA \\
\supit{3}Department of Anatomy and Structural Biology, Gruss Lipper Biophotonics Center, Albert Einstein College of Medicine, Bronx, NY 10461, USA
}
}

\authorinfo{* Correspondence should be addressed to: preibischs@janelia.hhmi.org and tomancak@mpi-cbg.de}

\begin{document} 

\maketitle 

\setcounter{page}{1}
\pagenumbering{roman}
\pagenumbering{arabic}

\abstract

Light sheet fluorescence microscopy is able to image large specimen with high resolution by imaging the samples from multiple angles. Multi-view deconvolution can significantly improve the resolution and contrast of the images, but its application has been limited due to the large size of the datasets. Here we present a Bayesian-based derivation of multi-view deconvolution that drastically improves the convergence time and provide a fast implementation utilizing graphics hardware.

\section*{MAIN DOCUMENT}

\begin{figure*}[h!]
\includegraphics[width=\textwidth]{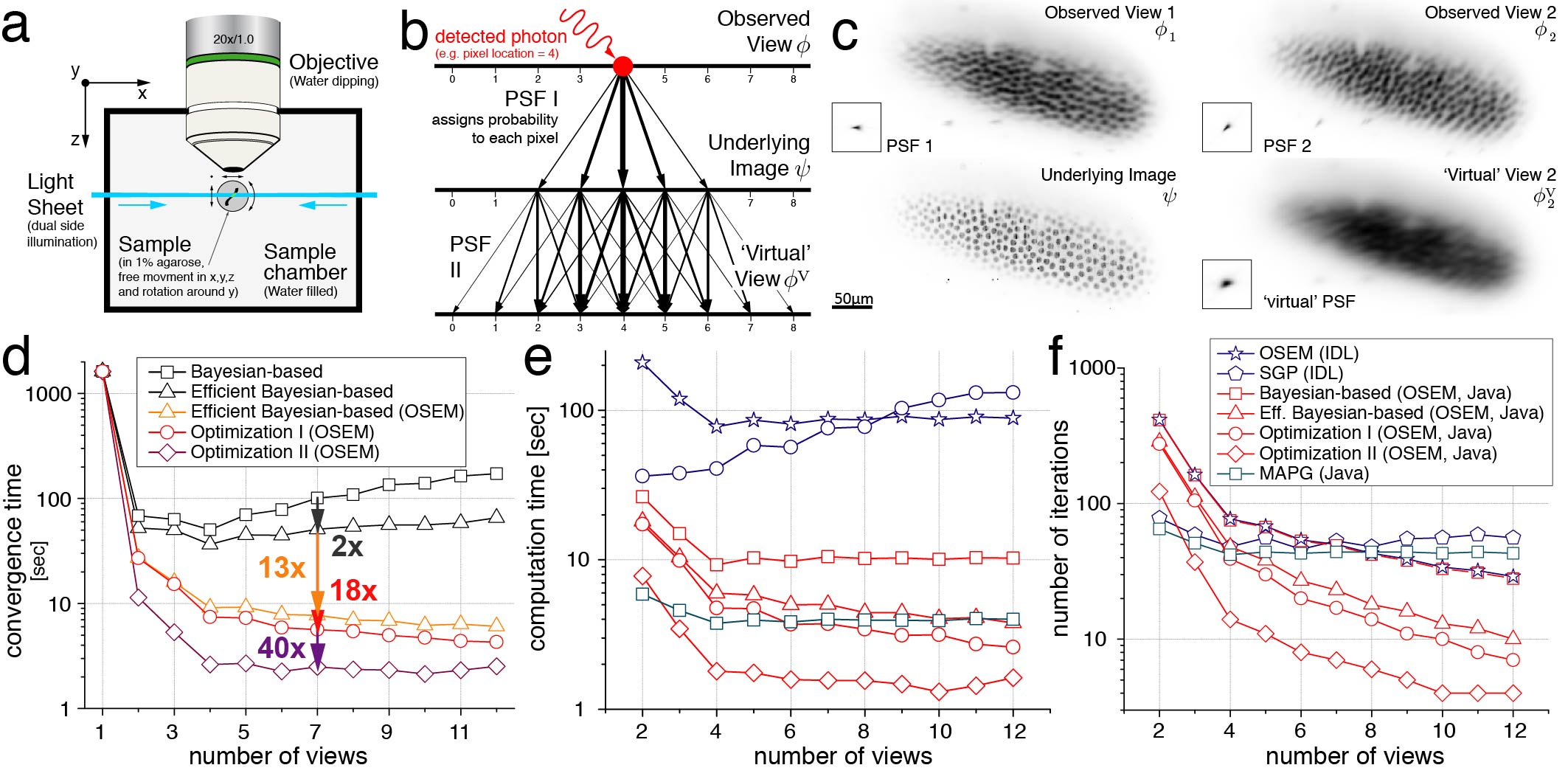}
\vspace{-2.0mm}
\caption{\hspace{-0.5mm} \emph{Principles and performance.} (\textbf{a}) The basic layout of a light sheet microscope capable of multi-view acquisitions. (\textbf{b}) Illustrates the idea of ‘virtual views’. A photon detected at a certain location in a view was emitted by a fluorophore in the sample; the point-spread function assigns a probability to every location in the underlying image having emitted that photon. Consecutively, the point-spread function of any other view assigns to each of its own locations the probability to detect a photon corresponding to the same fluorophore. (\textbf{c}) Shows an example of an entire ‘virtual view’ computed from observed view 1 and the knowledge of PSF1 and PSF 2. (\textbf{d}) Compares the convergence time of the different Bayesian-based methods. We used a known ground truth image (\SF \B{\ref{fig:viewsimages}}) and let all variations converge until they reach precisely the same quality. Note that the increase in computation time for an increasing number of views of the combined methods (black) is due to the fact that with an increasing number of views more computational effort is required to perform one update of the deconvolved image (\SF \B{\ref{fig:views}}) (\textbf{e}) Compares the convergence times for the same ground truth image of our Bayesian-based methods to other optimized multi-view deconvolution algorithms\cite{Shepp1982,Hudson1994,Verveer2007,Bonetti2009}. Note that part of the huge difference to OSEM and SGP is the result of not optimized IDL code. (\textbf{f}) Compares the corresponding number of iterations in comparison to other optimized multi-view deconvolution algorithms. Note that the Java and IDL implementation of OSEM perform almost identically. 
}
\label{fig:main1}
\end{figure*}

Modern light sheet microscopes\cite{HuiskenAl2004,keller2008,Truong2011} are able to acquire large, developing specimens with high temporal and spatial resolution typically by imaging them from multiple directions (\F \B{\ref{fig:main1}a}). The low photodamage offered by a light sheet microscope’s design allows the recording of massive, time-lapse datasets that have the potential to enable the reconstruction of entire lineage trees of the developing specimen. However, accurate segmentation and tracking of nuclei and cells in these datasets remain a challenge because image quality is limited by the optical properties of the imaging system and the compromises between acquisition speed and resolution. Deconvolution utilizes knowledge about the optical system to substantially increase spatial resolution and contrast after acquisition. An advantage unique to light sheet microscopy and in particular to Selective Plane Illumination Microscopy (SPIM), is the ability to observe the same location in the specimen from multiple angles which renders the ill-posed problem of deconvolution more tractable\cite{Swoger2007,Shepp1982,Hudson1994,Verveer2007,Bonetti2009,KrzicPhD,Temerinac2012}.

Richardson-Lucy (RL) deconvolution\cite{richardson1972,lucy1974} (\SN \B{\ref{sec:remarks}, \ref{sec:singleview}}) is a Bayesian-based derivation resulting in an iterative expectation-maximization (EM) algorithm\cite{Shepp1982,Dempster77} that is often chosen for its simplicity and performance. Multi-view deconvolution has previously been derived using the EM framework\cite{Shepp1982,KrzicPhD,Temerinac2012}, however the convergence time of the algorithm remains orders of magnitude longer than the time required to record the data. We address this problem by deriving an optimized formulation of Bayesian-based deconvolution for multiple view geometry that explicitly incorporates conditional probabilities between the views (\F \B{\ref{fig:main1}b,c}) and combine it with Ordered Subset Expectation Maximization (OSEM)\cite{Hudson1994} (\F \B{\ref{fig:main1}d}) achieving significantly faster convergence \mbox{(\F \B{\ref{fig:main1}d,e,f})}. 

Bayesian-based deconvolution models images and point spread functions (PSFs) as probability distributions. The goal is to estimate the most probable underlying distribution (deconvolved image) that explains best all observed distributions (views) given their conditional probabilities (PSFs). We first re-derived the original Richardson-Lucy deconvolution algorithm and subsequently extended it to multiple-view geometry yielding

\begin{eqnarray}
\label{eq:main1}
 f_{RL} & = & \int_{x_v} \frac{ \phi_v(x_v) }{\int_\xi \psi^r(\xi) P(x_v|\xi) d\xi} P(x_v|\xi) dx_v\\
\label{eq:main2}
 \psi^{r+1}(\xi) & = & \psi^r(\xi) \displaystyle\prod_{v \in V} f_{RL}
\end{eqnarray}

where $\psi(\xi)$ denotes the deconvolved image at iteration $r$, $\phi_v(x_v)$ the input views, both as functions of their respective pixel locations $\xi$ and $x_v$ , while $P(x_v|\xi)$ denotes the individual PSFs (\SN \B{\ref{sec:singleview}, \ref{sec:multiview}}). Equation \ref{eq:main1} denotes a classical RL update step for one view; equation \ref{eq:main2} illustrates the combination of all views into one update of the deconvolved image. Our equation suggests a multiplicative combination, in contrast to maximum-likelihood expectation-maximation\cite{Shepp1982} that combines RL updates by addition. We prove that equation \ref{eq:main2} also converges to the Maximum-Likelihood (ML) solution (\SN \B{\ref{sec:proofconvergence}}), while it is important to note that the ML solution is not necessarily the correct solution if disturbances like noise or misalignments are present in the input images. Importantly, previous extensions to multiple views\cite{Shepp1982,Hudson1994,Verveer2007,Bonetti2009,KrzicPhD,Temerinac2012} are based on the assumption that the individual views are independent observations (\SF \B{\ref{fig:principle-virtual-osem}}). Assuming independence between two views implies that by observing one view, nothing can be learned about the other view. We show that this independence assumption is not required to derive equation \ref{eq:main2}. Thus our solution represents the first complete derivation of Richardson-Lucy multi-view deconvolution based on probability theory and Bayes’ theorem.

As we do not need to consider views to be independent, we next asked if the conditional probabilities describing the relationship between two views can be modeled and used in order to improve convergence behavior (\SN \B{\ref{sec:efficientmv}}). Assuming that a single photon is observed in the first view, the PSF of this view and Bayes’ theorem can be used to assign a probability to every location in the deconvolved image having emitted this photon (\F \B{\ref{fig:main1}b}). Based on this probability distribution, the PSF of the second view directly yields the probability distribution describing where to expect a corresponding observation for the same fluorophore in the second view (\F \B{\ref{fig:main1}b}). Following this reasoning, we argue that it is possible to compute an approximate image (‘virtual’ view) of one view from another view provided that the PSF’s of both views are known (\F \B{\ref{fig:main1}c}).

We use these ‘virtual’ views to perform intermediate update steps at no additional computational cost, decreasing the computational effort approximately 2-fold (\F \B{\ref{fig:main1}d}) and \SN \B{\ref{sec:efficientmv}}). The multiplicative combination (equation \ref{eq:main2}) directly suggests a sequential approach, where each RL update (equation \ref{eq:main1}) is directly applied to $\psi(\xi)$ (\SF \B{\ref{fig:principle-virtual-osem}}). This sequential scheme is equivalent to the OSEM\cite{Hudson1994} algorithm and results in a 13-fold decrease in convergence time. This gain increases linearly with the number of views\cite{Hudson1994} (\F \B{\ref{fig:main1}d} and \SF \B{\ref{fig:views}}). The new algorithm also performs well in the presence of noise and imperfect point spread functions (\SF \B{\ref{fig:noiseplot},\ref{fig:snr},\ref{fig:psf}}). To further reduce convergence time we introduce ad-hoc simplifications (optimization I \& II) for the estimation of conditional probabilities that achieve up to 40-fold improvement compared to deconvolution methods that assume view independence (\F \B{\ref{fig:main1}d,e,f}, \SF \B{\ref{fig:views}} and \mbox{\SN \B{\ref{sec:benchmark}}}). If the input views show very low signal-to-noise ratio (atypical for SPIM) the speed-up is preserved but the quality of the deconvolved image is reduced. Our Bayesian-based derivation does not assume a specific noise model but it is in practice robust to Poisson noise, which is the dominating source of noise in light-sheet microscopy acquisitions (\SF \B{\ref{fig:benchmarks2},\ref{fig:noiseplot}}). As a compromise between quality and speed we use, if not stated otherwise, the intermediate optimization I for all deconvolution experiments on real datasets.

\begin{figure*}[h!]
\includegraphics[width=\textwidth]{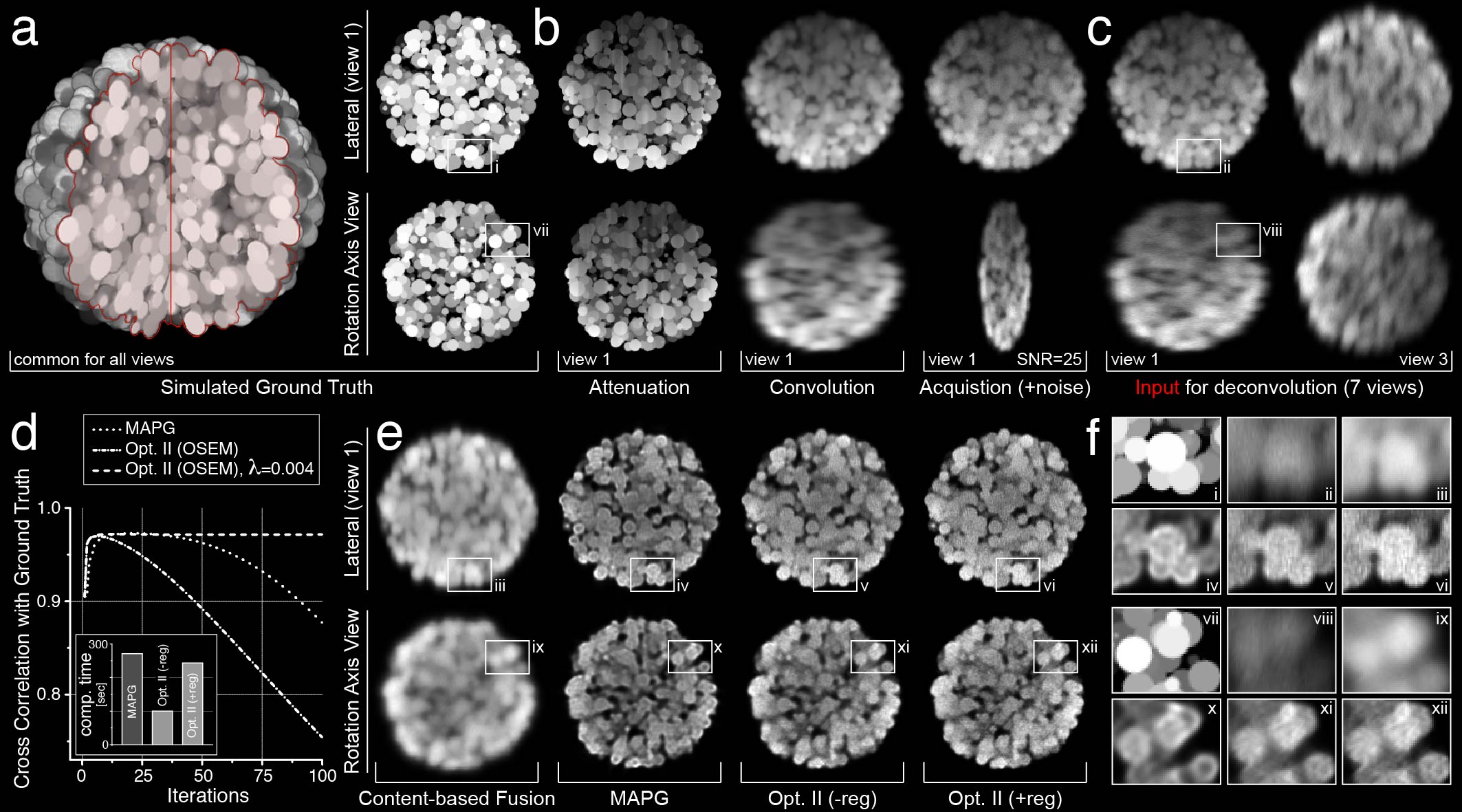}
\vspace{-2.0mm}
\caption{\hspace{-0.5mm} \emph{Deconvolution of simulated three dimensional multi-view data.} (\textbf{a}) On the left 3d rendering of a computer generated volume resembling a biological specimen. Red line marks the vedge removed from the volume to show the content inside. On the right sections through the generated volume in lateral direction (as seen by the SPIM camera, top) and along the rotation axis (bottom). (\textbf{b}) The same slice as in (\textbf{a}) with illumination attenuation applied (left), convolved with PSF of a SPIM microscope (middle) and image simulated using a poisson process (right). The bottom right panel shows the unscaled simulated light sheet sectioning data along the rotation axis. (\textbf{c}) Slices from view one and three of the seven views generated from (\textbf{a}) by applying processes pictured in (\textbf{b}) and rescaling to isotropic resolution. These seven volumes are the input to the fusion and deconvolution algorithms quantified in (\textbf{d}) and visualized in (\textbf{e}). (\textbf{d}) plots the cross-correlation of deconvolved and ground truth data as a function of the number of iterations for MAPG and our algorithm with and without regularization. The inset compares the computational time (both algorithms were implemented in Java to support partially overlapping datasets, \SF \B{\ref{fig:partial-overlap}}). (\textbf{e}) slices equivalent to (\textbf{c}) after content based fusion (first column), MAPG deconvolution (second column), our approach without regularization (third column) and with regularization (fourth column, lambda=0.004). (\textbf{f}) shows areas marked by boxes in (\textbf{b,c,e}) at higher magnification. 
}
\label{fig:main3}
\end{figure*}

We compared the performance of our method with previously published multi-view deconvolution algorithms\cite{Shepp1982,Hudson1994,Verveer2007,Bonetti2009,KrzicPhD,Temerinac2012} in terms of convergence behavior and runtime on the CPU (\F \B{\ref{fig:main1}e,f}, \F \B{\ref{fig:main3}d} and \SF \B{\ref{fig:views}b, \ref{fig:benchmarks2}a,b}). For typical SPIM multi-view scenarios consisting of around 7 views with a high signal-to-noise ratio our method requires 7 fold fewer iterations compared to OSEM\cite{Hudson1994}, Scaled Gradient Projection (SGP)\cite{Bonetti2009} and Maximum a posteriori with Gaussian Noise (MAPG)\cite{Verveer2007} while being 50 fold faster than the IDL implementation of SGP, 7 fold faster than OSEM and 3 fold faster than MAPG (implemented in Java). At the same time our optimization is able to improve the visual image quality of real and simulated datasets compared to MAPG (\F \B{\ref{fig:main3}e,f} and \SF \B{\ref{fig:benchmarks2}c-h}). Further speed up of 3 fold and reduced memory consumption is achieved by using CUDA implementation (\SF \B{\ref{fig:partial-overlap}e}). On real datasets that are often characterized by partial overlap of the input data, the actual overall computation time is increased due to missing data and the overheads imposed by boundary conditions (\SF \B{\ref{fig:partial-overlap}} and \SN \B{\ref{sec:partialoverlap}}). However, our approach is capable of dealing with partially overlapping acquisitions.

In order to evaluate the quality and performance of our algorithm on realistic three-dimensional multi-view image data we generated a simulated ground truth dataset resembling a biological specimen (\F \B{\ref{fig:main3}a}). We next simulated how this dataset looks like when imaged in a SPIM microscope from multiple angles by applying signal attenuation across the field of view, convolving the data with the PSF of the microscope, simulating the multi-view optical sectioning and using a Poisson process to generate the final pixel intensities (\F \B{\ref{fig:main3}b} and \SN \B{\ref{sec:simmv}}). We deconvolved the generated multi-view data (\F \B{\ref{fig:main3}c}) using our algorithm with and without regularization and compared the result to the content based fusion and the MAPG deconvolution (\F \B{\ref{fig:main3}d-f}). The results show that our algorithm reaches optimal reconstruction quality faster (\F \B{\ref{fig:main3}d}), introduces less artifacts compared to MAPG (\F \B{\ref{fig:main3}e,f}, note the artificial ring artifacts, and Supplementary video 2 and 3, note the artificial patterns in yz for MAPG) and that regularization is required to achieve convergence under realistic imaging conditions that we simulated (\F \B{\ref{fig:main3}d,f}).

Prerequisite for multi-view deconvolution of light sheet microscopy data are precisely aligned multi-view datasets and estimates of point spread functions for all views. We exploit the fact that for the purposes of registration we include sub-resolution fluorescent beads into the rigid agarose medium in which the specimen is embedded. The beads are initially used for multi-view registration of the SPIM data\cite{Preibisch2010} and subsequently to extract the PSF for each view for the purposes of multi-view deconvolution. We average the intensity of PSFs for each view for all the beads that were identified as corresponding during registration yielding a precise measure of the PSF for each view under the specific experimental condition. This synergy of registration and deconvolution ensures realistic representation of PSFs under any imaging condition. Alternatively, simulated PSFs or PSFs measured by other means can be provided as inputs to the deconvolution algorithm.

\begin{figure*}[h!]
\includegraphics[width=\textwidth]{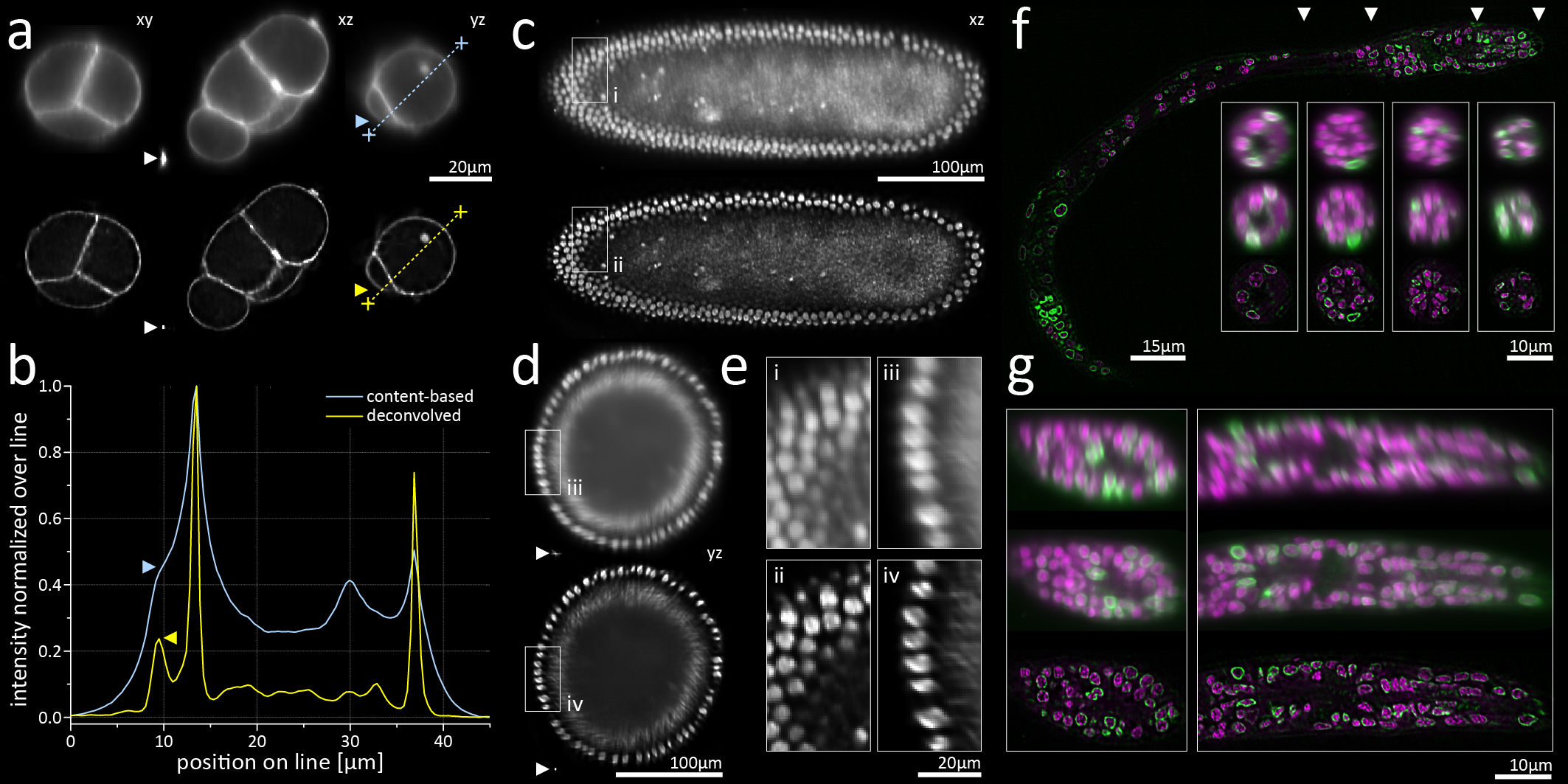}
\vspace{-2.0mm}
\caption{\hspace{-0.5mm} \emph{Application to biological data.} (\textbf{a}) Comparison of reconstruction results using content-based fusion (upper row) and multi-view deconvolution (lower row) on a 4-cell stage C. elegans embryo expressing PH-domain-GFP fusion marking the membranes. Dotted lines mark plots shown in (\textbf{b}), white arrows mark PSFs of a fluorescent bead before and after deconvolution. (\textbf{b}) Line plot through the volume along the rotation axis (yz), typically showing lowest resolution in light sheet acquisitions. Contrast along the line is locally normalized. Signal-to-noise is significantly enhanced, arrows mark points that illustrate increased resolution. (\textbf{c,d}) show cut planes through a blastoderm stage Drosophila embryo expressing His-YFP in all cells. White boxes mark areas magnified in (\textbf{e}). Detailed comparison of computation times for this dataset is shown in \F \B{\ref{fig:main1}e}. (\textbf{e}) Magnified view on small parts of the Drosophila embryo. Left panel shows one of the directly acquired views, right panel shows a view along the rotation axis usually characterized by the lowest resolution. (\textbf{f,g}) Comparison of the deconvolved image data to the input data of a fixed C. elegans larvae in L1 stage expressing LMN-1-GFP (green) and stained with Hoechst (magenta). (\textbf{f}) Single slice through the deconvolved dataset, arrows mark 4 locations of transversal cuts shown below. The cuts compare two orthogonal input views (0, 90 degrees) with the deconvolved data. Note that no input view offers high resolution in this orientation approximately along the rotation axis. (\textbf{g}) The first row of the left box shows a random slice of a view in axial orientation marking the worst possible resolution of the microscope. The second row shows an input view in lateral orientation, i.e. the best possible resolution achieved by the microscope. The third row shows the corresponding deconvolved image. The box on the right shows a random slice through the nervous system. Note that the alignment of the C. elegans L1 dataset was refined using nuclear positions as described in \SN \B{\ref{sec:l1}}. 
}
\label{fig:main2}
\end{figure*}

We applied our deconvolution approach to multi-view SPIM acquisitions of \emph{Drosophila} and \emph{C. elegans} embryos (\F \B{\ref{fig:main2}a-e}). We achieve a significant increase in contrast as well as resolution with respect to the content-based fusion\cite{Preibisch2010} (\F \B{\ref{fig:main2}b} and \SF \B{\ref{fig:resolution}}), while only a few iterations are required and computation times are typically in the range of a few minutes per multi-view acquisition (\ST \B{\ref{tab:experiments}}). A remaining challenge for creating a complete computational model of \emph{C. elegans} larvae in L1 stage is to be able to identify all nuclei in the nervous system\cite{long2009}. We applied the deconvolution to a 4-view acquisition of a fixed specimen expressing GFP tagged lamin (LMN-1::GFP) labeling the nuclear lamina and stained for DNA with Hoechst (\F \B{\ref{fig:main2}f,g}). Running the multi-view deconvolution for 100 iterations using optimization II, we achieve a significantly improved contrast and resolution compared the input data acquired with the Zeiss Lightsheet Z.1 microscope. Previous attempts to segment nuclei in confocal images of L1 worm resulted in unambiguous identification of 357 of the 558 nuclei\cite{long2009}. Our deconvolved dataset allows the manual segmentation of all 558 nuclei with uncertainty of about 5 nuclei among annotation trials.

The results of multi-view deconvolution on SPIM data are equivalent to the application of Structured Illumination in SPIM\cite{keller2010} providing a convenient post-processing alternative to increasing contrast (\SF \B{\ref{fig:SIM}}). Moreover, we show that multi-view deconvolution produces superior results when comparing an acquisition of the same sample with SPIM and two-photon microscope (\SF \B{\ref{fig:twophoton}}). Finally, the benefits of the multi-view deconvolution approach are not limited to SPIM as illustrated by the deconvolved multi-view Spinning Disc Confocal Microscope acquisition of \emph{C. elegans} L1 larva\cite{Preibisch2010} (\SF \B{\ref{fig:spinningdisc}}). Taken together these results illustrate that our multi-view deconvolution can be applied to increase the resolution of optical sectioning microscopy universally.

The increased contrast and resolution are especially visible on samples acquired with thicker light sheets such as in the case of OpenSPIM\cite{Pitrone2013} (\SF \B{\ref{fig:openspim}}). Out-of-focus light is significantly reduced and individual nuclei become separable even in orientations perpendicular to the rotation axis that have not been imaged by the microscope directly. Also very large datasets, such as the acquisition of Drosophila ovaries, with input data of over 5 billion voxels become computable in reasonable time (\SF \B{\ref{fig:eggchamber}} and \ST \B{\ref{tab:experiments}}). To further substantiate the utility of the algorithm we deconvolved an entire time-course of Drosophila embryonic development consisting of 236 time-points). The computation time was 24 hours using two Nvidia Quadro 4000 graphics cards, highlighting the performance and applicability of the method to very large datasets.

A major obstacle for widespread application of deconvolution approaches to multi-view light sheet microscopy data is lack of usable and scalable multi-view deconvolution software. We integrated our fast converging algorithm into Fiji’s\cite{fiji2012} multi-view processing pipeline as open-source plugin where it complements existing approaches to multi-view data registration (\url{http://fiji.sc/Multi-View_Deconvolution}). We provide an efficient implementation for GPU and CPU taking advantage of ImgLib\cite{PietzschAl12}. It offers processing times of a few minutes (\ST \B{\ref{tab:experiments}}), comparable to the acquisition rates of common light sheet microscopes such as OpenSPIM or Lightsheet Z.1 (Carl Zeiss Microimaging). The long-term time-lapse acquisitions derived from Lightsheet Z.1 are truly massive (2.16 TB for 6 view 715 timepoint \emph{Drosophila} embryogenesis recording) but our deconvolution plugin can process them in parallel on a computer cluster in real time, i.e. the same time it takes to acquire the data. To account for potential noise in the input images we added an option for Tikhonov regularization\cite{Tikhonov1977} (\SF \B{\ref{fig:noiseplot},\ref{fig:snr}}). The deconvolution can be processed on the entire image at once for optimal performance or in blocks to reduce the memory requirements. The only free parameter of the method that must be chosen by the user is the number of iterations for the deconvolution process (\SF \B{\ref{fig:views},\ref{fig:viewsimages}}). We facilitate this choice by providing a debug mode allowing the user to inspect all intermediate iterations and identify optimal tradeoff between quality and computation time. For a typical multi-view acquisition comprising 6–8 views we suggest between 10-15 iterations.

One of the challenges in image deconvolution is to arrive at the correct solution quickly without compromising quality. We have achieved significant improvement in convergence time over existing methods by exploiting conditional probabilities between views in a multi-view deconvolution scenario, while producing visually identical or improved results at SNR’s typical for light-sheet microscopy (\F \B{\ref{fig:main3}e,f} and \SF \B{\ref{fig:benchmarks2}c-h}). We have further implemented the algorithm as an open source GPU accelerated software in Fiji where it synergizes with other related plugins into an integrated solution for the processing of multi-view light sheet microscopy data of arbitrary size. 

\acknowledgments
We thank Tobias Pietzsch for very helpful discussions, proofreading and access to his unpublished software, Nathan Clack, Fernando Carrillo Oesterreich and Hugo Bowne-Anderson for discussions, Nicola Maghelli for two-photon imaging, Peter Verveer for his source code and helpful discussions, Michael Weber for imaging the Drosophila time series, Steffen Jaensch for preparing the C. elegans embryo, Jun Kelly Liu for the LW698 strain, Stephan Saalfeld for help with 3D rendering, P.J. Keller for supporting F.A. and the SI-SPIM dataset, Albert Cardona for access to his computer and Carl Zeiss Microimaging for providing us with the SPIM prototype. S.P. was supported by MPI-CBG, HHMI and the Human Frontier Science Program (HFSP) Postdoctoral Fellowship in R.H.S. lab, with additional support from NIH GM57071. F.A. was supported by HHMI. G.M. was supported by HHMI and MPI-CBG. P.T. was supported by The European Research Council Community′s Seventh Framework Program (FP7/2007-2013) grant agreement 260746.

\section*{AUTHOR CONTRIBUTIONS}
S.P. and F.A. derived the equations for multi-view deconvolution. S.P. implemented the software and performed all analysis, F.A. implemented the GPU code. E.S. generated and imaged H2Av-mRFPruby fly line. M.S. prepared and M.S. and S.P. imaged the \emph{C. elegans} L1 sample. S.P. \& P.T. conceived the idea and wrote the manuscript. R.H.S. provided support and encouragement, G.M. \& P.T. supervised the project.

\renewcommand{\figurename}{Supplementary Figure}
\setcounter{figure}{0} 

\pagebreak

\section*{SUPPLEMENTARY MATERIAL}
\hspace{20mm}

\begin{table}[h!]
\center
{
\fontsize{12pt}{12pt}\selectfont
\center
\begin{tabular}{lp{11cm}}
\textbf{\textcolor{red}{Supplementary File}} & \textbf{\textcolor{red}{Title}}\\ \\
\hline
\\
\textbf{Supplementary Figure \ref{fig:01}} & Illustration of conditional probabilities describing the dependencies of two views \tablespace \\ 
\textbf{Supplementary Figure \ref{fig:principle-virtual-osem}} & The principle of 'virtual' views and sequential updating \tablespace \\
\textbf{Supplementary Figure \ref{fig:assumptionConv}} &  Illustration of assumption required for incorporating 'virtual' views without additional computational effort \tablespace \\
\textbf{Supplementary Figure \ref{fig:views}} & Performance comparison of the multi-view deconvolution methods and dependence on the PSF \tablespace \\
\textbf{Supplementary Figure \ref{fig:viewsimages}} & Images used for analysis and visual performance \tablespace \\
\textbf{Supplementary Figure \ref{fig:benchmarks2}} & Comparison to other optimized multi-view deconvolutions \tablespace \\
\textbf{Supplementary Figure \ref{fig:noiseplot}} & Effect of noise on the deconvolution results \tablespace \\
\textbf{Supplementary Figure \ref{fig:snr}} & Intermediate stages of deconvolution results for varying SNR's and regularization \tablespace \\
\textbf{Supplementary Figure \ref{fig:psf}} & Quality of deconvolution for imprecise estimation of the PSF\tablespace\\
\textbf{Supplementary Figure \ref{fig:lightsheet}} & Variation of PSF across the light sheet in SPIM acquistions \tablespace \\
\textbf{Supplementary Figure \ref{fig:SIM}} & Comparison of Multi-View Deconvolution to Structured Illumination Light Sheet Data\tablespace \\
\textbf{Supplementary Figure \ref{fig:twophoton}} & Comparison of Multi-View Deconvolution to two-photon microscopy \tablespace \\
\textbf{Supplementary Figure \ref{fig:spinningdisc}} & Multi-View Deconvolution of Spinning-Disc Confocal Data \tablespace \\
\textbf{Supplementary Figure \ref{fig:resolution}} & Quantification of resolution enhancement by Multi-View Deconvolution \tablespace \\
\textbf{Supplementary Figure \ref{fig:openspim}} & Reconstruction quality of an OpenSPIM acquistion \tablespace \\
\textbf{Supplementary Figure \ref{fig:eggchamber}} & Quality of reconstruction of Drosophila ovaries \tablespace \\
\textbf{Supplementary Figure \ref{fig:partial-overlap}} & Effects of partial overlap and CUDA performance \tablespace \\
\textbf{Supplementary Table \ref{tab:experiments}} & Summary of datasets used in this publication \tablespace \\
\textbf{Supplementary Note} & Detailed derivations and discussion of the efficient Bayesian-based multi-view deconvolution \tablespace \\
\end{tabular}}
\caption{Overview of supplementary material.} 
\end{table}

\pagebreak
\section*{SUPPLEMENTARY FIGURES}

\hspace{20mm}

\subsection*{SUPPLEMENTARY FIGURE 1 | Illustration of conditional probabilities describing the dependencies of two views}

\vspace{1mm}

\begin{figure*}[h!]
\includegraphics[width=\textwidth]{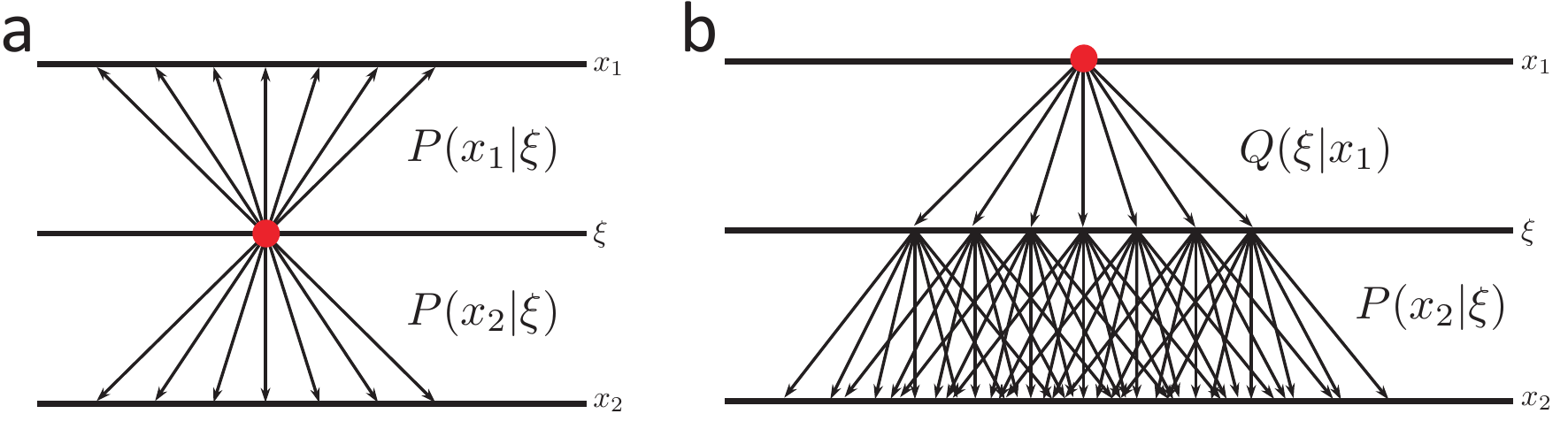}
\vspace{-2.0mm}
\caption{\hspace{-0.5mm} \emph{Illustration of conditional probabilities describing the dependencies of two views.} \mbox{(\textbf{a}) illustrates} the conditional independence of two observed distributions $\phi_1(x_1)$ and $\phi_2(x_2)$ if it is known that the event $\xi=\xi'$ on the underlying distribution $\psi(\xi)$ occured. Given $\xi=\xi'$, both distributions are conditionally independent, the probability where to expect an observation only depends on $\xi=\xi'$ and the respective individual point spread function $P(x_1|\xi)$ and $P(x_2|\xi)$, i.e. $P(x_1|\xi,x_2) = P(x_1|\xi)$ and $P(x_2|\xi,x_1) = P(x_2|\xi)$. \mbox{(\textbf{b}) illustrates} the relationship between an observed distribution $\phi_2(x_2)$ and $\phi_1(x_1)$ if the event $x_1=x_1'$ occured. Solely the 'inverse' point spread function $Q(\xi|x_1)$ defines the probability for any event $\xi=\xi'$ to have caused the observation $x_1=x_1'$. The point spread function $P(x_2|\xi)$ consecutively defines the probability where to expect a corresponding observation $x_2=x_2'$ given the probability distribution $\psi(\xi)$. 
}
\label{fig:01}
\end{figure*}

\vspace{10mm}

\subsection*{SUPPLEMENTARY FIGURE 2 | The principle of 'virtual' views and sequential updating}

\vspace{1mm}

\begin{figure*}[h!]
\includegraphics[width=\textwidth]{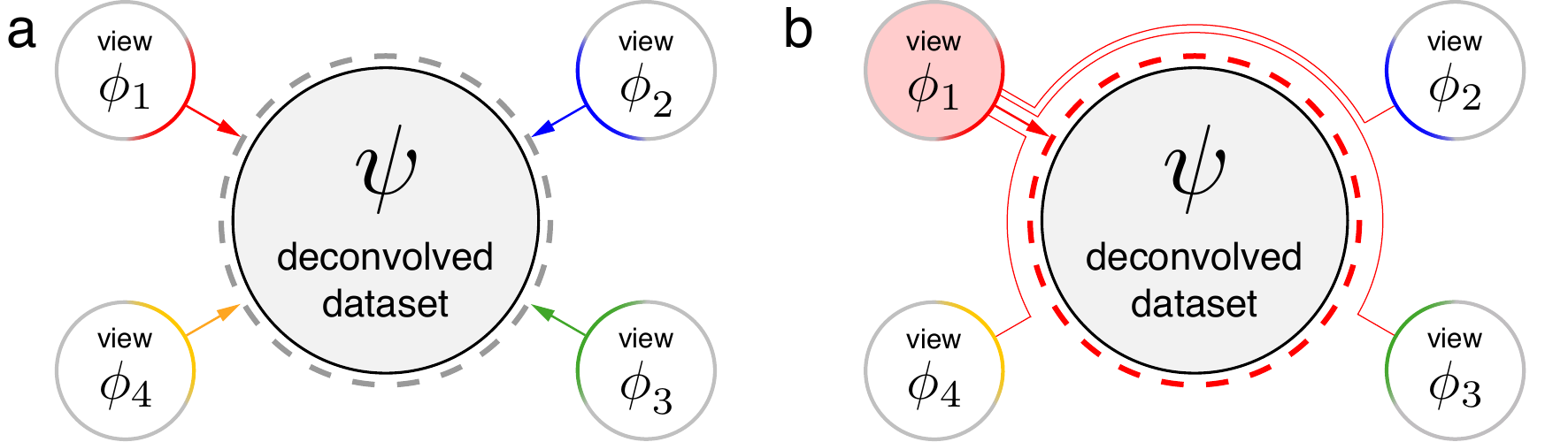}
\vspace{-2.0mm}
\caption{\hspace{-0.5mm}\emph{The principle of 'virtual' views and sequential updating.} (\textbf{a}) The ‘classical’ multi-view deconvolution\cite{Shepp1982, KrzicPhD, Temerinac2012, Bonetti2009} where an update step is computed individually for each view and subsequently combined into one update of the deconvolved image. (\textbf{b}) Our new derivation considering conditional probabilities between views. Each individual update step takes into account all other views using ‘virtual views’ and additionally updates the deconvolved image individually, i.e. updates are performed sequentially\cite{Hudson1994} and not combined. 
}\label{fig:principle-virtual-osem}
\end{figure*}

\pagebreak

\subsection*{SUPPLEMENTARY FIGURE 3 | Illustration of assumption required for incorporating 'virtual' views without additional computational effort}

\vspace{1mm}

\begin{figure*}[h!]
\includegraphics[width=\textwidth]{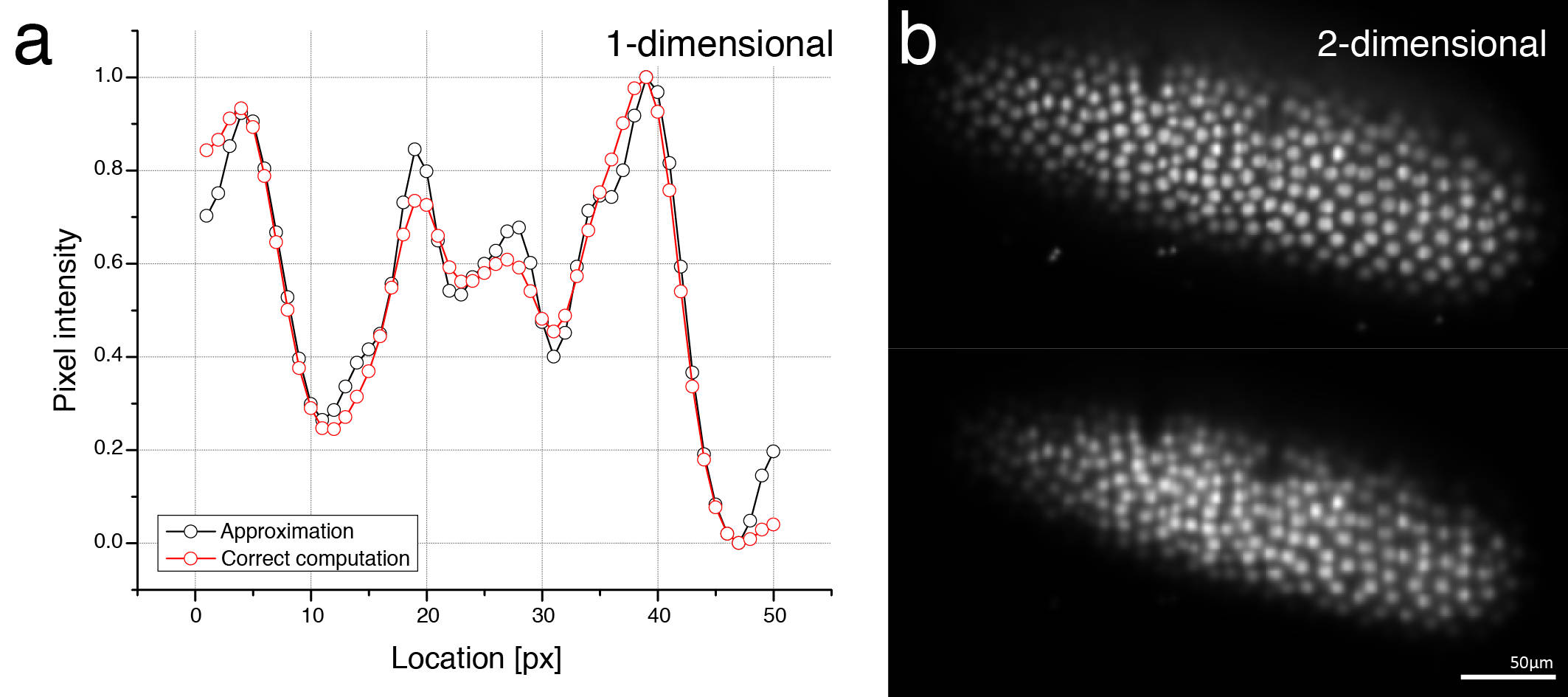}
\vspace{-2.0mm}
\caption{\hspace{-0.5mm}\emph{Illustration of assumption in equation \ref{eq:eq210}.} (\textbf{a}) shows the difference in the result when computing $( f \ast g ) \cdot ( f \ast h )$ in red and the approximation $f \ast ( g \cdot h )$ in black for a random one-dimensional input sequence ($f$) and two kernels with $\sigma$=3 ($g$) and $\sigma$=2 ($h$) after normalization. (\textbf{b}) shows the difference when using the two-dimensional image from \fig \ref{fig:viewsimages}a as input ($f$) and the first two point spread functions from \fig \ref{fig:viewsimages}e as kernels ($g,h$). The upper panel pictures the approximation, the lower panel the correct computation. Note that for (\textbf{a,b}) the approximation is slightly less blurred. Note that the beads are also visible in the lower panel when adjusting the brightness/contrast.
}\label{fig:assumptionConv}
\end{figure*}

\pagebreak

\subsection*{SUPPLEMENTARY FIGURE 4 | Performance comparison of the multi-view deconvolution methods and dependence on the PSF}

\vspace{1mm}

\begin{figure*}[h!]
\includegraphics[width=\textwidth]{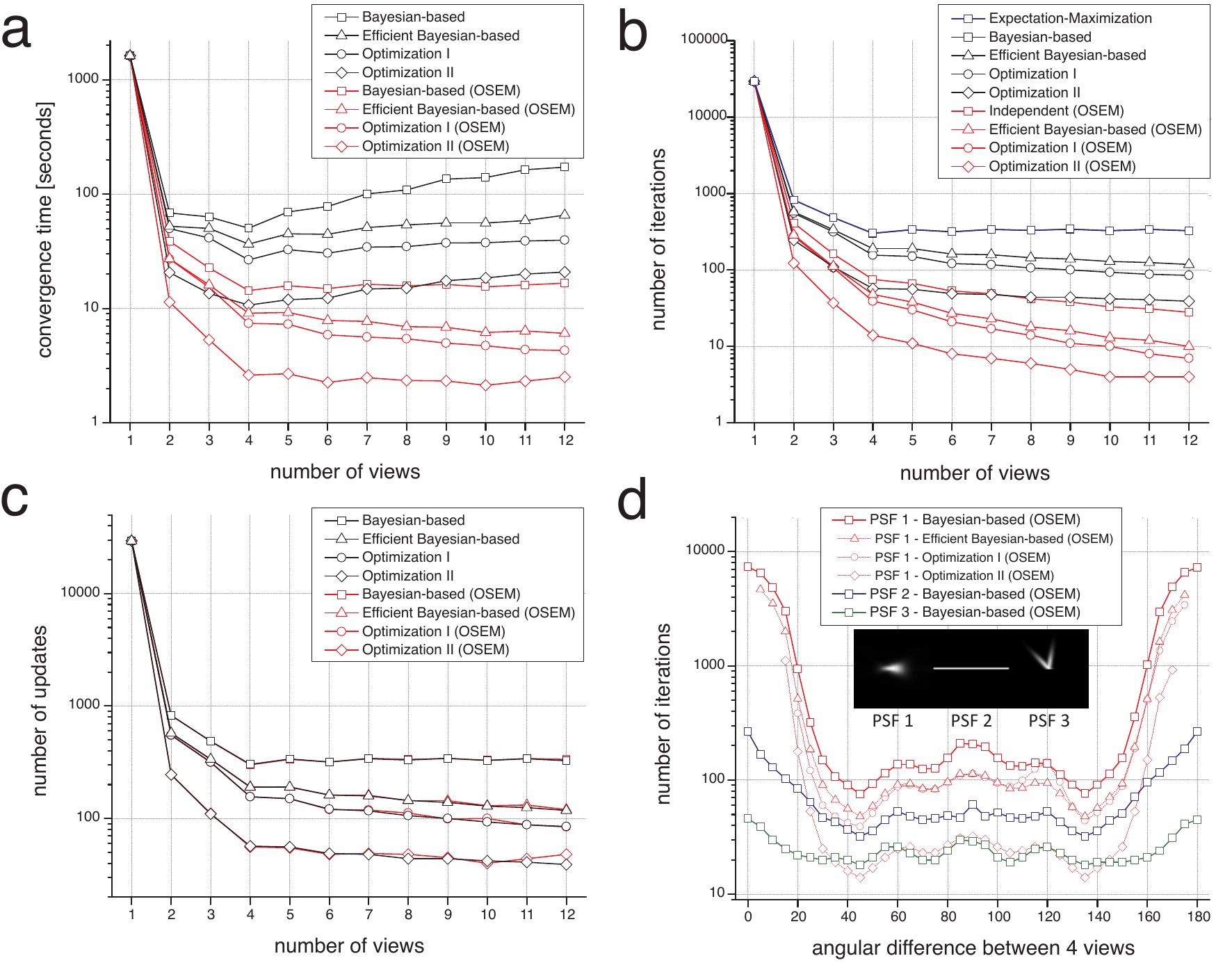}
\vspace{-2.0mm}
\caption{\hspace{-0.5mm}\emph{Performance comparison and dependence on the PSF} (\textbf{a}) The convergence time of the different algorithms until they reach the same average difference to the ground truth image shown in \fig \ref{fig:viewsimages}e. (\textbf{b}) The number of iterations required until all algorithms reach the same average difference to the ground truth image. One 'iteration' comprises all computional steps until each view contributed once to update the underlying distribution. Note that our Bayesian-based derivation and the Maximization-Likelihood Expectation-Maximization\cite{Shepp1982} method perform almost identical (\textbf{c}) The total number of updates of the underlying distribution until the same average difference is reached. (\textbf{d}) The number of iterations required until the same difference to the ground truth is achieved using 4 views. The number of iterations is plotted relative to the angular difference between the input PSFs. An angular difference of 0 degrees refers to 4 identical PSFs and therefore 4 identical input images, an example of an angular difference of 45 degrees is shown in \fig \ref{fig:viewsimages}e. Plots are shown for different types of PSFs. (\textbf{a-d}) y-axis has logarithmic scale, all computations were performed on a dual-core Intel Core i7 with 2.7Ghz.
}\label{fig:views}
\end{figure*}

\pagebreak

\subsection*{SUPPLEMENTARY FIGURE 5 | Images used for analysis and visual performance}

\vspace{1mm}

\begin{figure*}[h!]
\includegraphics[width=\textwidth]{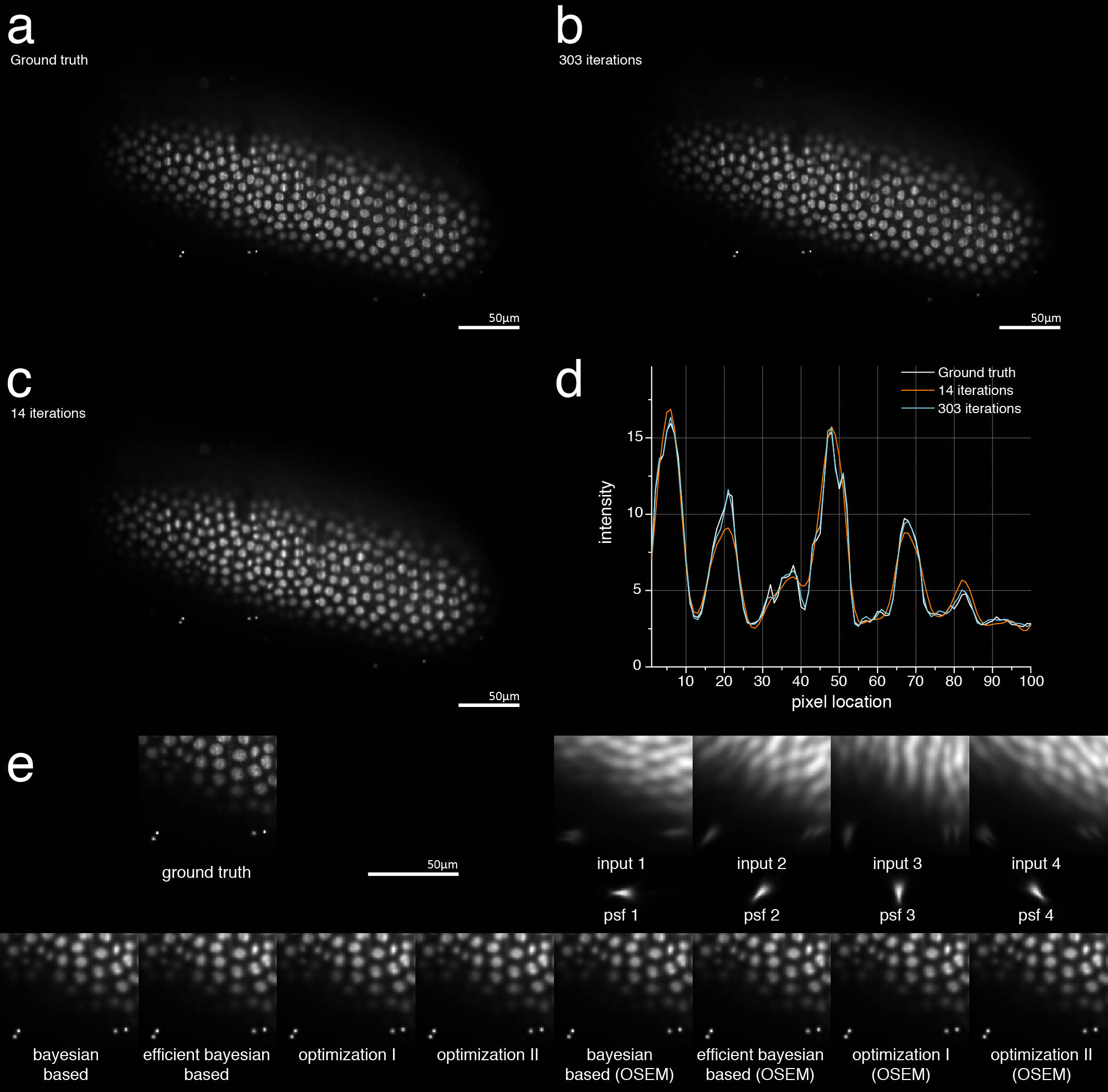}
\vspace{-2.0mm}
\caption{\hspace{-0.5mm}\emph{Images used for analysis and visual performance.} (\textbf{a}) The entire ground truth image used for all analyses shown in the supplement. (\textbf{b}) Reconstruction quality after 301 iterations using optimization II and sequential updates on 4 input views and PSF's as shown in (e). (\textbf{c}) Reconstruction quality after 14 iterations for the same input as (b). (\textbf{d}) Line-plot through the image highlighting the deconvolution quality after 301 (b) and 14 (c) iterations compared to the ground truth (a). (\textbf{e}) Magnificantion of a small region of the ground truth image (a), the 4 input PSF's and 4 input datasets as well as the results for all algorithms as used in \fig \ref{fig:views}a-c for performance measurements.
}\label{fig:viewsimages}
\end{figure*}

\pagebreak

\subsection*{SUPPLEMENTARY FIGURE 6 | Comparison to other optimized multi-view deconvolutions}

\vspace{1mm}

\begin{figure*}[h!]
\includegraphics[width=\textwidth]{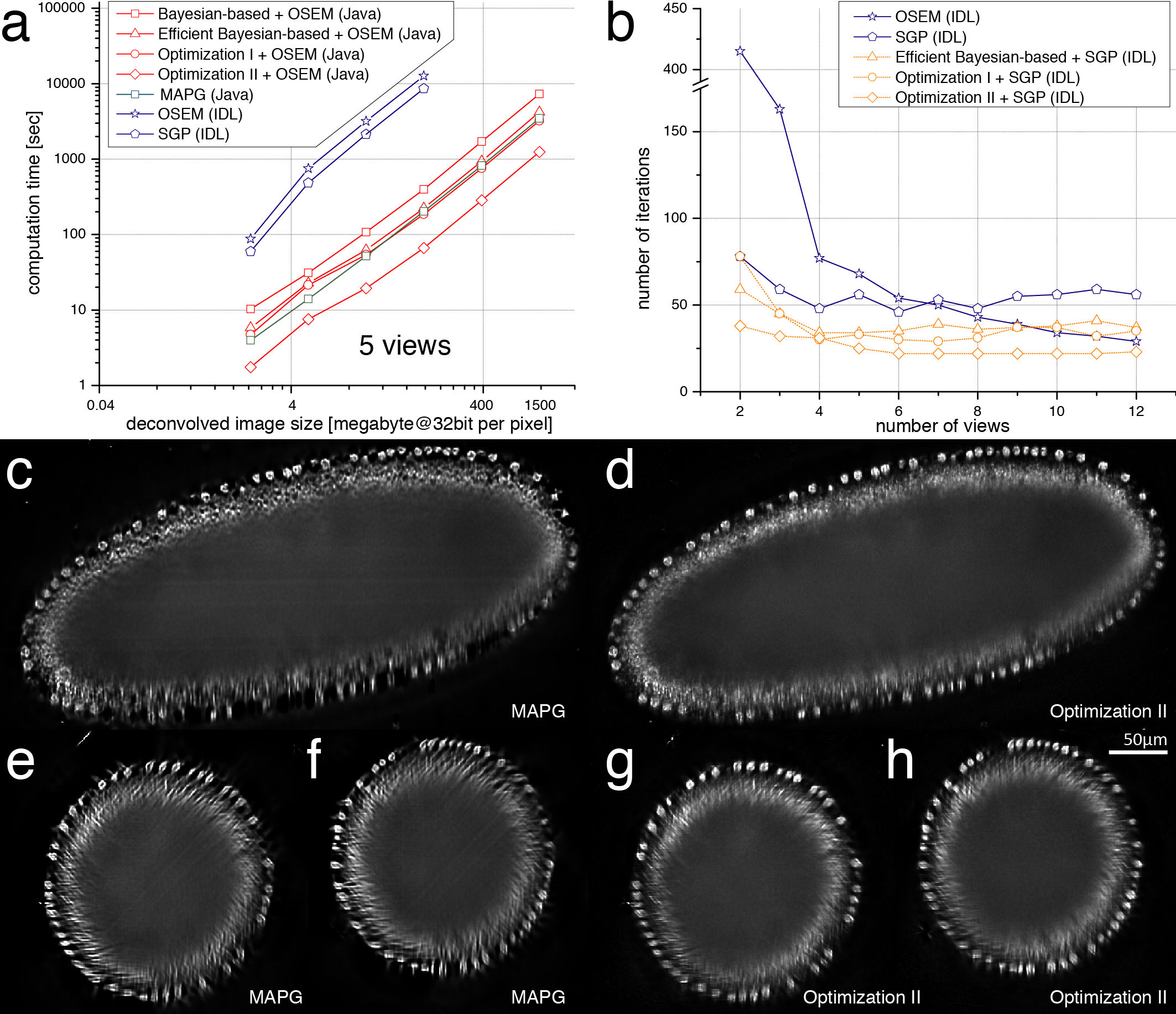}
\vspace{-2.0mm}
\caption{\hspace{-0.5mm}\emph{Comparison to other optimized multi-view deconvolution schemes.} (\textbf{a,b}) 
Compares optimized versions of multi-view deconvolution, including the IDL implementations of Scaled Gradient Projection (SGP)\cite{Bonetti2009}, Ordered Subset Expectation Maximization (OSEM)\cite{Hudson1994}, Maximum a posteriori with Gaussian Noise (MAPG)\cite{Verveer2007}, and our derivations combined with OSEM (see also main text figure 1e,f). All computations were performed on a machine with 128 GB of RAM and two 2.7 GHz Intel E5-2680 processors. (\textbf{a}) Correlates computation time and image size until the deconvolved image reached the same difference to the known ground truth image. All algorithms perform relatively proportional, however the IDL implementations run out of memory. (\textbf{b}) illustrates that our optimizations can also be combined with SGP in order to achieve a faster convergence. (\textbf{c-h}) compare the reconstruction quality of MAPG and Optimization II using the 7-view acquisition of the \emph{Drosophila} embryo expressing His-YFP (main text figure 3c,d,e). Without ground truth we chose a stage of similar sharpness (26 iterations of MAPG and 9 iterations of Optimization II, approximately in correspondence with main figure 1f) not using any regularization. Optimization II achieves a visually higher image quality, while MAPG shows some artifacts and enhances the stripe pattern arising from partially overlapping input images. (\textbf{c,d}) show a slice in lateral orientation of one of the input views, (\textbf{e-h}) show slices perpendicular to the rotation axis.
}\label{fig:benchmarks2}
\end{figure*}
\pagebreak

\subsection*{SUPPLEMENTARY FIGURE 7 | Effect of noise on the deconvolution results}

\vspace{1mm}

\begin{figure*}[h!]
\includegraphics[width=\textwidth]{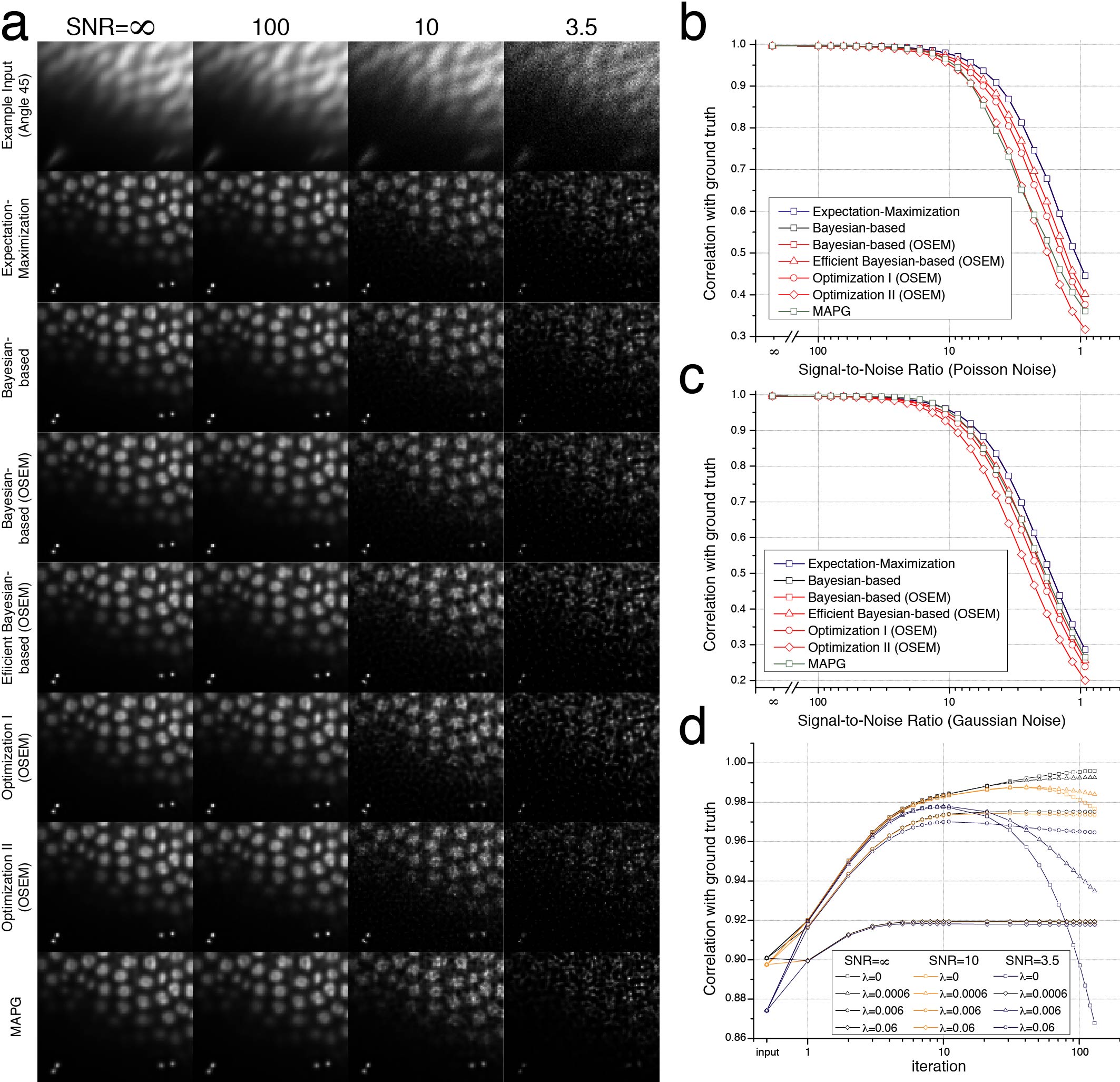}
\vspace{-2.0mm}
\caption{\hspace{-0.5mm}\emph{Effect of noise on the deconvolution results.} (\textbf{a}) Deconvolved images corresponding to the points in graph (b) to illustrate the resulting image quality corresponding to a certain correlation coefficient. (\textbf{b,c}) The resulting cross-correlation between the ground truth image and the deconvolved image depending on the signal-to-noise ratio in the input images. (\textbf{b}) Poisson noise, (\textbf{c}) Gaussian noise. (\textbf{d}) The cross correlation between the ground truth image and the deconvolved image at certain iteration steps during the deconvolution shown for different signal-to-noise ratios (SNR=$\infty$ [no noise], SNR=10, SNR=3.5) and varying parameters of the Tikhonov regularization ($\lambda$=0 [no regularization], $\lambda$=0.0006, $\lambda$=0.006, $\lambda$=0.06). \Fig \ref{fig:snr} shows the corresponding images for all data points in this plot. This graph is based on the Bayesian-based derivation using sequential updates in order to be able to illustrate the behaviour in early stages of the devonvolution.
}\label{fig:noiseplot}
\end{figure*}

\pagebreak

\subsection*{SUPPLEMENTARY FIGURE 8 | Intermediate stages of deconvolution results for varying SNR's and regularization}

\vspace{1mm}

\begin{figure*}[h!]
\includegraphics[width=\textwidth]{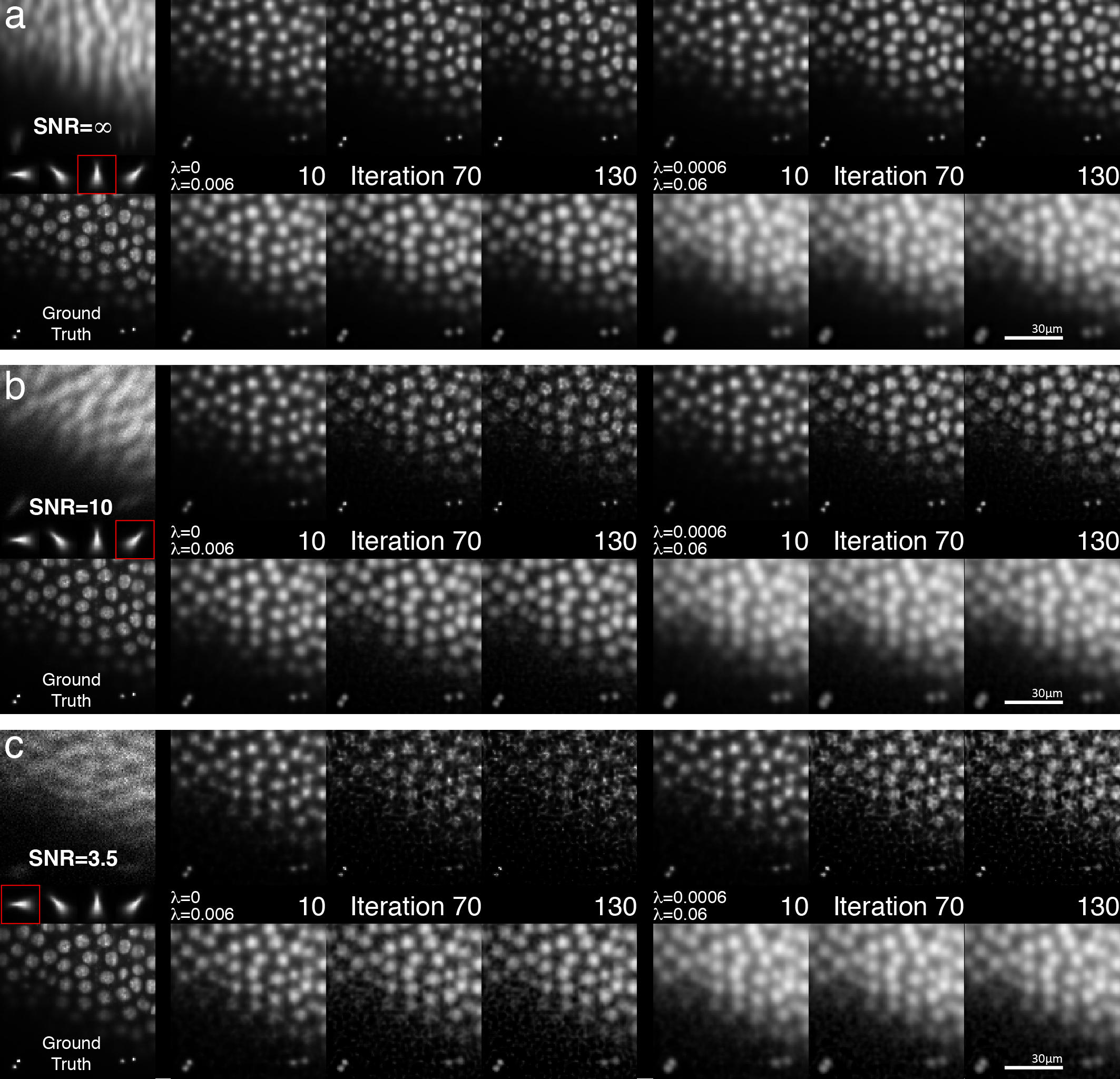}
\vspace{-2.0mm}
\caption{\hspace{-0.5mm}\emph{Intermediate stages of deconvolution results for varying SNR's and regularization}. \mbox{(\textbf{a-c})} 1\textsuperscript{st} row shows input data for the PSF in the red box, PSF's and ground truth, the other rows show the images at iteration 10, 70 and 130 for varying parameters of the Tikhonov regularization ($\lambda$=0 [no regularization], $\lambda$=0.0006, $\lambda$=0.006, $\lambda$=0.06).
(\textbf{a}) Results and input for SNR=$\infty$ (no noise). Here, $\lambda$=0 shows best results. (\textbf{b}) Results and input for SNR=10 (Poisson noise). Small structures like the fluorescent beads close to each other remain separable. (\textbf{c}) Results and input for SNR=3.5 (Poisson noise). Note that although the beads cannot be resolved anymore in the input data, the deconvolution produces a reasonable result, visually best for a $\lambda$ between 0.0006 and 0.006.
}\label{fig:snr}
\end{figure*}

\pagebreak

\subsection*{SUPPLEMENTARY FIGURE 9 | Quality of deconvolution for imprecise estimation of the PSF}

\vspace{1mm}

\begin{figure*}[h!]
\begin{center}
\includegraphics[width=9cm]{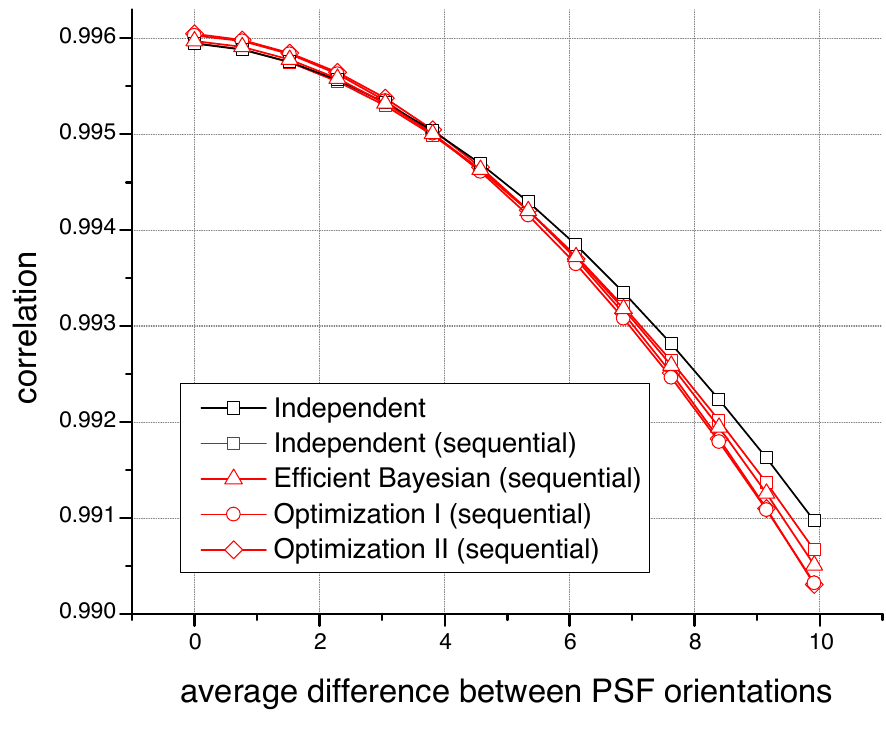}
\caption{\hspace{-0.5mm}\emph{Quality of deconvolution for imprecise estimation of the PSF}. The cross-correlation between the deconvolved image and the ground truth images when the PSF's used for deconvolution were rotated by random angles relative to the PSF's used to create the input images.
}\label{fig:psf}
\end{center}
\end{figure*}

\vspace{20mm}

\subsection*{SUPPLEMENTARY FIGURE 10 | Variation of PSF across the light sheet in SPIM acquistions}

\vspace{1mm}

\begin{figure*}[h!]
\begin{center}
\includegraphics[width=\textwidth]{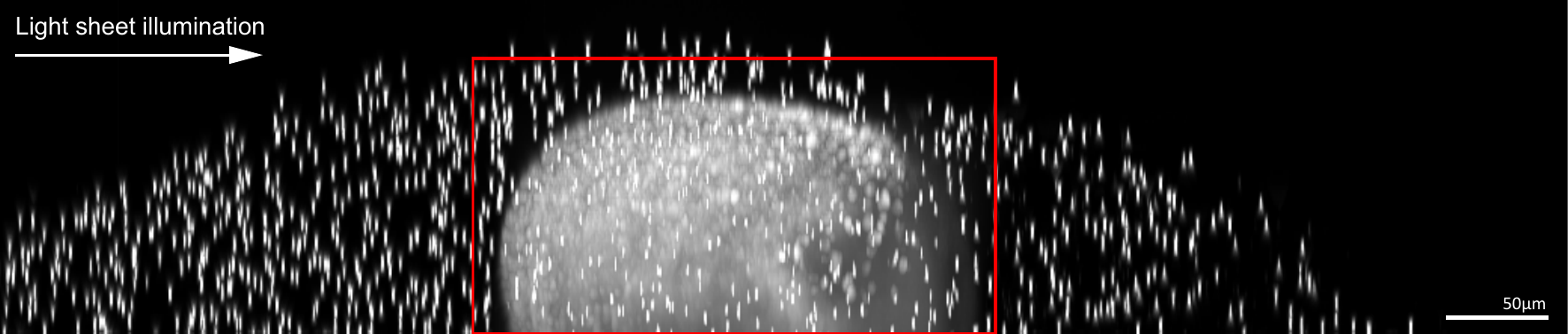}
\vspace{-2.0mm}
\caption{\hspace{-0.5mm}\emph{Variation of the PSF across the light sheet in SPIM acquistions}. The maximum intensity projection perpendicular to the light sheet of a \emph{Drosophila} embryo expressing His-YFP in all nuclei. The fluorescent beads have a diameter of 500nm. The arrow shows the illumination direction of the light sheet. The fluorescent beads should reflect the concave shape of a light sheet. The red box illustrates the area that is approximately used for deconvolution.
}\label{fig:lightsheet}
\end{center}
\end{figure*}

\pagebreak

\subsection*{SUPPLEMENTARY FIGURE 11 | Comparison of Multi-View Deconvolution to Structured Illumination Light Sheet Data}

\vspace{1mm}

\begin{figure*}[h!]
\begin{center}
\includegraphics[width=\textwidth]{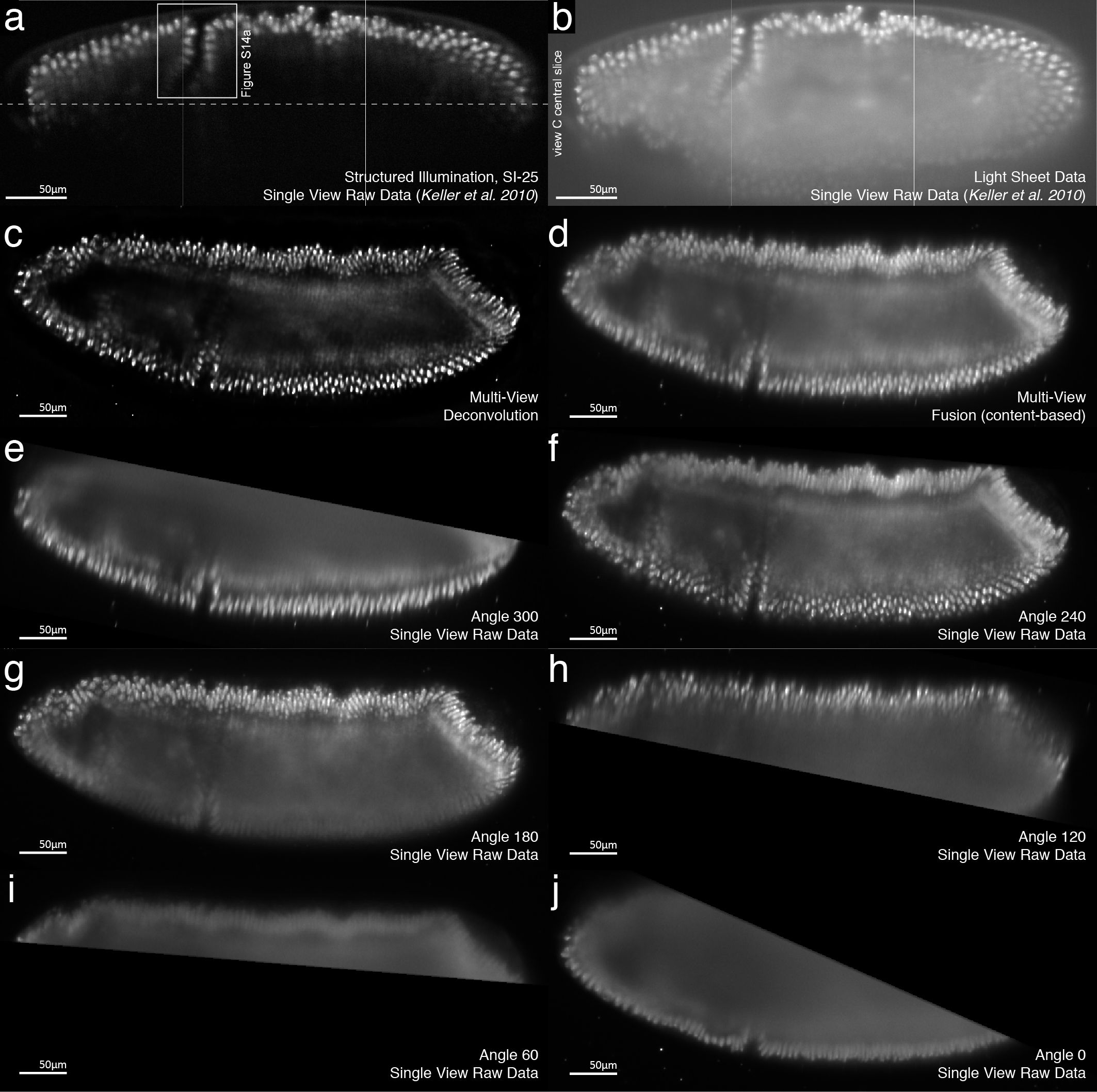}
\vspace{-2.0mm}
\caption{\hspace{-0.5mm} \emph{Comparison of Multi-View Deconvolution to Structured Illumination Light Sheet Data.} (\textbf{a}) Slice through a \emph{Drosophila} embryo expressing a nuclear marker acquired with DSLM and structured illumination (SI). (\textbf{b}) Corresponding slice acquired with standard light sheet microscopy. (\textbf{a}) and (\textbf{b}) taken from Keller et al.\cite{keller2010}. (\textbf{c-j}) Slice through a \emph{Drosophila} embryo in a similar stage of embryonic development expressing His-YFP. (\textbf{c}) shows the result of the multi-view deconvolution, (\textbf{d}) the result of the content-based fusion and (\textbf{e-j}) shows a slice through the aligned\cite{Preibisch2010} raw data as acquired by the Zeiss demonstrator B.
}\label{fig:SIM}
\end{center}
\end{figure*}

\pagebreak

\subsection*{SUPPLEMENTARY FIGURE 12 | Comparison of Multi-View Deconvolution to 2p Microscopy}

\vspace{1mm}

\begin{figure*}[h!]
\begin{center}
\includegraphics[width=\textwidth]{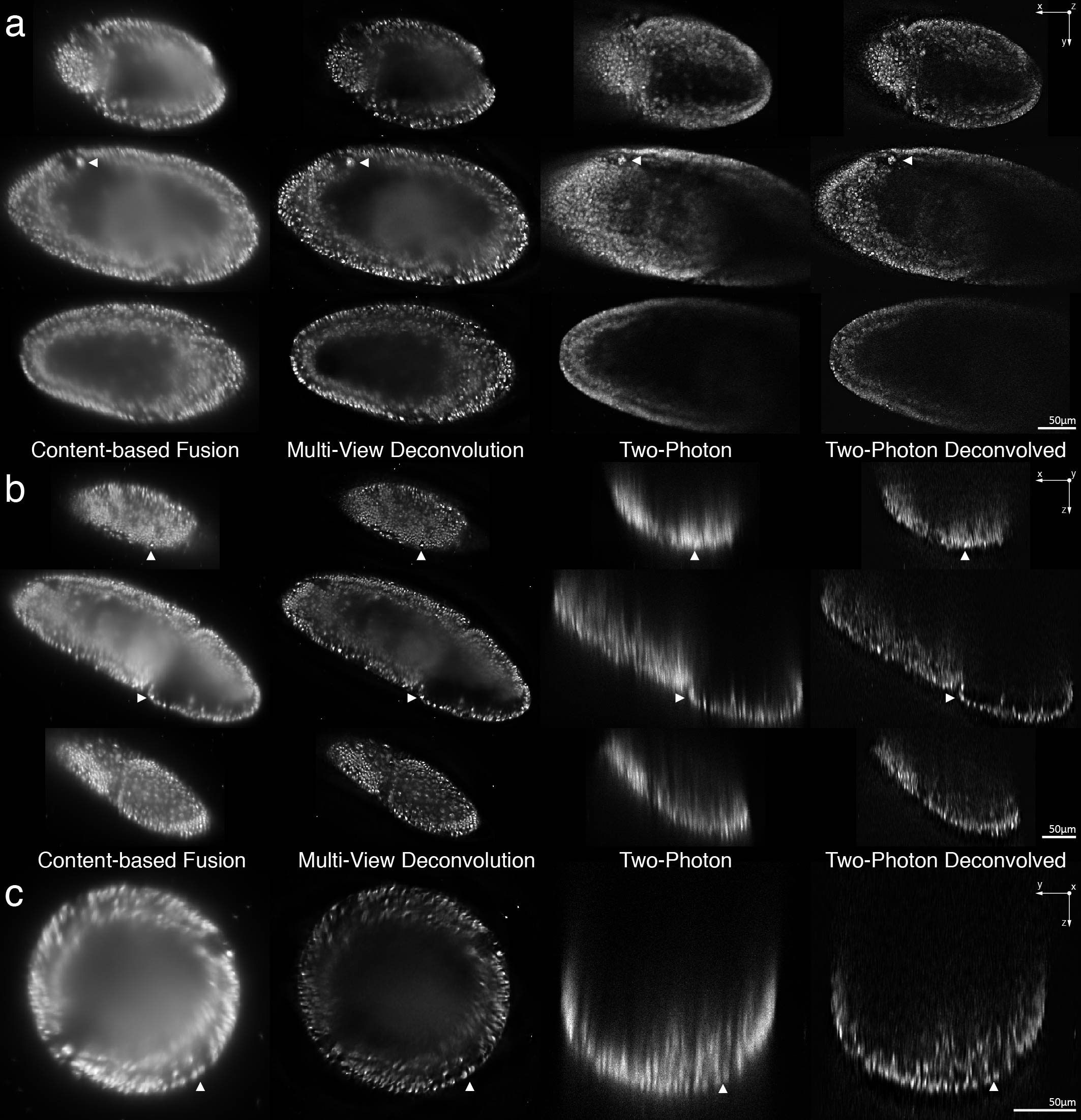}
\vspace{-2.0mm}
\caption{\hspace{-0.5mm} \emph{Comparing multi-view deconvolution to two-photon (2p) microscopy.} (\textbf{a-c}) slices through a fixed \emph{Drosophila} embryo stained with Sytox green labeling nuclei. Same specimen was acquired with the Zeiss SPIM prototype (20x/0.5NA water dipping obj.) and directly afterwards with a 2p microscope (20x/0.8NA air obj.). We compare the quality of content-based fusion, multi-view deconvolution, raw 2p stack and single view deconvolution of the 2p acquisition. (\textbf{a}) lateral (xy), (\textbf{b}) axial (xz), (\textbf{c}) axial (yz) orientation of the 2p stack, SPIM data is aligned relative to it using the beads in the agarose. Arrows mark corresponding nuclei.
}\label{fig:twophoton}
\end{center}
\end{figure*}

\pagebreak

\subsection*{SUPPLEMENTARY FIGURE 13 | Multi-View Deconvolution of Spinning-Disc Confocal Data}

\vspace{1mm}

\begin{figure*}[h!]
\begin{center}
\includegraphics[width=\textwidth]{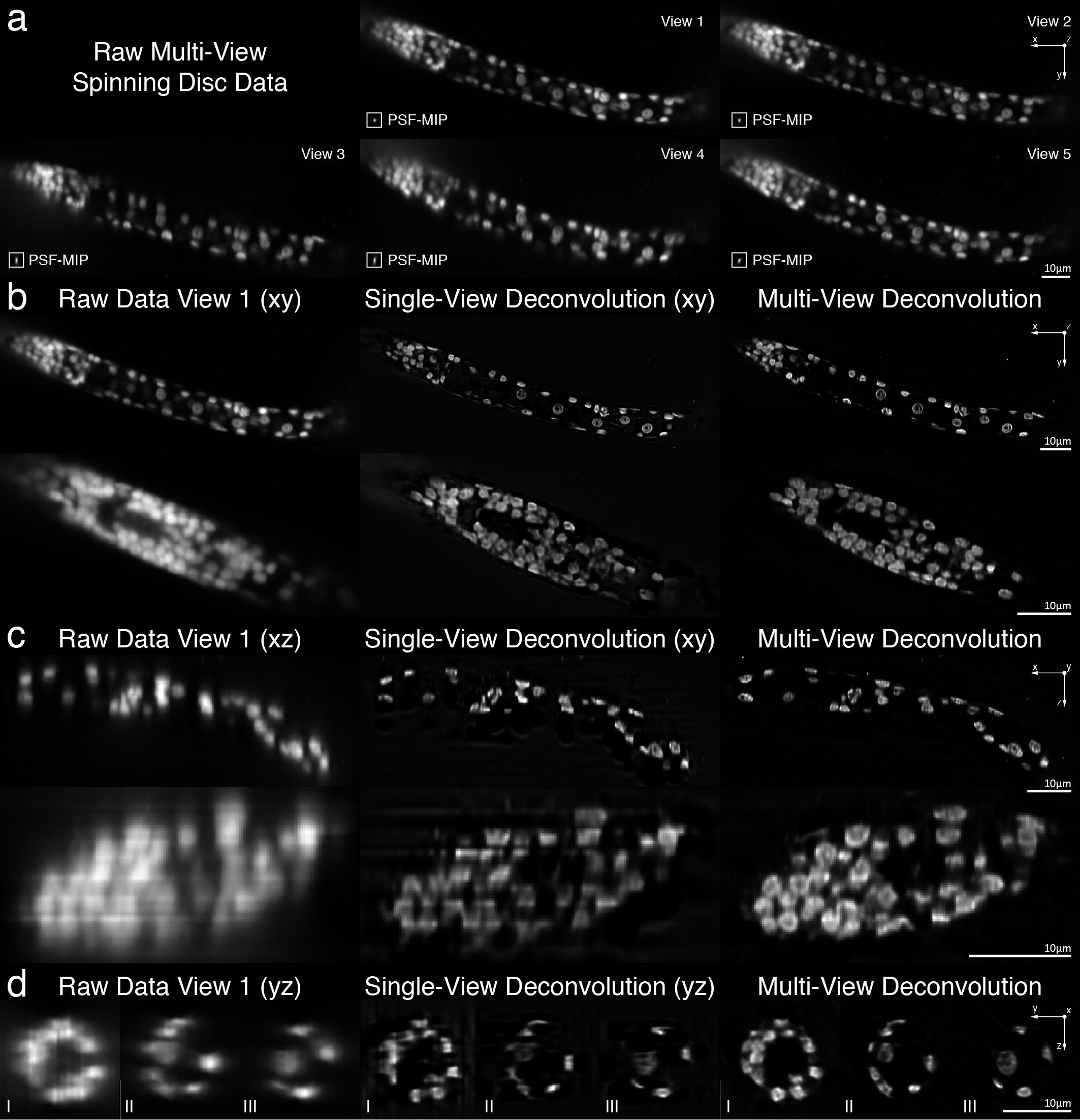}
\vspace{-2.0mm}
\caption{\hspace{-0.5mm} \emph{Multi-View Deconvolution of a Spinning-Disc Confocal Dataset.} (\textbf{a-d}) show slices through a fixed C. elegans in L1 stage stained with Sytox green labeling nuclei. The specimen was acquired on a spinning disc confocal microscope (20x/0.5NA water dipping objective). The sample was embedded in agarose and rotated using a self-build device\cite{Preibisch2010}. (\textbf{a}) Slice through the aligned input views; insets show averaged MIP of the PSF. (\textbf{b-d}) slices with different orientations through the larva comparing the quality of the first view of the input data, the single-view deconvolution of view 1 and the multi-view deconvolution of the entire dataset.
}\label{fig:spinningdisc}
\end{center}
\end{figure*}

\pagebreak

\subsection*{SUPPLEMENTARY FIGURE 14 | Quantification of resolution enhancement by Multi-View Deconvolution}

\vspace{1mm}

\begin{figure*}[h!]
\begin{center}
\includegraphics[width=\textwidth]{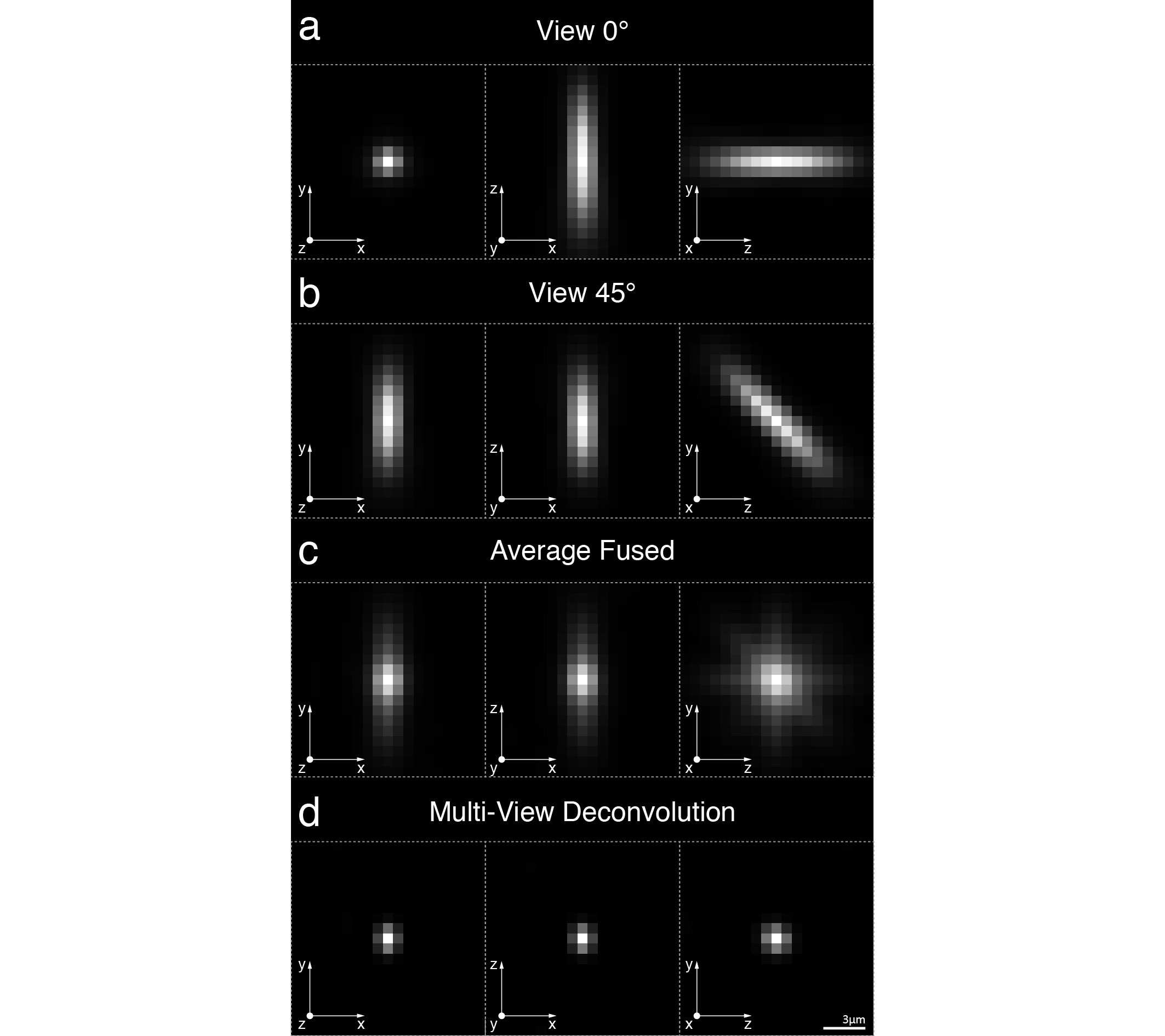}
\vspace{-2.0mm}
\caption{\hspace{-0.5mm} \emph{Quantification of resolution enhancement by Multi-View Deconvolution.} (\textbf{a-d}) compare the average of all fluorescent beads matched by the bead-based registration\cite{Preibisch2010} for two input views (\textbf{a,b}), after multi-view fusion (\textbf{c}), and after multi-view deconvolution (\textbf{d}). The resolution enhancement is apparent, especially along the rotation axis (third column, yz) between (\textbf{c}) and (\textbf{d}). The dataset used for this analysis is the 7-view acquisition of a developing \emph{Drosophila} embryo (see main text figure 3c-e), deconvolved for 15 iterations with $\lambda$=0.0006  using Optimization I.
}\label{fig:resolution}
\end{center}
\end{figure*}

\pagebreak

\subsection*{SUPPLEMENTARY FIGURE 15 | Reconstruction quality of an OpenSPIM acquistion}

\vspace{1mm}

\begin{figure*}[h!]
\begin{center}
\includegraphics[width=\textwidth]{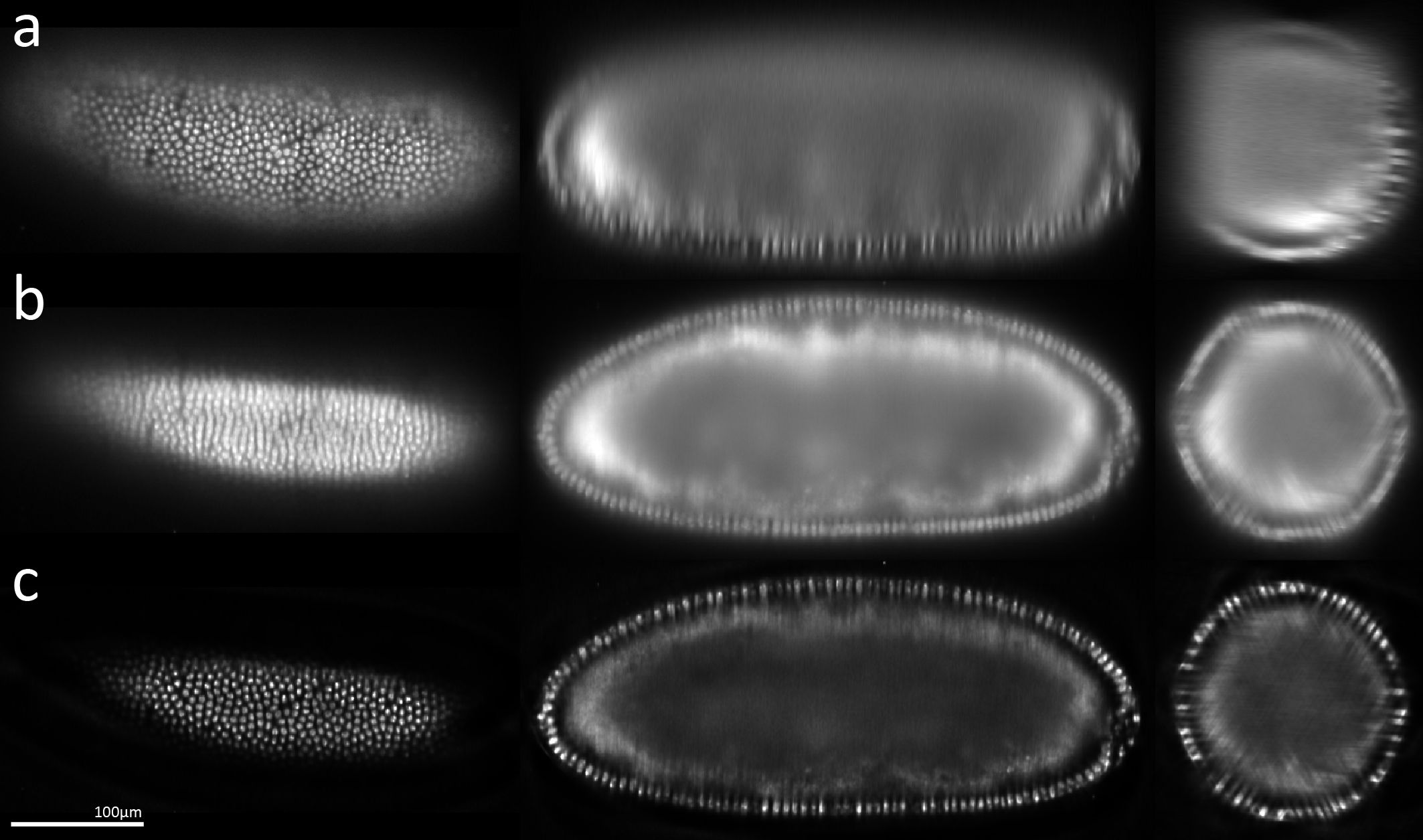}
\vspace{-2.0mm}
\caption{\hspace{-0.5mm} \emph{Comparison of reconstruction quality on the OpenSPIM.} (\textbf{a}) Quality of one of the input views as acquired by the OpenSPIM microscope. (\textbf{b}) Quality of the content-based fusion of the registered dataset. (\textbf{c}) Quality of the deconvolution of the registered dataset. (\textbf{a-c}) The first column shows a slice in the lateral orientation of the input dataset, the second column shows an orthogonal slice, the third column shows a slice perpendicular to the rotation axis.  All slices are in the exactly same position and show the identical portion of each volume and are directly comparable.  The light sheet thickness of the OpenSPIM is larger than of Zeiss prototype, therefore more out-of-focus light is visible and (a,b) are more blurred. Therefore the effect of deconvolution is especially visible, most dominantly in the third column showing the slice perpendicular to the rotation axis. The dataset has a size of 793$\times$384$\times$370~px, acquired with in 6 views totalling around 680 million pixels and 2.6 gigabytes of data. Computation time for 12 iterations was 12 minutes on two Nvidia Quadro 4000 GPU's using optimization I.
}\label{fig:openspim}
\end{center}
\end{figure*}

\pagebreak

\subsection*{SUPPLEMENTARY FIGURE 16 | Quality of reconstruction of Drosophila ovaries}

\vspace{1mm}

\begin{figure*}[h!]
\begin{center}
\includegraphics[width=\textwidth]{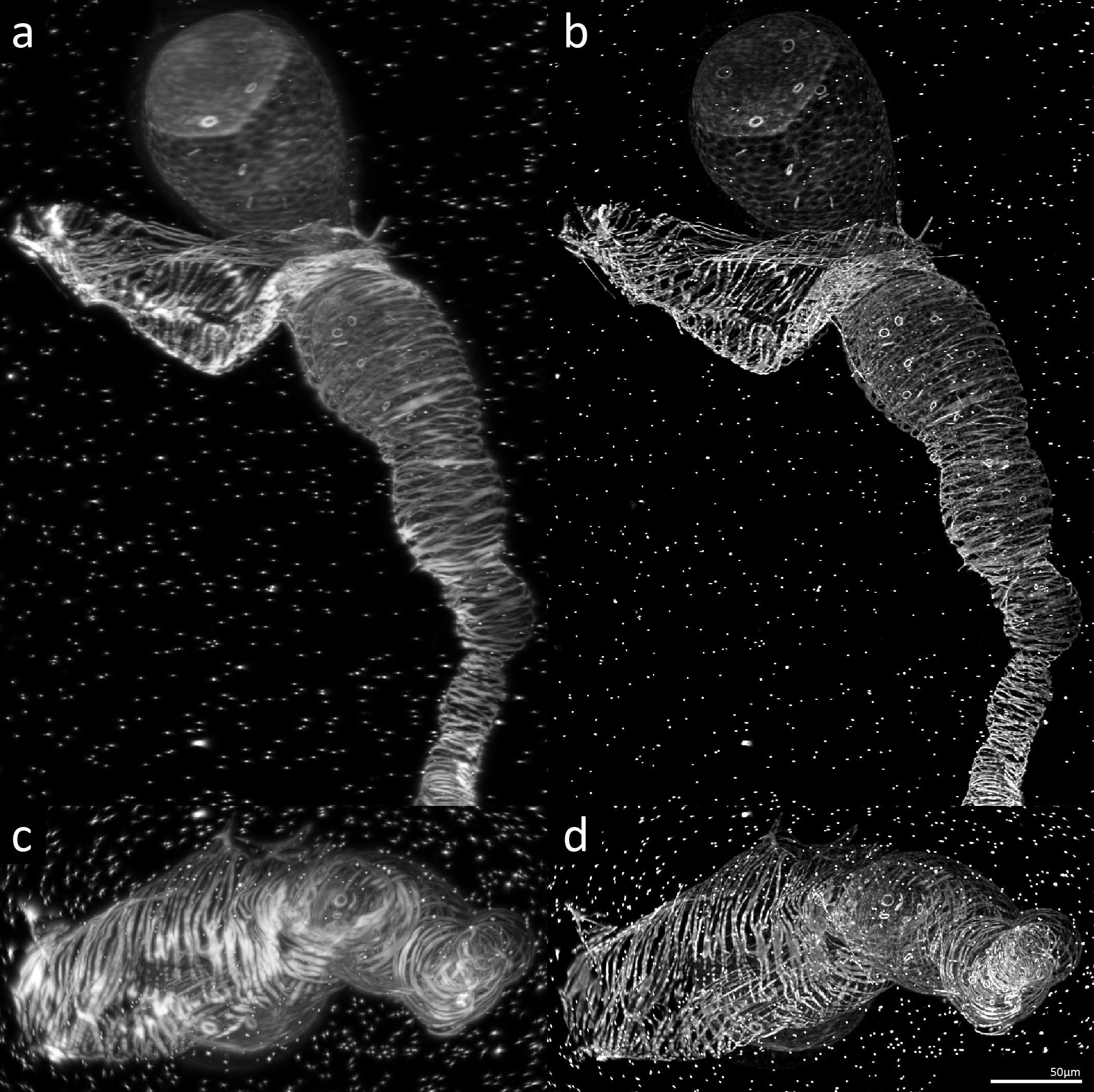}
\vspace{-2.0mm}
\caption{\hspace{-0.5mm} \emph{Comparison of reconstruction quality of \emph{Drosophila} ovaries acquired on the Zeiss SPIM prototype using maximum intensity projections.} (\textbf{a}) shows the content-based fusion along the orientation of one of the acquired views. (\textbf{b}) shows the same image deconvolved. (\textbf{c}) shows the projection along the rotation axis of the content-based fusion, (\textbf{d}) of the deconvolved dataset. The final dataset has a size of 822$\times$1211$\times$430~px, acquired in 12 views totalling an input size of around 5 billion pixels and 19 gigabytes of data (32 bit floating point data required for deconvolution). Computation time for 12 iterations was 36 minutes on two Nvidia Quadro 4000 GPU's using optimization I.
}\label{fig:eggchamber}
\end{center}
\end{figure*}

\pagebreak

\subsection*{SUPPLEMENTARY FIGURE 17 | Effects of partial overlap and CUDA performance}

\vspace{1mm}

\begin{figure*}[h!]
\begin{center}
\includegraphics[width=\textwidth]{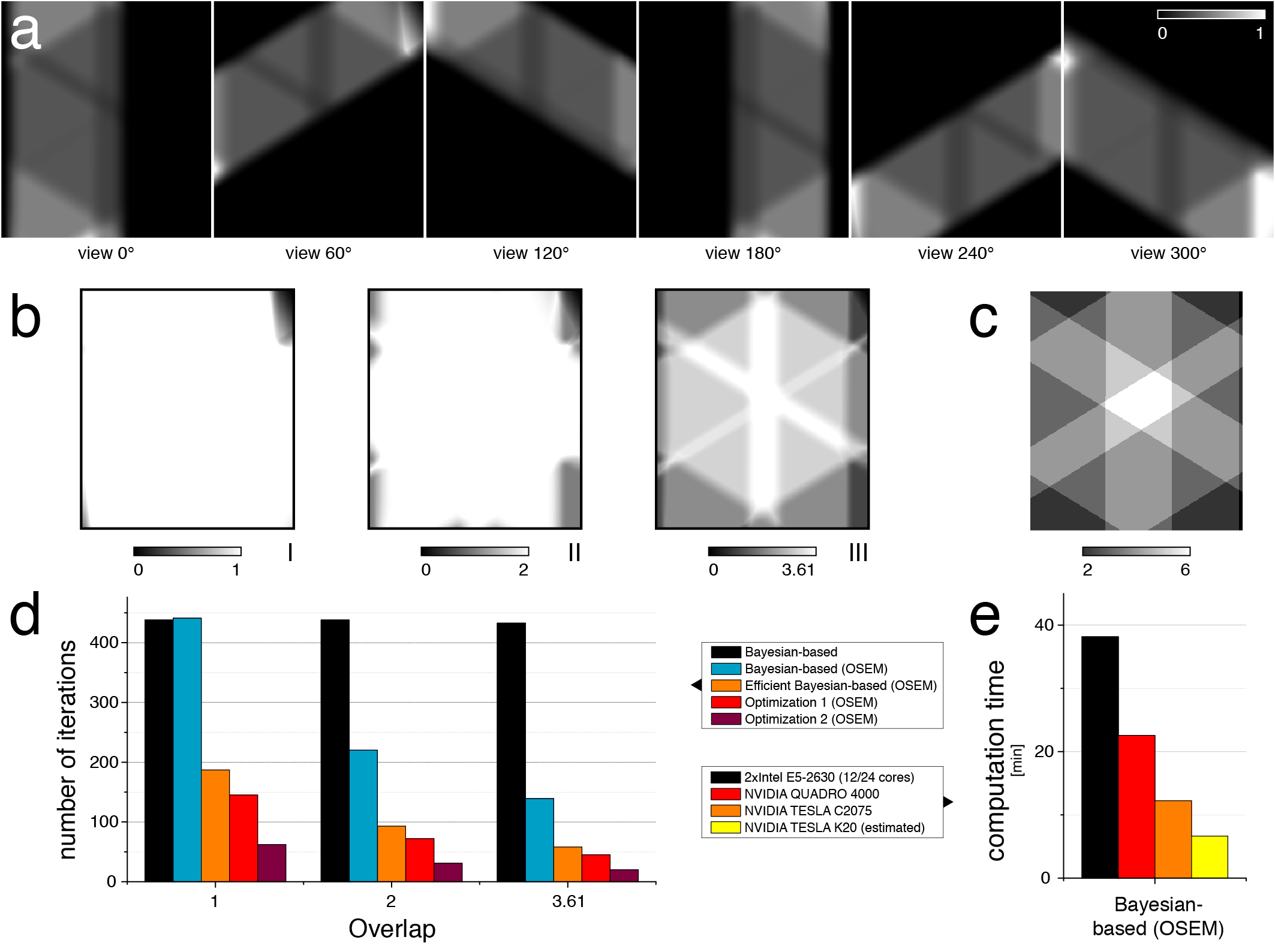}
\vspace{-4.0mm}
\caption{\hspace{-0.5mm} \emph{Effects of partial overlap and CUDA performance.} (\textbf{a}) shows for one slice perpendicular to the rotation axis the weights for each pixel of each view of a six-view SPIM acquistion (see \fig \ref{fig:SIM}e-j for image data). Every view only covers part of the entire volume. Close to the boundaries of each view we limit its contribution using a cosine blending function preventing artifacts due to sharp edges\cite{Preibisch2010}. For each individual pixel the sum of weights over all views is normalized to be $\leq$1. Black corresponds to a weight of 0 (this view is not contributing), white to a weight of 1 (only this view is contributing). (\textbf{b}) illustrates how much each pixel is deconvolved in every iteration when using different amounts of OSEM speedup (i.e. assuming a certain amount of overlapping views). Note that individual weights must not be $>$1. (\textbf{b}--I) normalizing the sum of all weights to $\leq$1 results in a uniformly deconvolved image except the corners where the underlying data is missing, however no speedup is achieved by OSEM (\textbf{d}~left).  Note that summing up all 6 images from (\textbf{a}) results in this image. (\textbf{b}--II) two views is the minimal number of overlapping views at every pixel (see \textbf{c}), so normalization to $\leq$2 still provides a pretty uniform deconvolution and a 2-fold speed up (\textbf{d} center). (\textbf{b}--III) normalizing to $\leq$3.61 (average number of overlapping views) results in more deconvolution of center parts, which is not desireable. Many parts of the image are not covered by enough views to achieve a sum of weights of 3.61.
(\textbf{d}) performance improvement of partially overlapping datasets using the weights pictured above and a cropped version of the ground truth image (\fig \ref{fig:viewsimages}). The effect is identical to perfectly overlapping views, but the effective number of overlapping views is reduced. Our new optimizations improve performance in any case. (\textbf{e}) the relative speed-up of deconvolution performance that can be achieved using our CUDA implementation.
}\label{fig:partial-overlap}
\end{center}
\end{figure*}

\pagebreak

\section*{SUPPLEMENTARY TABLES}

\hspace{20mm}

\subsection*{SUPPLEMENTARY TABLE 1 | Summary of datasets used in this publication}

\hspace{20mm}

\setcounter{table}{0} 
\newcounter{savecntr1}
\newcounter{restorecntr1}
\newcounter{savecntr2}
\newcounter{restorecntr2}

\begin{savenotes}
\begin{table}[h!]
\center
{
\fontsize{9pt}{10pt}\selectfont
\center
\begin{tabular}{p{5.5cm}p{3.0cm}p{3.6cm}p{3.5cm}}
\textbf{Dataset} & \textbf{Size, Lightsheet} & \textbf{Computation Time,} & \textbf{Machine}\\ 
 & \textbf{Thickness, SNR\footnote{The SNR is estimated by computing the average intensity of the signal, divided by the standard deviation of the signal in areas with homogenous sample intensity}}  &\textbf{Iterations, Method} &  \\
\\
\hline
\\
\emph{Drosophila} embryo expressing His-YFP in all cells  acquired with Zeiss SPIM prototype using a 20x/0.5 detection objective (\textbf{Fig. 2c-e}, \textbf{Supp. Fig. \ref{fig:resolution}})\setcounter{savecntr1}{\value{footnote}}\footnote{This SPIM acquisition was already used in Preibisch (2010)\cite{Preibisch2010} to illustrate the results of the bead-based registration and multi-view fusion; we use the underlying dataset again to illustrate the improved results of the multi-view deconvolution.} & \mbox{720$\times$380$\times$350~px}, \mbox{7~views}, \mbox{LS$\sim$5$\mu$m, SNR$\sim$30} & \mbox{7~minutes,~~~~~~~~~~~~~~~~} \mbox{12 iterations}, \mbox{optimization I}, $\lambda$~=~0.006 & \mbox{\textcolor{red}{2$\times$~Nvidia~Quadro~4000}\setcounter{savecntr2}{\value{footnote}}\footnote{Two graphics cards in one PC, which can process two 512$\times$512$\times$512 blocks in parallel}}, 64~GB RAM\\
\\
\emph{Drosophila} embryo expressing His-YFP in all cells acquired with the OpenSPIM using a 20x/0.5 detection objective (\textbf{Supp. Fig. \ref{fig:openspim}}) & \mbox{793$\times$384$\times$370~px}, \mbox{6~views}, \mbox{LS$\sim$10$\mu$m, SNR$\sim$15} & \mbox{12 minutes\footnote{Note that the increased computation time is due to larger anisotropy of the acquired stacks leading to larger effective PSF sizes, which increases  computational effort. The image could therefore not be split up into two \mbox{512$\times$512$\times$512} blocks.},~~~~~~~~~~~~~~~~} \mbox{12 iterations}, \mbox{optimization I}, $\lambda$~=~0.006 & \mbox{\textcolor{red}{2$\times$~Nvidia~Quadro~4000}}\setcounter{restorecntr2}{\value{footnote}}\setcounter{footnote}{\value{savecntr2}}\footnotemark \setcounter{footnote}{\value{restorecntr2}}, 64~GB RAM\\ 
\\
\emph{Drosophila} ovaries acquired on the Zeiss SPIM prototype using a 20x/0.5 detection objective (\textbf{Supp. Fig. \ref{fig:eggchamber}}) & \mbox{1211$\times$822$\times$430~px}, \mbox{12 views}, \mbox{LS$\sim$5$\mu$m, SNR$\sim$19} & \mbox{36 minutes,~~~~~~~~~~~~~~~~} \mbox{12 iterations}, \mbox{optimization I}, $\lambda$~=~0.006  & \mbox{\textcolor{red}{2$\times$~Nvidia~Quadro~4000}}\setcounter{restorecntr2}{\value{footnote}}\setcounter{footnote}{\value{savecntr2}}\footnotemark \setcounter{footnote}{\value{restorecntr2}}, 64~GB RAM \\
\\
\emph{Drosophila} embryo expressing His-YFP in all cells acquired with Zeiss SPIM prototype using a 20x/0.5 detection objective (\textbf{Supp. Video 2-4},\textbf{Supp. Fig. \ref{fig:SIM}}) & \mbox{792$\times$320$\times$310~px}, \mbox{6~views}, \mbox{236 timepoints}, \mbox{LS$\sim$5$\mu$m, SNR$\sim$26} & \mbox{24.3 hours,~~~~~~~~~~~~~~~~} \mbox{12 iterations}, \mbox{optimization I}, $\lambda$~=~0.006 & \mbox{\textcolor{red}{2$\times$~Nvidia~Quadro~4000}}\setcounter{restorecntr2}{\value{footnote}}\setcounter{footnote}{\value{savecntr2}}\footnotemark \setcounter{footnote}{\value{restorecntr2}}, 64~GB RAM\\
\\
\emph{Drosophila} embryo expressing Histone-H2Av-mRFPruby fusion in all cells imaged on Zeiss Lightsheet Z1 with a 20x/1.0 detection objective and dual-sided illumination & \mbox{928$\times$390$\times$390~px}, \mbox{6~views}, \mbox{715~timepoints}, \mbox{LS$\sim$5$\mu$m, SNR$\sim$21} & \mbox{35 hours,~~~~~~~~~~~~~~~~} \mbox{10 iterations}, \mbox{optimization I}, $\lambda$~=~0.0006 & \mbox{\textcolor{red}{4$\times$~Nvidia~TESLA}}\footnote{Run on a cluster with 4 nodes that are equipped with one Nvidia TESLA and 64~GB of system memory}, 64~GB RAM \\
\\
\emph{C. elegans} embryo in 4-cell stage expressing PH-domain-GFP fusion acquired with Zeiss SPIM prototype using a 40x/0.8 detection objective (\textbf{Fig. 2a,b})\setcounter{restorecntr1}{\value{footnote}}\setcounter{footnote}{\value{savecntr1}}\footnotemark \setcounter{footnote}{\value{restorecntr1}} & \mbox{180$\times$135$\times$180~px}, \mbox{6~views}, \mbox{LS$\sim$3.5$\mu$m, SNR$\sim$40} & \mbox{1 minute,~~~~~~~~~~~~~~~~} \mbox{20 iterations}, \mbox{optimization I}, $\lambda$~=~0.006 & \mbox{\textcolor{blue}{2$\times$~Intel~Xeon E5-2630}}, 64~GB RAM\\
\\
Fixed \emph{C. elegans} larvae in L1 stage expressing LMN-1::GFP and stained with Hoechst imaged on Zeiss Lightsheet Z1 with a 20x/1.0 detection objective (\textbf{Fig. 2f,g} and \textbf{Supp. Video 5-8}) & \mbox{1640$\times$1070$\times$345~px}, \mbox{4 views}, \mbox{2~channels}, \mbox{LS$\sim$2$\mu$m}, \mbox{SNR$\sim$62 (Hoechst)}, \mbox{SNR$\sim$24 (GFP)} & \mbox{2$\times$160 minutes,~~~~~~~~~~~~~~~~} \mbox{100 iterations}, \mbox{optimization II}, $\lambda$~=~0 & \mbox{\textcolor{blue}{2$\times$~Intel~Xeon E5-2690}}, 128~GB RAM \\
\\
\end{tabular}}
\caption{ }
\label{tab:experiments}
\end{table}
\end{savenotes}

\hspace{20mm}

\subsection*{SUPPLEMENTARY TABLE 1 (CONTINUED) | Summary of datasets used in this publication}

\hspace{20mm}

\begin{savenotes}
\begin{table}[h!]
\center
{
\fontsize{9pt}{10pt}\selectfont
\center
\begin{tabular}{p{5.5cm}p{3.0cm}p{3.6cm}p{3.5cm}}
\textbf{Dataset} & \textbf{Size, Lightsheet} & \textbf{Computation Time,} & \textbf{Machine}\\ 
 & \textbf{Thickness, SNR}  &\textbf{Iterations, Method} &  \\
\\
\hline
\\
\emph Fixed {C. elegans} in L1 stage stained with Sytox green acquired with \textcolor{red}{Spinning Disc Confocal} using a 20x/0.5 detection objective (\textbf{Supp. Fig. \ref{fig:spinningdisc}})\footnote{This multi-view spinning disc acquisition was already used in Preibisch (2010)\cite{Preibisch2010} to illustrate the applicability of the bead-based registration and multi-view fusion to other technologies than SPIM; we use the underlying dataset again to illustrate the improved results and applicability of the multi-view deconvolution.} & \mbox{1135$\times$400$\times$430~px}, \mbox{5~views}, \mbox{LS N/A, SNR$\sim$28} & \mbox{36 minutes,~~~~~~~~~~~~~~~~} \mbox{50 iterations}, \mbox{optimization II}, $\lambda$~=~0.0006 & \mbox{\textcolor{blue}{2$\times$~Intel~Xeon E5-2680}}, 128~GB RAM\\
\\
\emph Fixed {C. elegans} in L1 stage stained with Sytox green acquired with \textcolor{red}{Spinning Disc Confocal} using a 20x/0.5 detection objective (\textbf{Supp. Fig. \ref{fig:spinningdisc}})\footnote{This is the same dataset as in the row above, but showing the time it took to compute the single-view deconvolution.} & \mbox{1151$\times$426$\times$190~px}, \textcolor{red}{\mbox{1~view}}, \mbox{LS N/A, SNR$\sim$28} & \mbox{202 minutes,~~~~~~~~~~~~~~~~} \mbox{900 iterations}, \textcolor{red}{\mbox{Lucy-Richardson}}, $\lambda$~=~0.0006 & \mbox{\textcolor{blue}{2$\times$~Intel~Xeon E5620}}, 64~GB RAM\\
\\
\emph Fixed {Drosophila} embryo stained with Sytox green acquired on the Zeiss SPIM prototype using a 20x/0.5 detection objective (\textbf{Supp. Fig. \ref{fig:twophoton}}) & \mbox{642$\times$316$\times$391~px}, \mbox{9~views}, \mbox{LS$\sim$5$\mu$m, SNR$\sim$20} & \mbox{15 minutes,~~~~~~~~~~~~~~~~} \mbox{15 iterations}, \mbox{optimization I}, $\lambda$~=~0.006 & \mbox{\textcolor{blue}{2$\times$~Intel~Xeon E5-2680}}, 128~GB RAM\\
\\
\emph Fixed {Drosophila} embryo stained with Sytox green acquired on a \textcolor{red}{Two-Photon Microscope} using a 20x/0.8 detection objective (\textbf{Supp. Fig. \ref{fig:twophoton}}) & \mbox{856$\times$418$\times$561~px}, \textcolor{red}{\mbox{1~view}}, \mbox{LS N/A, SNR$\sim$7} & \mbox{160 minutes,~~~~~~~~~~~~~~~~} \mbox{300 iterations}, \textcolor{red}{\mbox{Lucy-Richardson}}, $\lambda$~=~0.006 & \mbox{\textcolor{blue}{2$\times$~Intel~Xeon E5620}}, 64~GB RAM\\
\\
\end{tabular}}
\caption{Supplementary Table 1: \emph{Summary of all datasets used in this publication}. Note that the multi-view deconvolution of the \emph{C. elegans} larvae in L1 stage (SPIM \& Spinning Disc Confocal) required an additional registration step, which is explained in section \ref{sec:l1}.}
\label{tab:experiments2}
\end{table}
\end{savenotes}

\pagebreak

\section*{SUPPLEMENTARY NOTE}

\hspace{20mm}

\section{REMARKS}
\label{sec:remarks}

This document closely follows the notation introduced in the paper of L. B. Lucy\cite{lucy1974} whenever possible. 
Note that for simplicity the derivations in this document only cover the one dimensional case. Nevertheless, all equations are valid for any \emph{n}-dimensional case.

\section{BAYESIAN-BASED SINGLE-VIEW DECONVOLUTION}
\label{sec:singleview}

This section re-derives the classical bayesian-based Richardson\cite{richardson1972}-Lucy\cite{lucy1974} deconvolution for single images, other derivations presented in this document build up on it. The goal is to estimate the frequency distribution of an underlying signal $\psi(\xi)$ from a finite number of measurements $x^{1'}, x^{2'}, ..., x^{N'}$. The resulting observed distribution $\phi(x)$ is defined as
\begin{equation}
\label{eq:eq1}
\phi(x) = \int_{\xi}{\psi(\xi)P(x|\xi)}d\xi
\end{equation}
where $P(x|\xi)$ is the probability of a measurement occuring at $x=x'$ when it is known that the event \mbox{$\xi=\xi'$} occured. In more practical image analysis terms equation \ref{eq:eq1} describes the one-dimensional convolution operation where $\phi(x)$ is the blurred image, $P(x|\xi)$ is the kernel and $\psi(\xi)$ is the undegraded (or deconvolved) image. All distributions are treated as probability distributions and fulfill the following constraints:
\begin{equation}
\label{eq:eq3}
\int_{\xi}{\psi(\xi)}d\xi = \int_{x}{\phi(x)}dx = \int_{x}{P(x|\xi)}dx = 1 ~~~and~~~ \psi(\xi) > 0, ~\phi(x) \geq 0, ~P(x|\xi) \geq 0
\end{equation}

\subsection{Derivation of the iterative deconvolution scheme}

\noindent The basis for the derivation of the bayesian-based deconvolution is the tautology 
\begin{equation}
\label{eq:eq5}
P(\xi=\xi' \wedge x=x') = P(x=x' \wedge \xi=\xi')
\end{equation}
It states that it is equally probable that the event $\xi'$ results in a measurement at $x'$ and that the measurement at $x'$ was caused by the event $\xi'$. Integrating equation \ref{eq:eq5} over the measured distribution yields the joint probability distribution
\begin{equation}
\int_x P(\xi \wedge x) dx = \int_x P(x \wedge \xi) dx
\end{equation}
which can be expressed using conditional probabilities 
\begin{equation}
\int_x P(\xi) P(x|\xi) dx = \int_x P(x) P(\xi|x) dx 
\end{equation}
and in correspondence to Lucy's notation looks like (equation \ref{eq:eq1})
\begin{equation}
\label{eq:eq7}
\int_x \psi(\xi) P(x|\xi) dx = \int_x \phi(x) Q(\xi|x) dx 
\end{equation}
where $P(\xi) \equiv \psi(\xi), P(x) \equiv \phi(x), P(\xi|x) \equiv Q(\xi|x)$. $Q(\xi|x)$ denotes what Lucy calls the 'inverse' conditional probability to $P(x|\xi)$. It defines the probability that an event at $\xi'$ occured, given a specific measurement at $x'$. As $\psi(\xi)$ does not depend on $x$, equation \ref{eq:eq7} can be rewritten as
\begin{equation}
\label{eq:eq9}
\psi(\xi) \overbrace{\int_x P(x|\xi) dx}^{=1} = \int_x \phi(x) Q(\xi|x) dx 
\end{equation}
hence (due to equation \ref{eq:eq3})
\begin{equation}
\label{eq:eq11}
\psi(\xi) = \int_x \phi(x) Q(\xi|x) dx 
\end{equation}
which corresponds to the inverse of the convolution in equation \ref{eq:eq1}. Although $Q(\xi|x)$ cannot be used to directly compute $\psi(\xi)$, \emph{Bayes' Theorem} and subsequently equation \ref{eq:eq1} can be used to reformulate it as
\begin{equation}
\label{eq:eq12}
Q(\xi|x) = \frac{\psi(\xi) P(x|\xi)}{\phi(x)} =  \frac{\psi(\xi) P(x|\xi)}{\int_\xi \psi(\xi) P(x|\xi) d\xi}
\end{equation}
Replacing $Q(\xi|x)$ in equation \ref{eq:eq11} yields
\begin{equation}
\label{eq:eq13}
\psi(\xi) = \int_x \phi(x) \frac{\psi(\xi) P(x|\xi)}{\int_\xi \psi(\xi) P(x|\xi) d\xi} dx
          = \psi(\xi) \int_x  \frac{ \phi(x) }{\int_\xi \psi(\xi) P(x|\xi) d\xi} P(x|\xi) dx 
\end{equation}
which exactly re-states the deconvolution scheme introduced by Lucy and Richardson. The fact that both sides of the equation contain the desired underlying (deconvolved) distribution $\psi(\xi)$ suggests an iterative scheme to converge towards the correct solution
\begin{equation}
\label{eq:eq15}
\psi^{r+1}(\xi) = \psi^r(\xi) \int_x \frac{ \phi(x) }{\int_\xi \psi^r(\xi) P(x|\xi) d\xi} P(x|\xi) dx 
\end{equation}
where $\psi^0(\xi)$ is simply a constant distribution with each value being the average intensity of the measured distribution $\phi(x)$.  

Equation \ref{eq:eq15} turns out to be a maximum-likelihood (ML) expection-maximization (EM) formulation\cite{Dempster77}, which works as follows. First, it computes for every pixel the convolution of the current guess of the deconvolved image $\psi^r(\xi)$ with the kernel (PSF) $P(x|\xi)$, i.e. $\phi^r(x) = \int_\xi \psi^r(\xi) P(x|\xi) d\xi$. In EM-terms $\phi^r(x)$ describes the \emph{expected value}.  The quotient between the input image $\phi(x)$ and the \emph{expected value} $\phi^r(x)$ yields the disparity for every pixel. These values are initially large but will become very small upon convergence. In an ideal scenario all values of $\phi^r(x)$ and $\phi(x)$ will be identical once the algorithm converged. This ratio is subsequently convolved with the point spread function $P(x|\xi)$ reflecting which pixels influence each other. In EM-terms this is called the \emph{maximization step}. This also preserves smoothness. These resulting values are then pixel-wise multiplied with the current guess of the deconvolved image $\psi^r(\xi)$, which we call an RL-update (Richardson-Lucy). It results in a new guess for the deconvolved image.

Starting from an initial guess of an image with constant values, this scheme will converge towards the correct solution if the guess of the point spread function is correct and if the observed distribution is not degraded by noise, transformations, etc.

\subsubsection{Integrating $\xi$ and $x$}
\label{sec:mirroredkernels}

Note that convolution of $\psi^r(\xi)$ with $P(x|\xi)$ requires integration over $\xi$, while the convolution of the quotient image with $P(x|\xi)$ integrates over $x$. Integration over $x$ can be formulated as convolution if $P(x|\xi)$ is constant by using inverted coordinates $P(-x|\xi)$. Note that it can be ignored if the kernel is symmetric $P(x|\xi)=P(-x|\xi)$. For single-view datasets this is often the case, whereas multi-view datasets typically have non-symmetric kernels due to their transformations resulting from image alignment.

\section{BAYESIAN-BASED MULTI-VIEW DECONVOLUTION}
\label{sec:multiview} 

This section shows for the first time the entire derivation of bayesian-based multi-view deconvolution using probabilty theory. Compared to the single-view case we have a set of views $V = \{ v_1 ... v_{N} : N = |V| \}$ comprising $N$ observed distributions $\phi_v(x_v)$ (input views acquired from different angles), $N$ point spread functions $P_v(x_v|\xi)$ corresponding to each view, and one underlying signal distribution $\psi(\xi)$ (deconvolved image). The observed distributions $\phi_v(x_v)$ are accordingly defined as
\begin{eqnarray}
 \phi_1(x_1) & = & \displaystyle \int_{\xi}{\psi(\xi)P(x_1|\xi)}d\xi \\
 \phi_2(x_2) & = & \displaystyle \int_{\xi}{\psi(\xi)P(x_2|\xi)}d\xi \\
 & ... & \\
 \phi_N(x_N) & = & \displaystyle \int_{\xi}{\psi(\xi)P(x_N|\xi)}d\xi
\end{eqnarray}
The basis for the derivation of the bayesian-based multi-view deconvolution is again a tautology based on the individual observations
\begin{equation}
\label{eq:eq22}
P(\xi=\xi' \wedge x_1=x_1' \wedge ... \wedge x_N=x_N') = P(x_1=x_1' \wedge ... \wedge x_N=x_N' \wedge \xi=\xi')
\end{equation}
Integrating equation \ref{eq:eq22} over the measured distributions yields the joint probability distribution
\begin{equation}
\int_{x_1}...\int_{x_N} P(\xi \wedge x_1 \wedge ... \wedge x_N)dx_1...dx_n = \int_{x_1}...\int_{x_N} P(x_1 \wedge ... \wedge x_N \wedge \xi) dx_1...dx_N
\end{equation}
shortly written as
\begin{equation}
\label{eq:eq23}
\int_{\overline{x}} P(\xi, x_1, ... ,x_N) d\overline{x} = \int_{\overline{x}} P(x_1, ... ,x_N, \xi) d\overline{x}
\end{equation}
By expressing the term using conditional probabilities one obtains
\begin{equation}
\label{eq:eq24}
\int_{\overline{x}} P(\xi)P(x_1|\xi)P(x_2|\xi,x_1) \, ... \, P(x_N|\xi,x_1,...,x_{N-1}) d\overline{x} = 
\int_{\overline{x}} P(x_1)P(x_2|x_1)P(x_3|x_1,x_2) \, ... \, P(\xi|x_1,...,x_N) d\overline{x}
\end{equation}
On the left side of the equation all terms are conditionally independent of any $x_v$ given $\xi$. This results from the fact that if an event $\xi=\xi'$ occured, each individual measurement $x_v$ depends only on $\xi'$ and the respective point spread function $P(x_v|\xi)$ (\fig \ref{fig:01}a for illustration). Equation \ref{eq:eq24} therefore reduces to
\begin{equation}
\label{eq:eq26}
P(\xi) \overbrace{\int_{\overline{x}} P(x_1|\xi)P(x_2|\xi) \, ... \, P(x_N|\xi) d\overline{x}}^{=1} = 
\int_{\overline{x}} P(x_1)P(x_2|x_1)P(x_3|x_1,x_2) \, ... \, P(x_N|x_1,...,x_{N-1})P(\xi|x_1,...,x_N) d\overline{x}
\end{equation}
Assuming independence of the observed distributions $P(x_v)$ equation \ref{eq:eq26} further simplifies to
\begin{equation}
\label{eq:eq28}
P(\xi) = \int_{\overline{x}} P(x_1)P(x_2)P(x_3) \, ... \, P(x_N) P(\xi|x_1,...,x_N) d\overline{x}
\end{equation}
Although independence between the views is assumed\cite{KrzicPhD, PreibischPhD, Temerinac2012}, the underlying distribution $P(\xi)$ still depends on the observed distributions $P(x_v)$ through $P(\xi|x_1,...,x_N)$. 

\emph{Note: In section \ref{sec:proofmv} we will show that that the derivation of bayesian-based multi-view deconvolution can be be achieved without assuming independence of the observed distributions $P(x_v)$. Based on that derivation we argue that there is a relationship between the $P(x_v)$'s (\fig \ref{fig:01}b for illustration) and that it can be incorporated into the derivation to achieve faster convergence as shown in sections \ref{sec:efficientmv} and \ref{sec:simplifications}.}

We cannot approximate $P(\xi|x_1,...,x_N)$ directly and therefore need to reformulate it in order to express it using individual $P(\xi|x_v)$, which can subsequently be used to formulate the deconvolution task as shown in section \ref{sec:singleview}.  Note that according to Lucy's notation $P(\xi|x_1,...,x_N) \equiv Q(\xi|x_1,...,x_N)$ and $P(\xi|x_v) \equiv Q(\xi|x_v)$.
\begin{eqnarray}
\label{eq:eq32}
P(\xi|x_1,...,x_N) & = & \frac{P(\xi,x_1,...,x_N)}{P(x_1,...,x_N)} \\
\label{eq:eq34}
P(\xi|x_1,...,x_N) & = & \frac{P(\xi)P(x_1|\xi)P(x_2|\xi,x_1) ... P(x_N|\xi,x_1,...,x_{N-1})}
{P(x_1)P(x_2|x_1)...P(x_N|x_1,...,x_{N-1})}
\end{eqnarray}
Due to the conditional independence of the $P(x_v)$ given $\xi$ (equation \ref{eq:eq24} $\to$ \ref{eq:eq26} and \fig \ref{fig:01}a) and the assumption of independence between the $P(x_v)$ (equation \ref{eq:eq26} $\to$ \ref{eq:eq28}) equation \ref{eq:eq34} simplifies to
\begin{equation}
\label{eq:eq35}
P(\xi|x_1,...,x_N) = \frac{P(\xi)P(x_1|\xi) ... P(x_N|\xi)}{P(x_1)...P(x_N)}
\end{equation}
Using \emph{Bayes' Theorem} to replace all
\begin{equation}
P(x_v|\xi) = \frac{P(x_v)P(\xi|x_v)}{P(\xi)}
\end{equation}
yields
\begin{eqnarray}
\label{eq:eq36}
P(\xi|x_1,...,x_N) & = & \frac{P(\xi)~P(\xi|x_1)P(x_1) ... P(\xi|x_N)P(x_N)}{P(x_1)...P(x_N)~{P(\xi)}^N} \\
\label{eq:eq38}
P(\xi|x_1,...,x_N) & = & \frac{P(\xi)~P(\xi|x_1) ... P(\xi|x_N)}{{P(\xi)}^N} \\
\label{eq:eq40}
P(\xi|x_1,...,x_N) & = & \frac{P(\xi|x_1) ... P(\xi|x_N)}{{P(\xi)}^{N-1}}
\end{eqnarray}
Substituting equation \ref{eq:eq40} in equation \ref{eq:eq28} yields
\begin{equation}
\label{eq:eq44}
P(\xi) = \frac{\displaystyle\int_{\overline{x}} P(x_1) \, ... \, P(x_N) P(\xi|x_1) ... P(\xi|x_N) d\overline{x}}{ {P(\xi)}^{N-1} }
\end{equation}
and rewritten in Lucy's notation
\begin{eqnarray}
\label{eq:eq46}
\psi(\xi) & = & \frac{\displaystyle\int_{\overline{x}} \phi_1(x_1) \, ... \, \phi_N(x_N) Q(\xi|x_1) ... Q(\xi|x_N) d\overline{x}}{ {\psi(\xi)}^{N-1} } \\
\psi(\xi) & = & \frac{\displaystyle\int_{x_1} \phi_1(x_1)Q(\xi|x_1) dx_1 \, ... \, \int_{x_N} \phi_N(x_N)Q(\xi|x_N) dx_N }{ {\psi(\xi)}^{N-1} } \\
\psi(\xi) & = & \frac{\displaystyle\prod_{v \in V}\int_{x_v} \phi_v(x_v)Q(\xi|x_v) dx_v }{ {\psi(\xi)}^{N-1} } 
\end{eqnarray}
As in the single view case we replace $Q(\xi|x_v)$ with equation \ref{eq:eq12}
\begin{eqnarray}
 \psi(\xi) & = & \frac{ \displaystyle\prod_{v \in V} \int_{x_v} \phi_v(x_v) \frac{\psi(\xi) P(x_v|\xi)}{\int_\xi \psi(\xi) P(x_v|\xi) d\xi} dx_v }{ {\psi(\xi)}^{N-1} } \\      
 \psi(\xi) & = & \frac{ \psi(\xi)^{\xcancel{N}} \displaystyle\prod_{v \in V} \int_{x_v} \phi_v(x_v) \frac{ P(x_v|\xi)}{\int_\xi \psi(\xi) P(x_v|\xi) d\xi} dx_v }{ {\xcancel{ \psi(\xi)}^{N-1} } } \\      
 \psi(\xi) & = & \psi(\xi) \displaystyle\prod_{v \in V} \int_{x_v} \phi_v(x_v) \frac{ P(x_v|\xi)}{\int_\xi \psi(\xi) P(x_v|\xi) d\xi} dx_v
\end{eqnarray}
As in the single view case, both sides of the equation contain the desired deconvolved distribution $\psi(\xi)$. This again suggests the final iterative scheme 
\begin{equation}
\label{eq:eq50}
 \psi^{r+1}(\xi) = \psi^r(\xi) \displaystyle\prod_{v \in V} \int_{x_v} \frac{ \phi_v(x_v) }{\int_\xi \psi^r(\xi) P(x_v|\xi) d\xi} P(x_v|\xi) dx_v
\end{equation}
where $\psi^{0}(\xi)$ is considered a distribution with a constant value. Note that the final derived equation \ref{eq:eq50} ends up being the per pixel multiplication of the single view RL-updates from equation \ref{eq:eq15}. 

It is important to note that the maximum-likelihood expectation-maximization based derivation\cite{Shepp1982} yields an additive combination of the individual RL-updates, while our derivation based probability theory and \emph{Bayes' Theorem} ends up being a multiplicative combination. However, our derivation enables us to prove (section \ref{sec:proofmv}) that this formulation can be achieved without assuming independence of the observed distributions (input views), which allows us to introduce optimizations to the derviation (sections \ref{sec:efficientmv} and \ref{sec:simplifications}). We additionally proof of in section \ref{sec:proofconvergence} the convergence of our multiplicative derivation to the maximum-likelihood solution.

\section{PROOF OF CONVERGENCE}
\label{sec:proofconvergence}

This section proofs that our Bayesian-based derivation of multi-view deconvolution (equation \ref{eq:eq50}) converges to the maximum-likelihood (ML) solution using noise-free data.  We chose to adapt the proof developed for \emph{Ordered Subset Expectation Maximization} (OS-EM)\cite{Hudson1994} due to its similarity to our derivation (see section \ref{sec:optimizeconvergence}).  

\subsection{PROOF FOR NOISE-FREE DATA}

Assuming the existence of a feasible solution $\psi^{*}$ it has been shown\cite{Hudson1994} that the likelihood $L^r := L( \psi^r; \psi^{*} )$ of the solution $\psi$ at iteration $r$ can be computed as
\begin{equation}
\label{eq:eq1proof}
 L( \psi^r; \psi^{*} ) = - \displaystyle\int_{\xi} \psi^{*}(\xi) \log \frac{\psi^{*}(\xi)}{\psi^{r}(\xi)} d\xi
\end{equation}
Following the argumentations of Shepp and Vardi\cite{Shepp1982}, Kaufmann\cite{Kaufmann1987} and Hudson and Larkin\cite{Hudson1994} convergence of the algorithm is proven if the likelihood of the solution $\psi$ increases with every iteration since $L$ is bounded by $0$. In other words
\begin{eqnarray}
\label{eq:eq2proof}
  \Delta L & = & L^{r+1} - L^r \\
  & \geq & 0
\end{eqnarray}
We will now prove that $\Delta L$ is indeed always greater or equal to zero. Replacing equation \ref{eq:eq1proof} in equation \ref{eq:eq2proof} yields
\begin{eqnarray}
\label{eq:eq3proof}
 \Delta L & = & \displaystyle\int_{\xi}\psi^{*}(\xi) \log \frac{\psi^{*}(\xi)}{\psi^{r}(\xi)} - \psi^{*}(\xi) \log \frac{\psi^{*}(\xi)}{\psi^{r+1}(\xi)} d\xi\\
\label{eq:eq4proof}
 & = & \displaystyle\int_{\xi}\psi^{*}(\xi) \log \frac{\psi^{r+1}(\xi)}{\psi^{r}(\xi)} d\xi
\end{eqnarray}
Next, we substitute $\psi^{r+1}(\xi)$ with our derivation of Bayesian-based multi-view deconvolution (equation \ref{eq:eq50}). Note that for simplicity we replace $\int_\xi \psi^r(\xi) P(x_v|\xi) d\xi$ with $\phi_v^r(x_v)$, which refers to the current estimate of the observed distribution given the current guess of the underlying distribution $\psi^r$ (or in EM terms the \emph{expected value}).
\begin{eqnarray}
\label{eq:eq6proof}
 \Delta L & = & \displaystyle\int_{\xi}\psi^{*}(\xi) \log \frac{ 
    \psi^r(\xi) \left( \displaystyle\prod_{v \in V} \int_{x_v} \frac{ \phi_v(x_v) }{ \phi_v^r(x_v) } P(x_v|\xi) dx_v \right )^{\frac{1}{|V|}}
  }{\psi^{r}(\xi)}d\xi
\\
\label{eq:eq7proof}
    & = & \displaystyle\int_{\xi}\psi^{*}(\xi) \log 
    \left( \displaystyle\prod_{v \in V} \int_{x_v} \frac{ \phi_v(x_v) }{\phi_v^r(x_v)} P(x_v|\xi) dx_v \right )^
    {\frac{1}{|V|}}d\xi
\\
\label{eq:eq8proof}
    & = & \frac{1}{|V|}\displaystyle\sum_{v \in V}\displaystyle\int_{\xi}\psi^{*}(\xi) \log 
     \int_{x_v} \frac{ \phi_v(x_v) }{\phi_v^r(x_v)} P(x_v|\xi) dx_v d\xi
\end{eqnarray}
As equation \ref{eq:eq50} expresses a proportion, it is necessary to normalize for the number of observed distributions $|V|$ and apply the $|V|$'th root, i.e. compute the geometric mean.  Note that this normalization is the equivalent to the division by $|V|$ as applied in the ML-EM derivations\cite{Shepp1982, Temerinac2012} that use the arithmetic mean of the individual RL-updates in order to update underlying distribution.

Using Jensen's inequality equation \ref{eq:eq8proof} can be reformulated (equation \ref{eq:eq9proof}) and further simplified
\begin{eqnarray}
\label{eq:eq9proof}
  \Delta L & \geq & \frac{1}{|V|}\displaystyle\sum_{v \in V}\displaystyle\int_{\xi}\psi^{*}(\xi) 
  \int_{x_v} \log \left(\frac{ \phi_v(x_v) }{\phi_v^r(x_v)} \right) P(x_v|\xi) dx_v d\xi
\\
\label{eq:eq10proof}
  & = & \frac{1}{|V|}\displaystyle\sum_{v \in V} \int_{x_v} \log \left(\frac{ \phi_v(x_v) }{\phi_v^r(x_v)} \right)\displaystyle\int_{\xi}\psi^{*}(\xi) P(x_v|\xi) d\xi dx_v 
\end{eqnarray}
Substituting equation \ref{eq:eq1} in equation \ref{eq:eq10proof} yields
\begin{equation}
\label{eq:eq12proof}
  \Delta L \geq \frac{1}{|V|}\displaystyle\sum_{v \in V} \int_{x_v} \log \left(\frac{ \phi_v(x_v) }{\phi_v^r(x_v)} \right) \phi_v(x_v) dx_v 
\end{equation}
It follows directly that in order to prove that $\Delta L \geq 0$, it is sufficient to prove that
\begin{equation}
\label{eq:eq14proof}
  \int_{x_v} \log \left(\frac{ \phi_v(x_v) }{\phi_v^r(x_v)} \right) \phi_v(x_v) dx_v \geq 0
\end{equation}
Based on the inequality $\log x \geq 1-x^{-1}$ for $x > 0$ proven by \emph{Adolf Hurwitz}, we need to show that
\begin{eqnarray}
\label{eq:eq16proof}
  \int_{x_v} \log \left(\frac{ \phi_v(x_v) }{\phi_v^r(x_v)} \right) \phi_v(x_v) dx_v & \geq & 
  \int_{x_v} \left(1 - \frac{ \phi_v^r(x_v) }{\phi_v(x_v)} \right) \phi_v(x_v) dx_v \geq 0 
\\
\label{eq:eq17proof}
  & = & \int_{x_v} \phi_v(x_v) dx_v - \int_{x_v} \phi_v^r(x_v) dx_v 
\\
\label{eq:eq18proof}
  & = & \int_{x_v} \phi_v(x_v) dx_v - \int_{x_v} \int_\xi \psi^r(\xi) P(x_v|\xi) d\xi dx_v
\\
\label{eq:eq19proof}
  & = & \int_{x_v} \phi_v(x_v) dx_v - \int_\xi \psi^r(\xi)  \overbrace{\int_{x_v} P(x_v|\xi) dx_v}^\text{=1} d\xi
\\
\label{eq:eq20proof}
  & = & \int_{x_v} \phi_v(x_v) dx_v - \int_\xi \psi^r(\xi) d\xi
\\
\label{eq:eq21proof}
 & \geq & 0 \;\; \Longleftrightarrow \; \int_{x_v} \phi_v(x_v) dx_v \geq \int_\xi \psi^r(\xi) d\xi
\end{eqnarray}
In other words, convergence is proven if we show that energy of the underlying distribution $\psi(\xi)$ (deconvolved image) is never greater than energy of each observed distribution $\phi_v(x_v)$ (input views). Replacing $\psi^r(\xi)$ with our Bayesian-based derivation (equation \ref{eq:eq50}), shows that proving the condition in equation \ref{eq:eq21proof} is equivalent to proving
\begin{equation}
\label{eq:eq24proof}
  \int_{x_v} \phi_v(x_v) dx_v \geq \displaystyle\int_{\xi}  
  \psi^r(\xi) \left( \displaystyle\prod_{v \in V} \int_{x_v} \frac{ \phi_v(x_v) }{ \phi_v^r(x_v) } P(x_v|\xi) dx_v \right )^{\frac{1}{|V|}} d\xi 
\end{equation}
Note that as this inequality has to hold for any iteration $r$, we refrain from writing $r-1$ for simplicity. As the arithmetic average is always greater or equal than the geometric average\cite{taschenbuch2004} it follows that proving equation \ref{eq:eq24proof} is equivalent to
\begin{equation}
\label{eq:eq26aproof}
  \int_{x_v} \phi_v(x_v) dx_v 
  \geq \displaystyle\int_{\xi}  
  \psi^r(\xi) \frac{1}{|V|} \displaystyle\sum_{v \in V} \int_{x_v} \frac{ \phi_v(x_v) }{ \phi_v^r(x_v) } P(x_v|\xi) dx_v d\xi 
  \geq \displaystyle\int_{\xi}  
  \psi^r(\xi) \left( \displaystyle\prod_{v \in V} \int_{x_v} \frac{ \phi_v(x_v) }{ \phi_v^r(x_v) } P(x_v|\xi) dx_v \right )^{\frac{1}{|V|}} d\xi
\end{equation}
and we therefore need to prove the inequality only for the arithmetic average. We simplify equation \ref{eq:eq26aproof} as follows
\begin{eqnarray}
\label{eq:eq26bproof}
  \int_{x_v} \phi_v(x_v) dx_v & \geq & \frac{1}{|V|} \displaystyle\sum_{v \in V} \displaystyle\int_{\xi}  
  \psi^r(\xi) \int_{x_v} \frac{ \phi_v(x_v) }{ \phi_v^r(x_v) } P(x_v|\xi) dx_v d\xi 
\\
\label{eq:eq28proof}
  & = & \frac{1}{|V|} \displaystyle\sum_{v \in V} 
  \int_{x_v} \frac{ \phi_v(x_v) }{ \phi_v^r(x_v) } \displaystyle\int_{\xi} \psi^r(\xi) P(x_v|\xi) d\xi dx_v
\\
\label{eq:eq30proof}
  & = & \frac{1}{|V|} \displaystyle\sum_{v \in V} 
  \int_{x_v} \frac{ \phi_v(x_v) }{ \phi_v^r(x_v) } \phi_v^r(x_v) dx_v
\\
\label{eq:eq32proof}
  & = & \frac{1}{|V|} \displaystyle\sum_{v \in V} 
  \int_{x_v} \phi_v(x_v) dx_v
\end{eqnarray}
As the integral of all input views is identical (equation \ref{eq:eq3}), equation \ref{eq:eq32proof} is always true, and we proved that our Bayesian-based derivation of multi-view deconvolution always converges to the Maximum Likelihood.

\section{DERIVATION OF BAYESIAN-BASED MULTI-VIEW DECONVOLUTION WITHOUT ASSUMING INDEPENDENCE OF THE VIEWS}
\label{sec:proofmv}

Previous derivations of the Richardson-Lucy multi-view deconvolution\cite{Shepp1982, KrzicPhD, PreibischPhD, Temerinac2012} assumed independence of the individual views in order to derive variants of equation \ref{eq:eq50}. The following derivation shows that it is actually not necessary to assume independence of the observed distributions $\phi_v(x_v)$ (equation \ref{eq:eq26} $\to$ \ref{eq:eq28} and equation \ref{eq:eq34} $\to$ \ref{eq:eq35}) in order to derive the formulation for bayesian-based multi-view deconvolution shown in equation \ref{eq:eq50}.
 
We therefore rewrite equation \ref{eq:eq28} without assuming independence (which is then identical to equation \ref{eq:eq26}) and obtain
\begin{equation}
\label{eq:eq52}
 P(\xi) = \int_{\overline{x}} P(x_1)P(x_2|x_1) \, ... \, P(x_N|x_1,...,x_{N-1})P(\xi|x_1,...,x_N) d\overline{x}
\end{equation}
We consequently also do not assume independence in equation \ref{eq:eq35}, which intends to replace $P(\xi|x_1,...,x_N)$, and obtain
\begin{equation}
\label{eq:eq54}
 P(\xi|x_1,...,x_N) = \frac{P(\xi)P(x_1|\xi) \, ... \, P(x_N|\xi)}{P(x_1)P(x_2|x_1) \, ... \, P(x_N|x_1,...,x_{N-1})}
\end{equation}
Replacing equation \ref{eq:eq54} in \ref{eq:eq52} yields
\begin{equation}
\label{eq:eq55}
 P(\xi) = \int_{\overline{x}} P(x_1)P(x_2|x_1) \, ... \, P(x_N|x_1,...,x_{N-1}) \frac{P(\xi)P(x_1|\xi) \, ... \, P(x_N|\xi)}{P(x_1)P(x_2|x_1) \, ... \, P(x_N|x_1,...,x_{N-1})} d\overline{x}
\end{equation}
Cancelling out all terms below the fraction bar (from equation \ref{eq:eq54}) with the terms in front of the fraction bar (from equation \ref{eq:eq52}) results in
\begin{equation}
\label{eq:eq56}
 P(\xi) = \int_{\overline{x}} P(\xi)P(x_1|\xi) \, ... \, P(x_N|\xi) d\overline{x}
\end{equation}
Using again \emph{Bayes' Theorem} to replace all
\begin{equation}
 P(x_v|\xi) = \frac{P(x_v)P(\xi|x_v)}{P(\xi)}
\end{equation}
yields
\begin{eqnarray}
 P(\xi) & = & \frac{\displaystyle\int_{\overline{x}} P(\xi)P(x_1)P(\xi|x_1) \, ... \, P(x_N)P(\xi|x_N) d\overline{x}} {P(\xi)^N} \\
 P(\xi) & = & \frac{\displaystyle\int_{\overline{x}} P(x_1) \, ... \, P(x_N)P(\xi|x_1) \, ... \, P(\xi|x_N)d\overline{x}} {P(\xi)^{N-1}} 
\end{eqnarray}
which is identical to equation \ref{eq:eq44}. This proofs that we can derive the final equation \ref{eq:eq50} without assuming independence of the observed distributions.

\section{EXPRESSION IN CONVOLUTION ALGEBRA}
\label{sec:convalgebraMV}
 
In order to be able to efficiently compute equation \ref{eq:eq50} the integrals need to be expressed as convolutions, which can be computed in \emph{Fourier Space} using the \emph{Convolution Theorem}. Expressing equation \ref{eq:eq50} in convolution algebra (see also equation~\ref{eq:eq1}) requires two assumptions. Firstly, we assume the point spread functions $P(x_v|\xi)$ to be constant for every location in space. Secondly, we assume that the different coordinate systems $\xi$ and $x_1 \, ... \, x_N$ are identical, i.e. they are related by an identity transformation. We can assume that, as prior to the deconvolution the datasets have been aligned using the bead-based registration algorithm\cite{Preibisch2010}. The reformulation yields
\begin{equation}
\label{eq:eq51}
 \psi^{r+1} = \psi^r \displaystyle\prod_{v \in V} \frac{ \phi_v }{\psi^r \ast P_v} \ast P_v^{*}
\end{equation}
where $\ast$ refers to the convolution operator, $\cdot$ and $\prod_{}$ to scalar multiplication, $-$ to scalar division and
\begin{eqnarray}
 P_v & \equiv & P(x_v|\xi) \\ 
 \phi_v & \equiv & \phi_v(x_v)  \\ 
 \psi^r & \equiv & \psi^r(\xi)
\end{eqnarray}
Note that $P^{*}_v$ refers to the mirrored version of kernel $P_v$ (see section \ref{sec:mirroredkernels} for the explanation). 

\section{EFFICIENT BAYESIAN-BASED MULTI-VIEW DECONVOLUTION}
\label{sec:efficientmv}
Section \ref{sec:proofmv} shows that the derivation of bayesian-based multi-view deconvolution does not require the assumption that the observed distributions (views) are independent. We want to take advantage of that and incorporate the relationship between them into the deconvolution process to reduce convergence time. In order to express these dependencies we need to understand and model the conditional probabilities $P(x_w|x_v)$ describing how one view $\phi_w(x_w)$ depends on another view $\phi_v(x_v)$.

\subsection{MODELING CONDITIONAL PROBABILITIES}
\label{sec:conditionalestimation}
Let us assume that we made an observation $x_v=x_v'$ (see also \fig \ref{fig:01}b). The 'inverse' point spread function $Q(\xi|x_v)$ defines a probability for each location of the underlying distribution that it caused the event $\xi=\xi'$ that lead to this observation. Based on this probability distribution, the point spread function of any other observation $P(x_w|\xi)$ can be used to consecutively assign a probability to every of its locations defining how probable it is to expect an observation $x_w=x_w'$ corresponding to $x_v=x_v'$.  Assuming the point spread function $P(x_w|\xi)$ is known, this illustrates that we are able to estimate the conditional probability $P(x_w|x_v=x_v')$ for every location $x_w=x_w'$ as well as we can estimate the 'inverse' point spread function $Q(\xi|x_v)$.

However, we want to be able to compute an entire 'virtual' distribution, which is based on not only one singluar event $x_v=x_v'$, but an entire observed distribution $\phi_v(x_v)$. Such a 'virtual' distribution is solely based on the conditional probabilities $P(x_w|x_v)$ and summarizes our knowledge about a distribution $\phi_w(x_w)$ by just observing $\phi_v(x_v)$ and knowing the (inverse) point spread functions $Q(\xi|x_v)$ and $P(x_w|\xi)$. We denote a 'virtual' distribution $\phi_v^{V_w}(x_w)$; the subscript $v$ denotes the observed distribution it is based on, $w$ defines the distribution that is estimated and $V$ labels it as 'virtual' distribution.

The derviation of the formulation for a 'virtual' distribution is based on equations \ref{eq:eq1} and \ref{eq:eq11}. The 'inverse' point spread function $Q(\xi|x_w)$ relates $\phi_v(x_v)$ to the underlying signal distribution $\psi(\xi)$, and the point spread function $P(x_w|\xi)$ consecutively relates it to the conditionally dependent signal distribution $\phi_w(x_w)$ (see also \fig \ref{fig:01}b)
\begin{eqnarray}
\label{eq:eq150}
 \psi(\xi) & = & \int_{x_v} \phi_v(x_v) Q(\xi|x_v) dx_v \\
\label{eq:eq151}
 \phi_w(x_w) & = & \int_{\xi} \psi(\xi) P(x_w|\xi) d{\xi}
\end{eqnarray}
Substituting equation \ref{eq:eq150} in equation \ref{eq:eq151} yields
\begin{equation}
\label{eq:eq153}
 \phi_w(x_w) = \int_{\xi} \int_{x_v} \phi_v(x_v) Q(\xi|x_v) dx_v P(x_w|\xi) d{\xi} 
\end{equation}
As discussed in sections \ref{sec:singleview} and \ref{sec:multiview}, we cannot use $Q(\xi|x_v)$ directly to compute $\psi(\xi)$.  Using \emph{Bayes' theorem} it can be rewritten as
\begin{equation}
\label{eq:eq154}
 Q(\xi|x_v) = \frac{\psi(\xi) P(x_v|\xi)}{\phi_v(x_v)}
\end{equation}
Assuming $\phi_v(x_v)$ and $\psi(\xi)$ constant (or rather identical) simplifies equation \ref{eq:eq154} to
\begin{equation}
\label{eq:eq155}
 Q(\xi|x_v) = P(x_v|\xi)
\end{equation}
This assumption reflects that initially we do not have any prior knowledge of $Q(\xi|x_v)$ and therefore need to set it equal to the PSF $P(x_v|\xi)$, which states the worst-case scenario. In other words, the PSF constitutes an \emph{upper bound} for all possible locations of the underlying distribution $\psi(\xi)$ that could contribute to the observed distribution given an observation at a specific location $x_v=x_v'$. Thus, this assumption renders the estimate $\phi_v^{V_w}(x_w)$ less precise (equation \ref{eq:eq157} and main text figure 1c), while not omitting any of the possible solutions. Note that it would be possible to improve the guess of $Q(\xi|x_v)$ after every iteration. However, it would require a convolution with a different PSF at every location, which is currently computationally not feasible and is therefore omitted.
Replacing equation \ref{eq:eq155} in equation \ref{eq:eq153} yields
\begin{equation}
\label{eq:eq157}
 \phi_w(x_w) \approx \phi_v^{V_w}(x_w) = \int_{\xi} \int_{x_v} \phi_v(x_v) P(x_v|\xi) dx_v P(x_w|\xi) d{\xi} 
\end{equation}
Equation \ref{eq:eq157} enables the estimation of entire 'virtual' distributions $\phi_v^{V_w}(x_w)$, see and main text figure 1c for a visualization. These 'virtual' distributions constitute an \emph{upper boundary} describing how a distribution $\phi_w(x_w) \approx \phi_v^{V_w}(x_w)$ could look like while only knowing $\phi_v(x_v)$ and the two PSF's $P(x_v|\xi)$ and $P(x_w|\xi)$. We denote it \emph{upper boundary} as it describes the combination of all possiblities of how a observed distribution $\phi_w(x_w)$ can look like.

\subsection{INCORPORATING VIRTUAL VIEWS INTO THE DECONVOLUTION SCHEME}

In order to incorporate this knowledge into the deconvolution process (equation \ref{eq:eq50}), we perform updates not only based on the observed distributions $\phi_v(x_v)$ but also all possible virtual distributions $\phi_v^{V_w}(x_w)$ as modelled by equation \ref{eq:eq157} and shown in and main text figure 1c.  Based on all observed distributions
\begin{equation}
 V=\{ \phi_1(x_1), \phi_2(x_2), ... \phi_N(x_N) \}
\end{equation}
we can estimate the following 'virtual' distributions
\begin{equation}
  W =\{ \phi_1^{V_2}(x_2), \phi_1^{V_3}(x_3), ... , \phi_1^{V_N}(x_N), \phi_2^{V_1}(x_1), \phi_2^{V_3}(x_3), ... , \phi_2^{V_N}(x_N), ... ,
      \phi_N^{V_{N-1}}(x_{N-1}) \} \\
\end{equation}
where
\begin{equation}
  |W| = (N-1)^N ~ : ~ N=|V|
\end{equation}
Note that if only one input view exists, $W = \emptyset$. We define subsets $W_v \subseteq W$, which depend on specific observed distributions $\phi_v(x_v)$ as follows
\begin{eqnarray}
 W_1 & = & \{ \phi_1^{V_2}(x_2), \phi_1^{V_3}(x_3), ... , \phi_1^{V_N}(x_N) \} \\
 W_2 & = & \{ \phi_2^{V_1}(x_1), \phi_2^{V_3}(x_3), ... , \phi_2^{V_N}(x_N) \} \\
 & ... & \\
 W_N & = & \{ \phi_N^{V_1}(x_1), \phi_N^{V_2}(x_2), ... , \phi_N^{V_{N-1}}(x_{N-1}) \}
\end{eqnarray}
where
\begin{equation}
 W = \bigcup_{v \in V} W_v 
\end{equation}
Incorporating the virtual distributions into the multi-view deconvolution (equation \ref{eq:eq50}) yields
\begin{equation}
\label{eq:eq200}
 \psi^{r+1}(\xi) = \psi^r(\xi) \displaystyle\prod_{v \in V} \int_{x_v} \frac{ \phi_v(x_v) }{\int_\xi \psi^r(\xi) P(x_v|\xi) d\xi} P(x_v|\xi) dx_v
                   \displaystyle\prod_{w \in W_v} \int_{x_w} \frac{ \phi_v^{V_w}(x_w) }{\int_\xi \psi^r(\xi) P(x_w|\xi) d\xi} P(x_w|\xi) dx_w
\end{equation}
This formulation is simply a combination of observed and 'virtual' distributions, which does not yield any advantages in terms of computational complexity yet. During the following steps we will show that using a single assumption we are able to combine the update steps of the observed and 'virtual' distributions into one single update step for each observed distribution.

For simplicity we focus on one oberved distribution $\phi_v(x_v)$ and its corresponding subset of 'virtual' distributions $W_v$. Note that the following assumptions and simplifications apply to all subsets individually.
\begin{equation}
\label{eq:eq202}
 \int_{x_v} \frac{ \phi_v(x_v) }{\int_\xi \psi^r(\xi) P(x_v|\xi) d\xi} P(x_v|\xi) dx_v
 \displaystyle\prod_{w \in W_v} \int_{x_w} \frac{ \phi_v^{V_w}(x_w) }{\int_\xi \psi^r(\xi) P(x_w|\xi) d\xi} P(x_w|\xi) dx_w
\end{equation}
First, the 'virtual' distributions $\phi_v^{V_w}(x_w)$ are replaced with equation \ref{eq:eq157} which yields
\begin{equation}
\label{eq:eq204}
 \int_{x_v} \frac{ \phi_v(x_v) }{\int_\xi \psi^r(\xi) P(x_v|\xi) d\xi} P(x_v|\xi) dx_v
 \displaystyle\prod_{w \in W_v} \int_{x_w} \frac{ \int_{\xi} \int_{x_v} \phi_v(x_v) P(x_v|\xi) dx_v P(x_w|\xi) d{\xi} }{\int_\xi \psi^r(\xi) P(x_w|\xi) d\xi} P(x_w|\xi) dx_w
\end{equation}
Note that $\int_\xi \psi^r(\xi) P(x_w|\xi) d\xi$ corresponds to our current guess of the observed distribution $\phi_w(x_w)$, which is based on the current guess of the underlying distribution $\psi^r(\xi)$ and the point spread function $P(x_w|\xi)$ (equation \ref{eq:eq1}). In order to transform it into a 'virtually' observed distribution compatible with $\phi_v^{V_w}(x_w)$, we also apply equation~\ref{eq:eq157}, i.e. we compute it from the current guess of the observed distribution $\phi_v(x_v)$ yielding
\begin{equation}
\label{eq:eq206}
 \int_{x_v} \frac{ \phi_v(x_v) }{\int_\xi \psi^r(\xi) P(x_v|\xi) d\xi} P(x_v|\xi) dx_v
 \displaystyle\prod_{w \in W_v} \int_{x_w} 
 \frac
 {\int_{\xi} \int_{x_v} \phi_v(x_v) P(x_v|\xi) dx_v P(x_w|\xi) d{\xi} }
 {\int_{\xi} \int_{x_v} \int_\xi \psi^r(\xi) P(x_v|\xi) d\xi P(x_v|\xi) dx_v P(x_w|\xi) d{\xi}}
 P(x_w|\xi) dx_w
\end{equation}
To better illustrate the final simplifications we transform equation \ref{eq:eq206} into convolution algebra (section \ref{sec:convalgebraMV}). The reformulation yields
\begin{equation}
\label{eq:eq208}
 \frac{ \phi_v }{\psi^r \ast P_v} \ast P^{*}_v
 \displaystyle\prod_{w \in W_v} 
 \frac
 { \phi_v \ast \overbracket[0.5pt]{P_v^{*} \ast P_w} }
 { \psi^r \ast P_v \ast \underbracket[0.5pt]{P^{*}_v \ast P_w} } \ast P_w^{*}
\end{equation}
Additional simplification of equation \ref{eq:eq208} requires an assumption in convolution algebra that we incorporate twice. Given three functions $f$, $g$ and $h$ we assume
\begin{equation}
\label{eq:eq210}
 ( f \ast g ) \cdot ( f \ast h ) \approx f \ast ( g \cdot h ) 
\end{equation}
We illustrate in \fig \ref{fig:assumptionConv} on a one-dimensional and two-dimensional example that for \emph{Gaussian}-like distributions this assumption may hold true after normalization of both sides of the equation. Note that the measured PSF's usually resemble a distribution similiar to a gaussian (\fig \ref{fig:viewsimages} and \ref{fig:lightsheet}). 

The numerator and the denominator of the ratio of the 'virtual' distribution in equation \ref{eq:eq208} both contain two consecutive convolutions with $P_v^{*}$ and $P_w$ as indicated by brackets. Based on equation \ref{eq:eq210} we assume
\begin{equation}
 \label{eq:eq212}
 \frac{(g \ast f)}{(h \ast f)} \approx \left( \frac{g}{h} \right) \ast f
\end{equation}
where
\begin{eqnarray}
  f & \equiv & P_v^{*} \ast P_w\\
  g & \equiv & \phi_v \\
  h & \equiv & \psi^r \ast P_v
\end{eqnarray}
and
\begin{equation}
 \label{eq:eq214a}
 \frac{(g \ast f)}{(h \ast f)} = (g \ast f) \cdot \frac{1}{(h \ast f)} 
                               = (g \ast f) \cdot \left(\frac{1}{h} \ast f \right) 
                               = \overbrace{(f \ast g) \cdot \left(f \ast \frac{1}{h} \right) 
			       \approx f \ast \left(g \cdot \frac{1}{h} \right)}^{equation~\ref{eq:eq210}}
                               = f \ast \left(\frac{g}{h} \right) 
                               = \left( \frac{g}{h} \right) \ast f
\end{equation}
Based on this assumption we can rewrite equation \ref{eq:eq208} as
\begin{equation}
\label{eq:eq214b}
 \overbracket[0.5pt]{\frac{ \phi_v }{\psi^r \ast P_v}} \ast P^{*}_v
 \displaystyle\prod_{w \in W_v} 
 \overbracket[0.5pt]{\frac { \phi_v }{ \psi^r \ast P_v }} \ast P_v^{*} \ast P_w \ast P_w^{*}
\end{equation}
Note that this reformulation yields two identical terms as outlined by brackets describing the ratio between the observed distribution $\phi_v$ and its guess based on the current iteration of the deconvolved distribution $\psi^r \ast P_v$. To further simplify equation \ref{eq:eq214b} we apply the assumption (equation \ref{eq:eq210}) again where
\begin{eqnarray}
 f & \equiv & \frac{ \phi_v }{\psi^r \ast P_v} \\
 g & \equiv & P_v^{*} \\
 h & \equiv & P_v^{*} \ast P_w \ast P_w^{*}
\end{eqnarray}
which yields
\begin{equation}
\label{eq:eq218}
 \frac{ \phi_v }{\psi^r \ast P_v} \ast 
 \left( P^{*}_v \displaystyle\prod_{w \in W_v} P_v^{*} \ast P_w \ast P_w^{*} \right)
\end{equation}
In the context of all observed distributions, the final formula for efficient bayesian-based multi-view deconvolution reads
\begin{equation}
\label{eq:eq220}
 \psi^{r+1} = \psi^r \displaystyle\prod_{v \in V}
 \frac{ \phi_v }{\psi^r \ast P_v} \ast 
 \overbrace{
 \left( P^{*}_v \displaystyle\prod_{w \in W_v} P_v^{*} \ast P_w \ast P_w^{*} \right)
 }^{P_v^{compound}}
\end{equation}
Equation \ref{eq:eq220} incorporates all observerd and 'virtual' distributions that speed up convergence, but requires the exactly same number of computations as the normal multi-view deconvolution (equation \ref{eq:eq50}) derived in section \ref{sec:multiview}. The only additional computational overhead is the initial computation of the \emph{compound} kernels for each observed distribution
\begin{equation}
 \label{eq:eq222}
 P_v^{compound} = P^{*}_v \displaystyle\prod_{w \in W_v} \overbrace{P_v^{*} \ast P_w \ast P_w^{*}}^{P_{w_v}^{compound}}
\end{equation}
The compound kernel for a specific observed distribution $\phi_v$ is computed by scalar multiplication of its mirrored point spread function $P_v^{*}$ with all 'virtual' compound kernels $P_{w_v}^{compound}$ based on the corresponding 'virtual' distributions $\phi_v^{V_w}(x_w) \in W_v$. All individual 'virtual' compound kernels are computed by convolving $P_v^{*}$ with $P_w$ and sequentially with $P_w^{*}$. For most multi-view deconvolution scenarios the computational effort for the pre-computation of the compound kernels can be neglected as the PSF's a very small compared to the images and they need to be computed only once.

\section{AD-HOC OPTIMIZATIONS OF THE EFFICIENT BAYESIAN-BASED MULTI-VIEW DECONVOLUTION}
\label{sec:simplifications}

The efficient bayesian-based multi-view deconvolution derived in section \ref{sec:efficientmv} offers possibilites for optimizations as the assumption underlying the estimation of the conditional probabilities (section \ref{sec:conditionalestimation}) results in smoothed guess of the 'virtual' distributions (and main text figure 1c).  Therefore, the core idea underlying all subsequently presented alterations of equation \ref{eq:eq220} is to change how the 'virtual' distributions are computed. Due to the optimizations introduced in the last section, this translates to modification of the 'virtual' compound kernels
\begin{equation}
 P_{w_v}^{compound} = P_v^{*} \ast P_w \ast P_w^{*}
\end{equation}
The goal is to decrease convergence time while preserving reasonable deconvolution results. This can be achieved by sharpening the 'virtual' distribution (and main text figure 1c) without omitting possible solutions or rendering them too unlikely. 

\subsection{OPTIMIZATION I - REDUCED DEPENDENCE ON VIRTUALIZED VIEW}
\label{sec:optimization1}

The computation of the 'virtual' compound kernels contains two convolutions with the point spread function of the 'virtualized' observation, one with $P_w$ and one with $P_w^{*}$. We found that skipping the convolution with $P_w^{*}$ significantly reduces convergence time while producing almost identical results even in the presence of noise (section \ref{sec:benchmark} and \ref{sec:noise}). 
\begin{equation}
\label{eq:eq230}
 \psi^{r+1} = \psi^r \displaystyle\prod_{v \in V}
 \frac{ \phi_v }{\psi^r \ast P_v} \ast 
 \left( P^{*}_v \displaystyle\prod_{w \in W_v} P_v^{*} \ast P_w \right)
\end{equation}

\subsection{OPTIMIZATION II - NO DEPENDENCE ON VIRTUALIZED VIEW}
\label{sec:optimization2}

We determined empirically that further assuming $P_w$ to be constant still produces reasonable results while further reducing convergence time. We are aware that this is quite an ad-hoc assumption, but in the presence of low noise levels still yields adequate results (section \ref{sec:benchmark} and \ref{sec:noise}). 
\begin{equation}
\label{eq:eq240}
 \psi^{r+1} = \psi^r \displaystyle\prod_{v \in V}
 \frac{ \phi_v }{\psi^r \ast P_v} \ast 
 \displaystyle\prod_{v, w \in W_v} P_v^{*}
\end{equation}
Interestingly, this formulation shows some similarity to an optimization of the classic single-view Richardson-Lucy deconvolution, which incorporates an exponent into the entire 'correction~factor'\cite{Singh2008}, not only the PSF for the second convolution operation. Our derivation of the efficient bayesian-based multi-view scenario intrinsically provides the exponent for the second convolution that can be used to speed up computation and achieve reasonable results. 

\subsection{NO DEPENDENCE ON OBSERVED VIEW}

Only keeping the convolutions of the 'virtualized' observation, i.e. $P_w$ and $P_w^{*}$ yields a non-functional formulation. This is in agreement with the estimation of the conditional probabilities (section \ref{sec:conditionalestimation}).

\section{ALTERNATIVE ITERATION FOR FASTER CONVERGENCE}
\label{sec:optimizeconvergence}

To further optimize convergence time we investigated the equations \ref{eq:eq50} and \ref{eq:eq220} in detail. Both multi-view deconvolution formulas evaluate all views in order to compute one single update step of $\psi(\xi)$. It was already noted that in both cases each update step is simply the \textbf{\emph{multiplication}} of all contributions from each observed distribution. This directly suggests an alternative update scheme where the individual contributions from each observed distribution are directly multiplied to update $\psi(\xi)$ in order to save computation time. In this iteration scheme, equation \ref{eq:eq220} reads as follows
\begin{eqnarray}
 \psi^{r+1} &  = & \psi^r \frac{ \phi_1 }{\psi^r \ast P_1} \ast 
  \left( P^{*}_1 \displaystyle\prod_{w \in W_1} P_1^{*} \ast P_w \ast P_w^{*} \right) \\
 \psi^{r+2} & = & \psi^{r+1} \frac{ \phi_2 }{\psi^{r+1} \ast P_2} \ast 
  \left( P^{*}_2 \displaystyle\prod_{w \in W_2} P_2^{*} \ast P_w \ast P_w^{*} \right) \\
  & ... &  \\
 \psi^{r+N} & = & \psi^{r+N-1} \frac{ \phi_N }{\psi^{r+N-1} \ast P_N} \ast 
  \left( P^{*}_N \displaystyle\prod_{w \in W_N} P_N^{*} \ast P_w \ast P_w^{*} \right) \\
\end{eqnarray}
Note that for equation \ref{eq:eq50} (equation \ref{eq:eq51}) the iterative scheme looks identical when the multiplicative part of the compound kernel is left out; it actually corresponds to the sequential application of the standard Richardson-Lucy (RL) updates (equation \ref{eq:eq15}), which corresponds to the principle of \emph{Ordered Subset Expectation Maximization}\cite{Hudson1994} (OS-EM, see section \ref{sec:osem}).

\subsection{RELATIONSHIP TO ORDERED SUBSET EXPECTION MAXIMIZATION (OS-EM)}
\label{sec:osem}
The principle of OS-EM is the sequential application of \emph{subsets} of the observed data to the underlying distribution $\psi(\xi)$ using standard Richardson-Lucy (RL) updates (equation \ref{eq:eq15}). Note that in a multi-view deconvolution scenario, each observed distribution $\phi_v(x_v)$ is equivalent to the OS-EM definition of a \emph{balanced subset} as all elements of the underlying distribution are updated for each $\phi_v(x_v)$. 
\begin{eqnarray}
 \psi^{r+1}(\xi) & = & \psi^r(\xi) \int_{x_1} \frac{ \phi_1(x_1) }{\int_\xi \psi^r(\xi) P(x_1|\xi) d\xi} P(x_1|\xi) dx_1\\
 \psi^{r+2}(\xi) & = & \psi^{r+1}(\xi) \int_{x_2} \frac{ \phi_2(x_2) }{\int_\xi \psi^{r+1}(\xi) P(x_2|\xi) d\xi} P(x_2|\xi) dx_2\\
  & ... &  \\
 \psi^{r+N}(\xi) & = & \psi^{r+N-1}(\xi) \displaystyle\int_{x_N} \frac{ \phi_N(x_N) }{\int_\xi \psi^{r+N-1}(\xi) P(x_N|\xi) d\xi} P(x_N|\xi) dx_N 
\end{eqnarray}
As pointed out in the main text, the obvious relationship to OS-EM is that the sequential application of RL-updates is directly suggested by our multiplicative derivation (equation \ref{eq:eq50}), compared to the additive EM derivation\cite{Shepp1982}.

\section{BENCHMARKS}
\label{sec:benchmark}

We compare the performance of our new derivations against classical multi-view deconvolution (section \ref{sec:benchmarkclassic}) and against other optimized multi-view deconvolution schemes (section \ref{sec:benchmarkoptimized}) using ground truth images. Subsequently, we investigate the general image quality (section \ref{sec:generalimagequality}), the dependence on the PSF's (section \ref{sec:dependenceonPSFs}), analyze the effect of noise and regularization (section \ref{sec:noise}) and show the result of imperfect point spread functions (section \ref{sec:imperfectPSFs}).

The iteration behaviour of the deconvolution depends on the image content and the shape of the PSF (\fig \ref{fig:views}d). In order to make the simulations relatively realistic for microscopic multi-view acquisitions, we chose as ground truth image one plane of a SPIM acquistion of a \emph{Drosophila} embryo expressing His-YFP in all cells (\fig \ref{fig:viewsimages}a,e) that we blur with a maximum intensity projection of a PSF in axial direction (xz), extracted from an actual SPIM acquistion (\fig \ref{fig:viewsimages}e).

\subsection{CONVERGENCE TIME, NUMBER OF ITERATIONS \& UPDATES COMPARED TO CLASSICAL MULTI-VIEW DECONVOLUTION}
\label{sec:benchmarkclassic}

First, we compare the performance of the efficient bayesian-based multi-view deconvolution (section \ref{sec:efficientmv}, equation \ref{eq:eq220}) and its optimizations I \& II (sections \ref{sec:optimization1} and \ref{sec:optimization2}, equations \ref{eq:eq230} and \ref{eq:eq240}) against the Bayesian-based derivation (sections \ref{sec:multiview} and \ref{sec:proofmv}, equations \ref{eq:eq50} and \ref{eq:eq51}) and the original Maximum-Likelihood Expectation-Maximization derviation\cite{Shepp1982} for combined (sections \ref{sec:multiview} -- \ref{sec:simplifications}) and sequential (section \ref{sec:optimizeconvergence}, OSEM\cite{Hudson1994}) updates of the underlying distribution.

\Fig \ref{fig:views}a-c illustrate computation time, number of iterations and number of updates of the underlying distribution that are required by the different derivations to converge to a point, where they achieve exactly the same average difference between the deconvolved image and the ground truth. Detailed parts of the ground truth, PSFs, input images and results used for \fig \ref{fig:views} are exemplarily pictured for 4 views in \fig \ref{fig:viewsimages}e, illustrating that all algorithms actually converge to the same result. The entire ground truth picture is shown in \fig \ref{fig:viewsimages}a, the deconvolution result as achieved in the benchmarks is shown in \fig \ref{fig:viewsimages}b. 

\Fig \ref{fig:views}a shows that our efficient Bayesian-based deconvolution (equation \ref{eq:eq220}) outperforms the Bayesian-based deconvolution (equation \ref{eq:eq50}) by a factor of around 1.5--2.5, depending on the number of views involved. Optimization I is faster by a factor of 2--4, optimization II by a factor of 3--8. Sequential updates (OSEM) pictured in red additionally speed up the computation by a factor of approximately \emph{n}, where \emph{n} describes the number of views involved. This additional multiplicative speed-up is independent of the derivation used. \Fig \ref{fig:views}b illustrates that our Bayesian-based deconvolution behaves very similar to the Maximum-Likelihood Expectation-Maximization method\cite{Shepp1982}. Note the logarithmic scale of all y-axes in \fig \ref{fig:views}.

It is striking that for combined updates (black) the computation time first decreases, but quickly starts to increase with the number of views involved. In contrast, for sequential updates (OSEM) the computation time descreases and then plateaus. The increase in computation time becomes clear when investigating the required number of iterations\footnote{Note that we consider one iteration completed when all views contributed to update the underlying distribution once. In the case of combined updates this refers to one update of the underlying distribution, in case of sequential updates this refers to \emph{n} updates.} (\fig \ref{fig:views}b). The number of iteration for combined updates (black) almost plateaus at a certain level, however, with increasing number of views, the computational effort to compute one update increases linearly. This leads to an almost linear increase in convergence time with an increasing number of views when using combined updates. When using sequential updates (red), the underlying distribution is updated for each view individually, hence the number of required iterations continuously decreases and only the convergence time plateaus with an increasing number of views. \Fig \ref{fig:views}c supports this interpretation by illustrating that for each derivation the number of updates of the underlying distribution defines when the same quality of deconvolution is achieved.

In any case, having more than one view available for the deconvolution process decreases computation time and number of required updates significantly. This effect is especially prominent at a low number of views. For example adding a second view decreases the computation time in average 45--fold, a third view still on average another 1.5--fold. 

One can argue that using combined update steps allows better parallelization of the code as all view contributions can be computed at the same time, whereas sequential updating requires to compute one view after the other. In practice, computing the update step for an individual view is already almost perfectly multi-threadable. It requires two convolutions computed in Fourier space and several per-pixel operations. Even when several GPU's are available it can be parallelized as it can be split into blocks. Using sequential updates additionally offers the advantage that the memory required for the computation is significantly reduced.

\subsection{CONVERGENCE TIME, NUMBER OF ITERATIONS \& UPDATES COMPARED TO OTHER OPTIMIZED MULTI-VIEW DECONVOLUTION SCHEMES}
\label{sec:benchmarkoptimized}

Previously, other efficient methods for optimized multi-view deconvolution have been proposed. We compare our methods against Scaled-Gradient-Projection (SGP)\cite{Bonetti2009}, Ordered Subset Expectation Maximization (OSEM)\cite{Hudson1994} and Maximum a posteriori with Gaussian Noise (MAPG)\cite{Verveer2007}. Again, we let the algorithms converge until they achieve the same average difference of 0.07 to the known ground truth image.

In order to compare the different algorithms we re-implemented MAPG in Java, based on the Python source code kindly provided by Dr. Peter Verveer.  In order to compare to SGP, we downloaded the IDL source code. In order to allow a reasonable convergence, it was necessary to add a constant background to the images before processing them with SGP. It was also necessary to change the size of the input data to a square, even dimension (614$\times$614) to not introduce a shift of one pixel in the deconvolved data by the IDL code. OSEM is identical to our sequential updates and therefore requires no additional implementation. Main text figure 1f illustrates that our Java implementation (Bayesian-based + OSEM) and the IDL OSEM implementation require almost the identical amount of iterations.

Regarding the number of iterations (main text figure 1f), the efficient Bayesian-based deconvolution and Optimization I perform better compared to all other efficient methods for more than 4 overlapping views, Optimization II already for more than 2 views. For example at 7 views, where OSEM (50), SGP (53) and MAPG (44) need around 50 iterations, the efficient Bayesian-based deconvolution requires 23 iteration, Optimization I 17 iterations, and Optimization II 7 iteration in order to converge to the same result.

Concerning computation time (main text figure 1e), any algorithm we implemented in Java completely outperforms any IDL implementation. For the almost identical implementation of OSEM Java is in average 8$\times$ faster than IDL on the same machine (2$\times$~Intel~Xeon E5-2680, 128 GB RAM), which slightly increases with the number of views (7.89$\times$ for 2 views, 8.7$\times$ for 11 views). Practically, Optimization II outperforms all methods, except MAPG at 2 views. At 7 views where SGP (IDL) and OSEM (IDL) require around 80 seconds to converge, MAPG converges in 4 seconds, the efficient Bayesian-based deconvolution in 5 seconds, Optimization I in 3.7 seconds and Optimization II in 1.6 seconds. 

Note that MAPG is conceptually different to all other deconvolution methods compared here. It assumes Gaussian noise and performs the deconvolution on a fused dataset, which results in a reduced reconstruction quality on real datasets (see also section \ref{sec:mapgcomparison}). It also means that its computation time is theoretically independent on the number of views, a property that is shared with the classical OSEM (\Fig \ref{fig:views}c). However, it is obvious that up to 4 views, the deconvolution performance significantly increases with an increasing number of views. We speculate that the reason for this is the coverage of frequencies in the Fourier spectrum. Each PSF view blurs the image in a certain direction, which means that certain frequencies are more preserved than others. For more than 4 views, it seems that most high frequencies are contributed by at least one of the views and therefore the performance does not increase any more for algorithms that do not take into account the relationships between the individual views. Note, that our optimized multi-view deconvolution methods still signifcantly increase their performance if more than 4 views contribute (main text figure 1d,e and \fig \ref{fig:views}a,b,c).

\Fig \ref{fig:benchmarks2}a plots the computation time versus the image size of the deconvolved image for a dataset consisting of 5 views. All methods behave more or less proportional, however, the IDL code is only able to process relatively small images.

\Fig \ref{fig:benchmarks2}b illustrates that our optimizations can theoretically also be combined with SGP, not only OSEM. The number of iterations is in average reduced 1.4-fold for the efficient Bayesian-based deconvolution, 2.5-fold for Optimization I, and 2.5-fold for Optimization II.

\subsection{VISUAL IMAGE QUALITY}
\label{sec:generalimagequality}

\Fig \ref{fig:viewsimages}c shows the result using optimization II and sequential updates (OSEM) after 14 iterations, the same quality as achieved by all algorithms as shown in \fig \ref{fig:viewsimages}e and used for the benchmarks in \fig \ref{fig:views}. In this case the quality of the deconvolved image is sufficient to separate small details like the fluorescent beads, which is not possible in the input images (\fig \ref{fig:viewsimages}e, right top). 301 iterations almost perfectly restore the image (\fig \ref{fig:viewsimages}b,d). In comparison, the Bayesian-based derivation (equation \ref{eq:eq50}) needs 301 iteration to simply arrive at the quality pictured in \fig \ref{fig:viewsimages}c,e.

\subsubsection{COMPARISON TO MAPG}
\label{sec:mapgcomparison}

\Fig \ref{fig:benchmarks2}c-h compares our fastest Optimization II to MAPG using the same 7-view acquisition of a \emph{Drosophila} embryo expressing His-YFP as in main text figure 3c-e. It shows that, despite being signifcantly faster then MAPG (main text figure 1e,f), Optimization II clearly outperforms MAPG in terms of overall image quality (\fig \ref{fig:benchmarks2}c-h). Using the same blending scheme (\fig \ref{fig:partial-overlap}) MAPG produces artifacts close to some of the nuclei (e.g. top left of \fig \ref{fig:benchmarks2}e) and enhances stripes inside the sample that arise from the beginning/ending of partially overlapping input views. Especially the lower part of \fig \ref{fig:benchmarks2}c shows reduced quality, which most likely arises from the fact that in that area one input views less contributes to the final deconvolved image (note that the 7 views are equally spaced in 45 degree steps from 0--270 degrees, however every pixel is covered by at least 2 views). Note that also Optimization II shows slightly reduced image quality in this area, but is able to compensate the reduced information content significantly better.

\subsection{GENERAL DEPENDENCE ON THE PSF's}
\label{sec:dependenceonPSFs}

For \fig \ref{fig:views}a-c the PSF's are arranged in a way so that the angular difference between them is maximal in the range from 0--180 degrees (\fig \ref{fig:viewsimages}e). \Fig \ref{fig:views}d visualizes for 4 views that the angular difference between the views significantly influences the convergence behaviour. Looking at two extreme cases explains this behaviour. In this synthetic environment a difference of 0 degrees between PSF's corresponds to 4 identical PSF's and therefore 4 identical input images. This constellation is identical to having just one view, which results in a very long convergence time (\fig \ref{fig:views}a). The same almost applies for 180 degrees as the PSF that was used is quite symmetrical. In those extreme cases our argument that we can learn something about a second view by looking at the first view (section \ref{sec:conditionalestimation}) does not hold. Therefore our efficient Bayesian-based deconvolution as well as the optimizations do not converge to the identical result and few datapoints close and equal to 0 and 180 degrees are omitted. Note that they still achieve a reasonable result, but simply cannot be plotted as this quality of reconstruction is not achieved.

In general, convergence time decreases as the level of overlap between the PSFs decreases. In case of non-isotropic, gaussian-like PSFs rotated around the center (as in multi-view microscopy), this translates to a decrease in convergence time with an increase in angular difference. From this we can derive that for overlapping multi-view acquisitions it should be advantageous to prefer an odd over an even number of equally spaced views.

\Fig \ref{fig:views}d also illustrates that convergence time significantly depends on the shape and size of the PSF. Different PSF's require different amount of iterations until they reach the same quality.  Intuitively this has to be true, as for example the most simple PSF consisting only of its central pixel does not require any deconvolution at all. Conversely, this also holds true for the images themselves; the iteration time required to reach a certain quality depends on the content. For example, the synthetic image used in Supplementary Movie 1 takes orders of magnitude longer to converge to same cross correlation of 0.99 to ground truth, compared to the image in \fig \ref{fig:viewsimages}a using the same PSF's, algorithm and iteration scheme.

\subsection{NOISE AND REGULARIZATION}
\label{sec:noise}

Although the signal-to-noise ratio is typically very high in light-sheet microscopy (see Supplementary Table \ref{tab:experiments}), it is a common problem and we therefore investigated the effect of noise on the performance of the different algorithms. As Poisson noise is the dominant source of noise in light microscopy we created our simulated input views using a Poisson process with variable SNR:
\begin{equation}
 SNR = \sqrt{N}
\end{equation}
where N is the number of photons collected. These images were then used to run the deconvolution process. The first row in \fig \ref{fig:noiseplot}a and first column in \fig \ref{fig:snr} show the resulting input data for varying noise levels. For \fig \ref{fig:snr}c we added Gaussian noise with an increasing mean to simulate the effects of Gaussian noise.

A comparison as in the previous section is unfortunately not possible as in the presence of noise none of the algorithms converges exactly towards the ground truth. Note that still very reasonable results are achieved as shown in \fig \ref{fig:noiseplot}a and \ref{fig:snr}. Therefore, we devised a different scenario to test the robustness to noise. For the case of no noise (SNR $=\infty$) we first identified the number of iterations required for each algorithm to reach the same quality (\fig \ref{fig:noiseplot}c, 1\textsuperscript{st} column). With increasing noise level we iterate the exact same number of iterations for each algorithm and analyze the output.

\Fig \ref{fig:noiseplot}a,b,c show that for the typical regime of SNR's in light sheet microscopy (see Supplementary Table \ref{tab:experiments}, estimation range from $15$ to $63$) all methods converge to visually identical results.

For low SNR's (independent of Poisson or Gaussian noise) the Bayesian-based deconvolution (equation \ref{eq:eq50}), the Maximum-Likelihood Expectation-Maximization (ML-EM) and the sequential updates (OSEM) score best with almost identical results. For Poisson noise, MAPG and Optimization II show comparable results with lower quality, Optimization I and the efficient Bayesian-based derivation lie in between. For Gaussian noise, MAPG, the Bayesian-based derivation and Optimization I produce very similar results while Optimization II shows lower quality.

To compensate for noise in the deconvolution we added the option of Tikhonov-regularization. \Fig \ref{fig:noiseplot}c illustrates the influence of the $\lambda$ parameter on the deconvolution results. \Fig \ref{fig:snr} shows corresponding images for all data points. We think that although the Tikhonov regularization slows down convergence (\fig \ref{fig:noiseplot}c), a low $\lambda$ might be a good choice even in environments of a high SNR (\fig \ref{fig:snr}).

\subsection{PSF-ESTIMATION}
\label{sec:imperfectPSFs}

Another common source of errors is an imprecise estimation of the PSF's. In the previous sections we always assumed to know the PSF exactly. In real life PSF's are either measured or theoretically computed and might therefore not precisely resemble the correct system PSF of the microscope due to misalignement, refractions, etc. 

In order to be able to estimate the effect of using imprecise PSF's for the deconvolution we randomly rotated the PSF's we used to create the input images before applying them to the deconvolution (\fig \ref{fig:psf}). We used the same scheme to analyze the results as discussed in section \ref{sec:noise}. Surprisingly, the effect on the deconvolution result is hardly noticable for all algorithms, even at an average rotation angle of 10 degrees. The deconvolved images are practically identical (therefore not shown), the maximal difference in the correlation coefficient is r=0.017. We suspect that this is a result of the almost Gaussian shape of the PSF's. Although the correct solution becomes less probable, it is still well within range.

We investigated the change of the PSF of a SPIM system that should occur due to concavity of the light sheet across the field of view.  Typical light-sheet microscopic acquisitions as shown in \fig \ref{fig:lightsheet} and \ref{fig:eggchamber} show no visible sign of change, even across the entire field of view.  Given the tolerance of the deconvolution regarding the shape of the PSF we concluded that it is not necessary to extract different PSFs at different physical locations. Note that the option to perform the deconvolution in blocks (section \ref{sec:impl}) would easily allow such an extension. We think that another real improvement in deconvolution quality could be achieved by being able to measure the correct PSF inside the sample, which could be combined with the work from Blume et al.\cite{blume2007}. Additionally to the experimental burden, it is unfortunately far from being computationally tractable.

\section{MULTI-VIEW DECONVOLUTION, RESOLUTION AND OTHER OPTIAL SECTIONING MICROSCOPY}

In order to better characterize the gain of multi-view deconvolution we performed several experiments and comparisons. In \fig \ref{fig:twophoton} we compare a multi-view acquisition done with SPIM to a single-view acquisition done by two-photon microscopy of the same sample. The fixed \emph{Drosophila} embryo stained with Sytox green was embedded into agarose and first imaged using a 20$\times$/0.5NA water dipping objective in the Zeiss SPIM prototype. After acquisition we cut the agarose and imaged the same sample with a two-photon microscope using a 20$\times$/0.8NA air objective. The datasets could be aligned using the fluorescent beads present in the SPIM and two-photon acquistion. \Fig \ref{fig:twophoton} compares the quality of the content-based multi-view fusion and multi-view deconvolution of the SPIM dataset to the two-photon stack and a Lucy-Richardson single-view deconvolution of the two-photon stack. While two-photon microscopy is able to detect more photons in the center of the embryo (\fig \ref{fig:twophoton}c), the multi-view deconvolution shows signifcantly better resolution and coverage of the sample.

\Fig \ref{fig:spinningdisc} illustrates that a multi-view deconvolution can principally be done by any optical sectioning microscope that is capable of sample rotation. We acquired a multi-view dataset using Spinning Disc Confocal microscopy and a self-build rotational device\cite{Preibisch2010}. We compare the quality of one individual input stack with the multi-view deconvolution and the single-view deconvolution of this stack. Although one view completely covers the sample, it is obvious that the multi-view deconvolution clearly improves the resolution compared to the single-view deconvolution (most obvious in \fig \ref{fig:spinningdisc}d, please zoom in to investigate details).

\Fig \ref{fig:SIM} visually compares raw light-sheet data, the result of content-based multi-view fusion and the multi-view deconvolution with Structured Illumination of DSLM data (SIM)\cite{keller2010}. It is obvious that SIM and multi-view deconvolution significantly increase contrast. We are, however, not able from the published data alone to make any statement on the increase of resolution by SIM compared to multi-view deconvolution.

To be able to quantify the gain in resolution we focused our analysis on the fluorescent beads (\fig \ref{fig:resolution}). We extracted all corresponding fluorescent beads from two input views, after multi-view fusion and after multi-view deconvolution. Comparing the input views and the multi-view fusion, it becomes apparent that the multi-view fusion reduces resolution in all dimensions except directly in the axial resolution of the input view. The multi-view deconvolution on the other hand increases resolution in all dimensions compared to the multi-view fused data. The multi-view deconvolution actually achieves almost isotropic resolution in all dimensions at least comparable to the resolution of each input stack in lateral direction.

\section{PARTIALLY OVERLAPPING DATASETS}
\label{sec:partialoverlap}

In practical multi-view deconvolution scenarios where large samples are acquired on light-sheet microscopes, it is often the case that not all views are entirely overlapping (e.g. main text figure 3c,d,e and \fig \ref{fig:SIM}, \ref{fig:openspim}, and \ref{fig:eggchamber}). The sequential update strategy (OSEM) intrinsically supports partially overlapping datasets as it allows to only update parts of the underlying distribution using subsets of the input data. It is, however, necessary to achieve a balanced update of all pixels of the underlying distribution (\fig \ref{fig:partial-overlap}a,b). To achieve that, we initially compute a weight image for each view. It consists of a blending function returning 1 in all central parts of a view; close to the boundaries weights are decreasing from 1 and 0 following a cosine function\cite{Preibisch2010}. This avoid artifacts due to hard edges arising from partial overlap (\fig \ref{fig:partial-overlap}a). 

\emph{It is important to note that for the normalizations individual weights are never increased above 1. It would otherwise lead to bigger steps than suggested by one Lucy-Richardson update step making the gradient descent of the deconvolution unstable.}

On each sequential update step the weight ($0...1$) of each each pixel of every view defines the fraction of change suggested by a Lucy-Richardson update step that will be applied to the deconvolved image. By default, the sum of weights of each individual pixel in the deconvolved image over all input view is normalized to $\leq$1 (\fig \ref{fig:partial-overlap}b--I). This provides the most uniform update of the deconvolved image possible. It is, however, identical to not using OSEM in terms of performance (\fig \ref{fig:partial-overlap}d~left) and provides no improvement over the Bayesian-based derivation. Note that it still reduces the memory requirements significantly.

In order to benefit from the OSEM speedup, it is necessary to find a reasonable number of overlapping views per pixel for a specific acquisition. It is not suggested to update every pixel with the full available weights of all input views as it leads to an uneven deconvolution, i.e. some areas will be sharper than others. We found that in most cases the minimal number of overlapping views (\fig \ref{fig:partial-overlap}b--II, c) will provide a reasonable tradeoff between speed-up and uniformity. Some areas close to the boundaries of the input views might still be less deconvolved, but only if those areas a close to the boundary of one of the input views.

In summary, the speedup achieved in partly overlapping datasets is the same as in completely overlapping datasets. However, less views overlap, which increases convergence time. Our optimizations increase the performance in any case (\fig \ref{fig:partial-overlap}d).

In order to let the user make a reasonable choice of the number of overlapping datasets we offer the safe choice of 1, the minimal number of views, the average number of views, or a user-defined value. The Fiji plugin also offers the option to output an image that shows the number of contributing view at every pixel in the deconvolved image. This offers the chance to select a reasonable value or maybe give hints in how to adjust the imaging strategy (number of views, size of stacks, etc.). Please note that also in real data, datasets are often completely overlapping (e.g. main text figure 3a,b,e,f,g and \fig \ref{fig:spinningdisc}).

\section{SIMULATION OF SPIM DATA}
\label{sec:simmv}

We simulate a three dimensional (3d) ground truth dataset that resembles a biological object such as an embryo or a spheroid (main text figure 2a, section \ref{sec:objsim}).  The simulated multi-view microscope rotates the sample around the x-axis, attenuates the signal, convolves the input, samples at lower axial resolution, and creates the final sampled intensities using a poisson process. A link to the \emph{Java source code} for the simulation and 3d volume rendering can be found in section \ref{sec:simrender}.

In order to simulate a specific input view for the multi-view deconvolution, we first rotate the ground truth image around the x-axis by $n^{\circ}$ (section \ref{sec:initialtransform}). This corresponds to the orientation in which the virtual microscope performs an acquisition. Secondly, we perform the acquisition by applying all degradations as outlined above (main text figure 2b, sections \ref{sec:attenuation}--\ref{sec:samplingandnoise}). Finally, we rotate the acquired 3d image back into the orientation of the ground truth image (section \ref{sec:mv-reg}). This corresponds to the task of multi-view registration in real multi-view datasets. Two examples of input views are pictured in main text figure 2c. 

\subsection{BIOLOGICAL OBJECT SIMULATION}
\label{sec:objsim}

We use ImgLib2\cite{PietzschAl12} to draw a 3d sphere consisting of many small 3d spheres that have random locations, size and intensity. We simulate at twice the resolution of the final ground truth image and downsample the result to avoid too artificial edges. 

\subsection{INITIAL TRANSFORMATION}
\label{sec:initialtransform}

The initial rotation around the x-axis orients the ground truth image so that the virtual microscope can perform an acquisition. However, every transformation of an image introduces artifacts due to interpolation.  While on a real microscope this initial transformation is performed physically and thus obviously does not introduce any artifacts, this initial transform is required for the simulation. To avoid that those artifacts are only present in the simulated views and not in the ground truth image (which might interfere with results), we also rotate the ground truth image by 15$^{\circ}$ around the rotation axis of the simulated multi-view microscope, i.e. all simulated input views are rotated by $(n+15)^{\circ}$ around the x-axis. The resulting ground truth image is depicted in main text figure 2a.

\subsection{SIGNAL ATTENUATION}
\label{sec:attenuation}

We simulate the signal degradation along the lightsheet using a simple physical model of light attenuation\cite{Uddin2011}. Based on an initial amount of laser power (or number of photons), the sample will absorb a certain percentage of photons at each spatial location, depending on the absorption rate $(\delta=0.01)$ and the probability density (intensity) of the ground truth image. An example of the resulting image is shown in main text figure 2b.

\subsection{CONVOLUTION}

To simulate excitation and emission PSF as well as light sheet thickness, we measured effective PSF's from fluorescent beads of real multi-view dataset taken with the Zeiss SPIM prototype and a 40x/0.8NA water dipping objective. We convolve the attenuated image with different PSF's for each view. An example of the resulting image is shown in main text figure 2b.

\subsection{SAMPLING AND NOISE} 
\label{sec:samplingandnoise}

To simulate the reduced axial resolution we sample only every third slice in axial (z) direction, but every pixel in lateral direction (xy). This corresponds to a typical multi-view acquisition (Supplementary Table \ref{tab:experiments}). The sampling process for each pixel is a poisson process, with the intensity of the convolved pixel being its average. An example of the resulting image is shown in main text figure 2b.

\subsection{MULTI-VIEW REGISTRATION}
\label{sec:mv-reg}

To align all simulated views, we first scale them to an isotropic volume and then rotate them back into the original orientation of the ground truth data. Linear interpolation is used for all transformations. 

\subsection{REMARKS ON RESULTS}

Main text figure 2e,f shows that MAPG and Optimization II outperform content-based fusion in terms of spatial resolution. However, MAPG shows strong ring-like artifacts at the outlines of the simulated spheres. Supplementary movie 3 shows additional artifical patterns produced by MAPG. Computation time is measured until the maximal cross correlation to the ground truth is achieved. Note that manual stopping of the deconvolution at earlier stages can reduce noise in the deconvolved image.

\section{IMPLEMENTATION DETAILS}

\subsection{DECONVOLUTION}
\label{sec:impl}

The multi-view deconvolution is implemented in Fiji\cite{fiji2012} using ImgLib\cite{PietzschAl12}. Performance critical tasks are the convolutions with the PSF's or the compound kernels. We implement them using Fourier convolutions and provide an alternative implementation of the Fourier convolution on the GPU. Note that it is not possible to implement the entire pipeline on the GPU due to the limited size of graphics card memory.  All significant parts of implementation including per-pixel operations, copy and paste of blocks and the Fast Fourier Transform are completely multi-threaded to allow maximal execution performance on the CPU and GPU. The source code is available on Github: \url{https://github.com/fiji/spimreconstruction}. The source code most relevant to the deconvolution can be found in the package \url{src.main.java.mpicbg.spim.postprocessing.deconvolution2}.

\subsection{SIMULATION AND RENDERING}
\label{sec:simrender}

The simulation of multi-view data (section \ref{sec:simmv}) and the 3d-rendering as shown in main text figure 2a are implemented in ImgLib2\cite{PietzschAl12}. The source code for the simulation is available on Github: \url{https://github.com/StephanPreibisch/multiview-simulation}. The source code for the 3d volume rendering can be found on Github as well \url{https://github.com/StephanPreibisch/volume-renderer}. Please note that this is a fork of the volume renderer written by Stephan Saalfeld \url{https://github.com/axtimwalde/volume-renderer}, the relevant class for rendering the sphere is \url{net.imglib2.render.volume.RenderNatureMethodsPaper.java}

\subsection{FIJI-PLUGIN}

The multi-view deconvolution is integrated into Fiji (\url{http://fiji.sc}), please make sure to update Fiji before you run the multi-view deconvolution. The typical workflow consists of three steps. 

\begin{enumerate}
 \item The first step is to run the bead-based registration\cite{Preibisch2010} on the data (\url{http://fiji.sc/SPIM_Bead_Registration}, \textcolor{blue}{Fiji -- Plugins -- SPIM Registration -- Bead-based registration}). 
 \item The second step is to perform a simple average multi-view fusion in order to define the correct bounding box on which the deconvolution should be performed (\url{http://fiji.sc/Multi-View_Fusion}, \textcolor{blue}{Fiji -- Plugins -- SPIM Registration -- Multi-view fusion}).
 \item The final step is to run the multi-view deconvolution using either the GPU or the CPU implementation (\url{http://fiji.sc/Multi-View_Deconvolution}, \textcolor{blue}{Fiji -- Plugins -- SPIM Registration -- Multi-view deconvolution}). 
\end{enumerate}

\noindent Detailed instructions on how to run the individual plugins can be found on their respective Fiji wiki pages, they are summarized on this page \url{http://fiji.sc/SPIM_Registration}. Note that due to the scripting capababilties of Fiji, the workflow can be automated and also for example be executed on a cluster (\url{http://fiji.sc/SPIM_Registration_on_cluster}).

\noindent \emph{Note: An example dataset is available for download on the Fiji page: } \url{http://fiji.sc/SPIM_Registration#Downloading_example_dataset}.

\subsection{GPU IMPLEMENTATION}

The GPU implementation based on CUDA (\url{http://www.nvidia.com/object/cuda_home_new.html}) alternatively executes the Fourier convolution on Nvidia hardware. The native code is called via \emph{Java Native Access}. The source code as well as pre-compiled libraries for CUDA5 for Windows 64bit and Linux 64bit are provided online (\url{http://fly.mpi-cbg.de/preibisch/nm/CUDA_code_conv3d.zip}). Note that for Windows the DLL has to be placed in the Fiji directory, for Linux in a subdirectory called \emph{lib/linux64} and that the current version of the Nvidia CUDA driver needs to be installed on the system.

\subsection{BULDING THE CUDA CODE}

Using the native CUDA code is unfortunately not as easy as using Fiji. If the provided pre-compiled libraries do not work, first make sure you have the current Nvidia CUDA driver (\url{https://developer.nvidia.com/cuda-downloads}) installed and the samples provided by Nvidia work. 

If Fiji still does not recognize the Nvidia CUDA capable devices, you might need to compile the CUDA source code (\url{http://fly.mpi-cbg.de/preibisch/nm/CUDA_code_conv3d.zip}). The supposedly simplest way is to use CMAKE, it is setup to compile directly. If, for some reason there are problems compiling it using CMAKE, you can try to compile it directly. Here is the command required to compile the CUDA library under linux, be sure to adapt the paths correctly.

\begin{lstlisting}[language=bash]
nvcc convolution3Dfft.cu --compiler-options '-fPIC' -shared -lcudart -lcufft 
-I/opt/cuda5/include/ -L/opt/cuda5/lib64 -lcuda -o libConvolution3D_fftCUDAlib.so
\end{lstlisting}

\section{NUCLEI-BASED ALIGNMENT OF THE C. ELEGANS L1 LARVAE}
\label{sec:l1}

In order to achieve a good deconvolution result, the individual views must be registered with very high precision. To achieve that, we match fluorescent beads that are embedded into the agarose with subpixel accuracy. However, in \emph{C. elegans} during larval stages, the cuticle itself acts as a lense refracting the light sheet, which results in a slight misalignement of data inside the specimen. We therefore apply a secondary alignment step, which identifies corresponding nuclei in between views using geometric local descriptor matching\cite{Preibisch2010}, and from that estimates an affine transformation model for each view correcting for the refraction due to the cuticle. The algorithm works similar to the bead-based registration\cite{Preibisch2010, PreibischPhD} and is implemented in Fiji as a plugin called \emph{Descriptor-based series registration} (Preibisch, unpublished software).

\section{OTHER RELATED LITERATURE}
The field of multi-view deconvolution is large and diverse; many areas of science contribute including medical science, astronomy, microscopy and the classical computer science. Within the focus of this publications it is not possible to discuss all aspects (e.g. multi-channel deconvolution). We therefore list other publications that contributed to various aspects of multi-image deconvolution\cite{Rajagopalan1998, Giannakis2000, Flusser2003, Vieilleville2011, heintzmann2002, ShawAl89,agard1984,agard1989,verveer1998,heintzmann2000,holmes1991,blume2007}.



\bibliography{../paper/report}   
\bibliographystyle{spiebib}   

\end{document}